\documentclass[final,3p,times]{elsarticle}

\usepackage{geometry}
\usepackage{amssymb}
\usepackage{amsmath}
\usepackage{graphicx}
\usepackage{subcaption}
\usepackage{soul,color,xcolor}
\definecolor{RevisionColor}{rgb}{0,0,0}
\usepackage{url}
\biboptions{numbers,sort&compress}
\makeatletter
\newcommand*{\revise}{\@ifnextchar\bgroup{\revise@}{\color{RevisionColor}}}
\DeclareRobustCommand{\revise}[1]{{\textcolor{RevisionColor}{#1}}}
\makeatother

\definecolor{RevisiontwoColor}{rgb}{0,0,0}

\makeatletter
\newcommand*{\revisetwo}{\@ifnextchar\bgroup{\revise2@}{\color{RevisiontwoColor}}}
\DeclareRobustCommand{\revisetwo}[1]{{\textcolor{RevisiontwoColor}{#1}}}
\makeatother

\definecolor{RevisionthreeColor}{rgb}{0,0,0}

\makeatletter
\newcommand*{\revisethree}{\@ifnextchar\bgroup{\revise3@}{\color{RevisionthreeColor}}}
\DeclareRobustCommand{\revisethree}[1]{{\textcolor{RevisionthreeColor}{#1}}}
\makeatother

\definecolor{RevisionfourColor}{rgb}{0,0,0}

\makeatletter
\newcommand*{\revisefour}{\@ifnextchar\bgroup{\revise4@}{\color{RevisionfourColor}}}
\DeclareRobustCommand{\revisefour}[1]{{\textcolor{RevisionfourColor}{#1}}}
\makeatother

\definecolor{RevisionfiveColor}{rgb}{0,0,0}

\makeatletter
\newcommand*{\revisefive}{\@ifnextchar\bgroup{\revise5@}{\color{RevisionfiveColor}}}
\DeclareRobustCommand{\revisefive}[1]{{\textcolor{RevisionfiveColor}{#1}}}
\makeatother

\usepackage{lineno}

\journal{}

\begin{document}

\begin{frontmatter}



\title{\revisefive{Probabilistic framework for infrastructure deformation anomaly warning using InSAR observations}}


\author[label1]{Hao Kuai} 
\author[label1]{Na Hao}
\author[label2]{Pantelis Karamitopoulos}
\author[label3]{Alice Cicirello}
\author[label1]{Giorgia Giardina}

\affiliation[label1]{organization={Department of Geoscience \& Engineering, Delft University of Technology},
            addressline={Stevinweg 1},
            city={Delft},
            postcode={2628 CN}, 
            state={Zuid-Holland},
            country={The Netherlands}}

\affiliation[label2]{organization={City of Amsterdam},
            addressline={Amstel 1},
            city={Amsterdam},
            postcode={1011 PN},
            state={Noord-Holland},
            country={The Netherlands}}

\affiliation[label3]{organization={Department of Engineering, University of Cambridge},
            addressline={Trumpington Street},
            city={Cambridge},
            postcode={CB2 1PZ}, 
            country={United Kingdom}}

\begin{abstract}
Interferometric Synthetic Aperture Radar (InSAR) offers a low-cost, wide-area solution for near continuous infrastructure monitoring. However, \revisetwo{translating InSAR observations into automated anomaly warnings remains challenging because displacements are relative}, affected by uncertainty, and acquired along line of sight (LOS). \revisefour{This paper introduces a probabilistic warning framework that accounts for these factors. LOS displacements are modelled through a Gaussian distribution. A limit state function is defined to describe the displacement exceedance, and the time-dependent probability of exceedance is evaluated. Instead of a fixed threshold, warnings are identified through abrupt increases in exceedance probability. The framework is tested using three case studies of quay walls in Amsterdam, The Netherlands, providing timely anomaly warnings and issued no warnings for the non-collapse case.} \revisetwo{Results demonstrate the potential of a probabilistic interpretation of InSAR time series for deformation anomaly warning.}
\end{abstract}



\end{frontmatter}



\section{Introduction}
\label{sec1}
Due to material ageing and severe environmental conditions, an increasing number of infrastructure assets are experiencing progressive deterioration \citep{sonyLiteratureReviewNextgeneration2019, lealfilhoAssessmentPrioritiesHandling2024}. \revisetwo{Consequently, structural health monitoring (SHM) has become essential for the monitoring of structural conditions and translation of data into warnings of potential structural failures that support intervention decisions \citep{limongelliSHMInformedManagement2020}.} \revisefour{Currently, the most commonly used SHM approaches rely on visual inspection and non-destructive testing \citep{kuaiCointegrationStrategyDamage2024}, as specified in national guidelines and standards in China \citep{ministryoftransportofthepeoplesrepublicofchinaStandardsTechnicalCondition2011}, the United States \citep{federalhighwayadministrationusaNationalBridgeInspection2022} and Italy \citep{ministryofinfrastructureandsustainablemobilityGuidelinesRiskClassification2022}.} Despite their widespread adoption, these methods are typically labour intensive and costly, which constrains their scalability and limits their effectiveness for large-scale monitoring applications \revise{\citep{leAutonomousRoboticSystem2017, salvatoreStructuralSeismicRisk2026}}.

Interferometric Synthetic Aperture Radar (InSAR) offers \revise{a valuable complement} to conventional monitoring approaches. By exploiting the phase difference between radar images acquired at different times, it enables the detection of surface displacements over large areas \revise{\citep{massonnetRadarInterferometryIts1998, gabrielMappingSmallElevation1989}}. \revise{As it does not require in-situ instrumentation and allows the simultaneous observation of multiple infrastructure assets, InSAR offers significant potential for efficient large-scale monitoring. Once displacement time series have been derived through appropriate processing, these InSAR products can support autonomous, simultaneous, and cost-effective monitoring of multiple infrastructure assets.}
To further enhance this capability, Multi-Temporal InSAR (MT-InSAR) was developed to overcome inherent limitations of conventional InSAR, including atmospheric effects and signal decorrelation \citep{ferrettiNonlinearSubsidenceRate2000, ferrettiPermanentScatterersSAR2001}. 

Using MT-InSAR, time series of satellite-acquired images can be processed to detect persistent scatterers (PSs), defined as pixels exhibiting stable backscattering characteristics over time \citep{ferrettiNonlinearSubsidenceRate2000}. By tracking these PSs, relative displacements can be estimated with millimeter-level accuracy \citep{ferrettiPermanentScatterersSAR2001, ferrettiSubmillimeterAccuracyInSAR2007}. 
\revisefour{Numerous studies have demonstrated the effectiveness of MT-InSAR-based techniques for wide-area deformation monitoring, including application to geohazards \citep{liuProminentCreepCharacteristics2024, zhuDetectionClassificationActive2026}, and different types of infrastructure, such as bridges \citep{mililloPreCollapseSpaceGeodetic2019, macchiaruloStructuralEvaluationUrban2024}, tunnels \citep{giardinaEvaluationInSARMonitoring2019, macchiaruloIntegratedInSARMonitoring2021}, buildings \citep{dicarloIntegrationMultitemporalSynthetic2021, maBuildingRiskMonitoring2022}, dams \citep{marchamalo-sacristanMTInSARDamModeling2023, tavakkoliestahbanatiPrecollapseSpaceborneDeformation2024} and quay walls \citep{luongoQuayWallsAmsterdam2025, kuaiIdentificationMechanismspecificPersistent2026}.}
While these studies demonstrate the capability of MT-InSAR to monitor infrastructure displacements, significant gaps remain regarding its transition into operational SHM. 

In the context of SHM, remote sensing techniques such as MT-InSAR do not directly measure or infer damage; instead, the acquired displacement signals must be interpreted to assess structural conditions and \revisefour{provide warnings of deformation anomaly \citep{brownjohnStructuralHealthMonitoring2006, wordenFundamentalAxiomsStructural2007}}. While the retrieval of displacement data is well established, the methods needed to convert these observations into condition assessments remain underdeveloped. Additionally, the cost-effectiveness of MT-InSAR can only be fully leveraged if these conversion methods are also automated. Otherwise, the processing of large datasets of satellite observations for individual structures introduces a significant computational bottleneck, undermining the advantages of wide-area coverage and low cost. This need for automation is becoming increasingly critical in the global context of ageing infrastructure, where the scale of assets requires automated SHM procedures \citep{liMechanicsinformedAutoencoderEnables2024}. 

Many of the existing conversion methods remain descriptive, focusing on documenting displacement patterns rather than translating these data into \revisetwo{warnings of anomalies}. For instance, while structural deformations like tunnelling-induced subsidence can be effectively tracked, the specific limit states at which such deformations necessitate maintenance \revisetwo{or further investigation} are often left undefined \citep{perissinShanghaiSubwayTunnels2012}. Furthermore, while recent advancements have attempted to bridge this gap by establishing displacement-based thresholds, these efforts remain largely deterministic and lack automation. Even when systems successfully reconstruct deformations and identify anomalies, deriving actionable maintenance information typically still relies heavily on expert interpretation \citep{farnetiMethodStructuralMonitoring2023}.

Specifically, the transition of MT-InSAR to reliable SHM remains hindered by three critical limitations. First, while MT-InSAR provides displacements along the satellite line-of-sight (LOS), effective damage detection requires measurements aligned with the structural principal axes; current LOS transformation procedures remain insufficiently automated. Second, the use of fixed thresholds as safety criteria fails to account for the stochastic nature of both MT-InSAR measurements and structural response. Third, MT-InSAR measurements are relative in time, being referenced to the first satellite acquisition: this prevents the detection of any deterioration occurring prior to the first acquisition.

\subsection{\revise{Limitations of LOS displacements unaligned with local structural axes}}
\label{sec1.1}

\begin{figure}
    \centering
\includegraphics[width=0.5\linewidth]{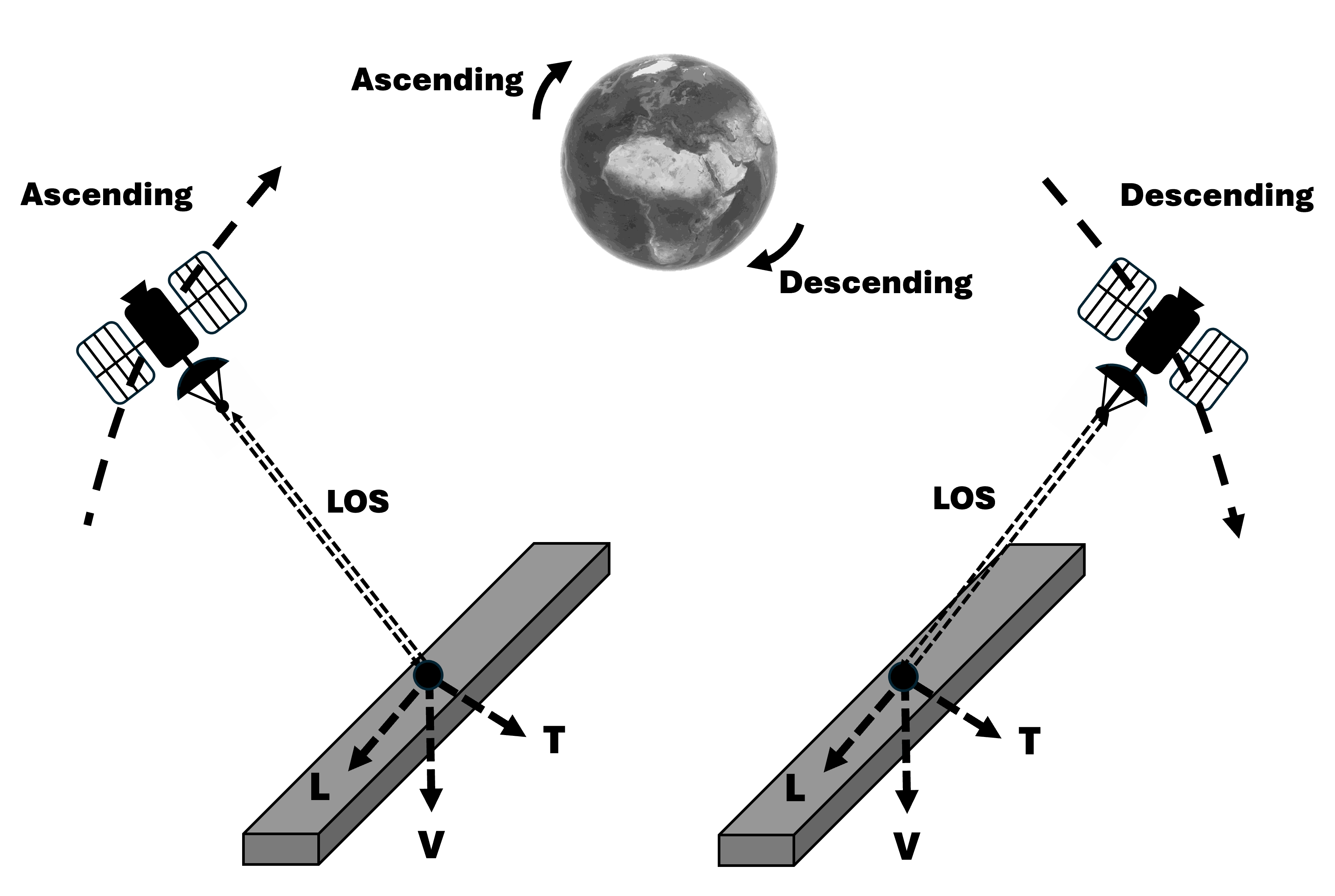}
    \caption{\revise{Geometric relationship between satellite LOS observations and the displacement vectors in the local structural coordinate system (vertical V, longitudinal L, and transversal T) for ascending and descending satellite tracks.}}
    \label{fig:displ_direct}
\end{figure}

As illustrated in Figure \ref{fig:displ_direct}, satellites follow near-polar orbits, observing target structures either from south to north (ascending track) or from north to south (descending track) \citep{macchiaruloMultiTemporalInSARTransport2023}. Regardless of the track, MT-InSAR \revise{derives} displacement only along the satellite’s LOS direction, which differs from the local structural coordinate system: vertical (V), longitudinal (L), and transversal (T). 

While LOS measurements can capture overall deformation trends, they are insufficient for accurately characterising complex structure condition since structural damage mechanisms and serviceability limits are defined with respect to specific principal axes. For example, pier settlement develops predominantly in the vertical direction, whereas LOS measurements correspond to a projection of the three-dimensional displacement field onto a single observation vector. Consequently, critical warning indicators may be obscured or misinterpreted due to the superposition of displacement components from different directions, preventing reliable damage detection unless displacements are resolved within the local structural coordinate system. Despite this limitation, LOS displacements have frequently been used directly, particularly in applications focused on identifying large deformation anomalies or in combination with complementary sensing techniques \citep{gagliardiIntegratedHealthMonitoring2022, luongoQuayWallsAmsterdam2025}. \revisefour{In other cases, LOS measurements have been projected onto a single assumed dominant direction of motion \citep{giardinaEvaluationInSARMonitoring2019, maBuildingRiskMonitoring2022}.}
For instance, when analysing ground settlement induced by tunnelling excavation, it is often reasonable to assume a dominant vertical displacement component \citep{macchiaruloIntegratedInSARMonitoring2021}. In other cases, however, no single dominant displacement component can be assumed, making such simplifications generally inapplicable.

A more advanced strategy involves the definition of so-called uniform motion regions (UMRs), within which all PSs are assumed to exhibit the same displacement pattern. Under this assumption, PSs from ascending and descending geometries can be grouped into the same UMR and treated as independent observations of the same displacement pattern from different viewing angles. By combining their multi-angle LOS displacements into a system of equations, the displacement along the local axes of the target structure can be retrieved \citep{brouwerAnalysisInSARDisplacement2021, macchiaruloStructuralEvaluationUrban2024}. 
\revise{However, current UMR definitions typically rely on prior knowledge of damage \citep{mililloPreCollapseSpaceGeodetic2019, tavakkoliestahbanatiPrecollapseSpaceborneDeformation2024} or the manual subdivision of structural components, such as dividing bridges into individual spans \citep{farnetiMethodStructuralMonitoring2023}. Some approaches further refine these regions based on specific deformation patterns, such as assigning separate regions for bridge supports and midspans to capture their distinct displacement patterns \citep{caloGISPluginAssessment2024}.} Although these approaches can provide a reasonable approximation of the true displacement of the target structure, the number and shape of UMRs are typically chosen without standardised criteria. This variability can make automation, reproducibility and scalability more challenging. Therefore, the development of a fully automated and objective approach for defining UMRs is essential to enable large-scale applications of MT-InSAR in SHM.

\subsection{Limitations of deterministic threshold-based damage detection}
\label{sec1.2}

\revisefour{Another limitation of current MT-InSAR based monitoring is the deterministic nature of most existing damage detection approaches \citep{farnetiMethodStructuralMonitoring2023, luongoQuayWallsAmsterdam2025}.} In many implementations, \revisetwo{warning of deformation anomaly} or decisions on interventions are based on pre-defined thresholds applied to monitored displacements or derived damage indicators. 
While such approaches are straightforward and operationally practical, they fail to account for uncertainties in MT-InSAR displacements \citep{kamariotisConsistentClassificationTreatment2024}. 

This limitation becomes even more critical when projecting LOS displacements onto the local coordinate system through the definition of UMRs. In existing approaches, all ascending and descending PSs are typically grouped within a given UMR, and their displacements are averaged to obtain a single representative value for projection. While this simplifies processing, using an average value can be problematic, as it is highly sensitive to outliers and measurement noise and may not accurately reflect the region's true displacements. Consequently, biases in the LOS displacement can propagate into the projected components. In this context, such deterministic treatments may lead to false warning under safe conditions or fail to provide timely warnings. Moving beyond deterministic detection is therefore essential to develop more robust and scalable monitoring strategies. 

\revisetwo{A probabilistic framework can directly address this challenge \citep{melchersStructuralReliabilityAnalysis2017}. Rather than relying solely on deterministic comparisons, these approaches treat the displacements of UMR as random variables described by appropriate probability distributions. An exceedance probability can then be evaluated using a limit state function, which identifies the boundary between the acceptable state of deformation and the threshold-exceeded state \citep{melchersStructuralReliabilityAnalysis2017}. \revisefour{Probabilistic methods, such as reliability analysis, have been successfully applied across a wide range of SHM applications \citep{dissanayakeReliabilityIndexStructural2008, fanTimevariantReliabilityPrediction2018}}. Recently, probabilistic methods have been explored for assessing geohazard-induced deformation by integrating MT-InSAR with complementary monitoring techniques \citep{caloProbabilisticbasedAssessmentSubsidence2024}. However, the application of MT-InSAR-derived displacement within a probabilistic framework for structural assessment remains insufficiently investigated.}

\subsection{Limitation of relative temporal referencing}
\label{sec1.3}

MT-InSAR derives displacement time series by processing multiple InSAR images acquired over the same area at different times \citep{ferrettiNonlinearSubsidenceRate2000, ferrettiPermanentScatterersSAR2001}. Because these displacements are estimated from phase differences between acquisitions, they are relative to the first satellite acquisition rather than absolute measurements \citep{macchiaruloMultiTemporalInSARTransport2023}. \revisefour{Widely used missions such us Sentinel-1 \citep{torresGMESSentinel1Mission2012, potinSentinel1MissionStatus2016} and TerraSAR-X \citep{werninghausTerraSARXMissionSystem2010} have been operational since the mid-2000s to 2010s.} As a result, for many infrastructures constructed prior to these missions, MT-InSAR observations do not capture the full deformation history over the structure’s service life.

\revisetwo{This limitation is particularly relevant for anomaly warning applications.} Structural damage is typically governed by the cumulative evolution of deformation or stress, meaning that pre-existing deterioration plays a critical role in current structural conditions. By treating the first acquisition as a "zero-deformation" baseline, MT-InSAR may underestimate the total cumulative displacements, especially for ageing infrastructure. Consequently, the uniform application of safety thresholds, such as the maximum allowable subsidence, may fail to capture the actual structural state. 
\revise{Detecting changes in MT-InSAR-derived displacement time series could be a potential approach to mitigate this limitation. However, how such changes can be reliably identified and interpreted for infrastructure monitoring remains insufficiently investigated.}

\subsection{\revisetwo{Probabilistic anomaly warning framework}}

\revisetwo{To overcome the aforementioned limitations, we propose a probabilistic warning framework for infrastructure deformation anomalies, designed for large-scale and automated SHM applications using MT-InSAR.} 
\revise{First}, \revisefour{an iterative} method to define UMR is introduced. The approach maximises the number of UMRs to preserve structural details, while strictly ensuring that both ascending and descending PSs are present within each UMR, which is a prerequisite for displacement projection. \revise{In contrast to existing studies that define UMRs based on known damaged areas, or manually selected structural components, the proposed method enables fully automated identification of UMRs.}

Second, the proposed framework introduces a probabilistic estimation for the displacement of UMRs. Existing approaches commonly use averaged LOS displacements within each UMR and treat the projected displacement components deterministically. \revisethree{In this framework, the LOS displacements of ascending and descending PSs within each UMR are modelled as Gaussian distributions to characterise the uncertainty due to a lack of knowledge of the fixed LOS displacement characterising a UMR.} Using the geometric relationship between LOS observations and the local coordinate system, these variables are combined to estimate the vertical displacement component. This results in a time-dependent probabilistic representation of vertical displacement.

\revisetwo{Third, a displacement-based assessment is conducted by comparing the probabilistic vertical displacement at each UMR at a given time with an acceptable displacement, yielding a time series of the probability of exceedance. This differs from conventional deterministic approaches that directly apply fixed limits to LOS displacement, projected displacement, or damage indicators. Given the relative nature of MT-InSAR displacements, predefined thresholds on exceedance probability are not reliable indicators of deformation anomaly. Instead, a change-point detection strategy is adopted. Under abnormal conditions, structural displacement becomes increasingly large in a relatively short period of time. Under the probabilistic framework, this can be described as a rapid increase in the probability of exceedance. The pruned exact linear time (PELT) algorithm is therefore employed to identify abrupt changes in the exceedance probability time series \citep{killickOptimalDetectionChangepoints2012}. \revisethree{The date on which a change is first detected is taken as the anomaly warning date.}}

\revise{The proposed framework was \revisethree{tested} using three documented quay wall cases in Amsterdam to demonstrate its practical applicability. Two of these quay walls collapsed: Grimburgwal (2020) and Entrepotdok (2017). In the third case, Houtmankade East, one part was found to be damaged and subsequently replaced, while the remaining part was undamaged. \revisethree{This case was used to provide preliminary evidence of the framework's ability to correctly distinguish damaged from undamaged sections without generating false alarms in the parts with no reported damage. }}

\section{Methodology}

\begin{figure}
    \centering
    \includegraphics[width=0.7\linewidth]{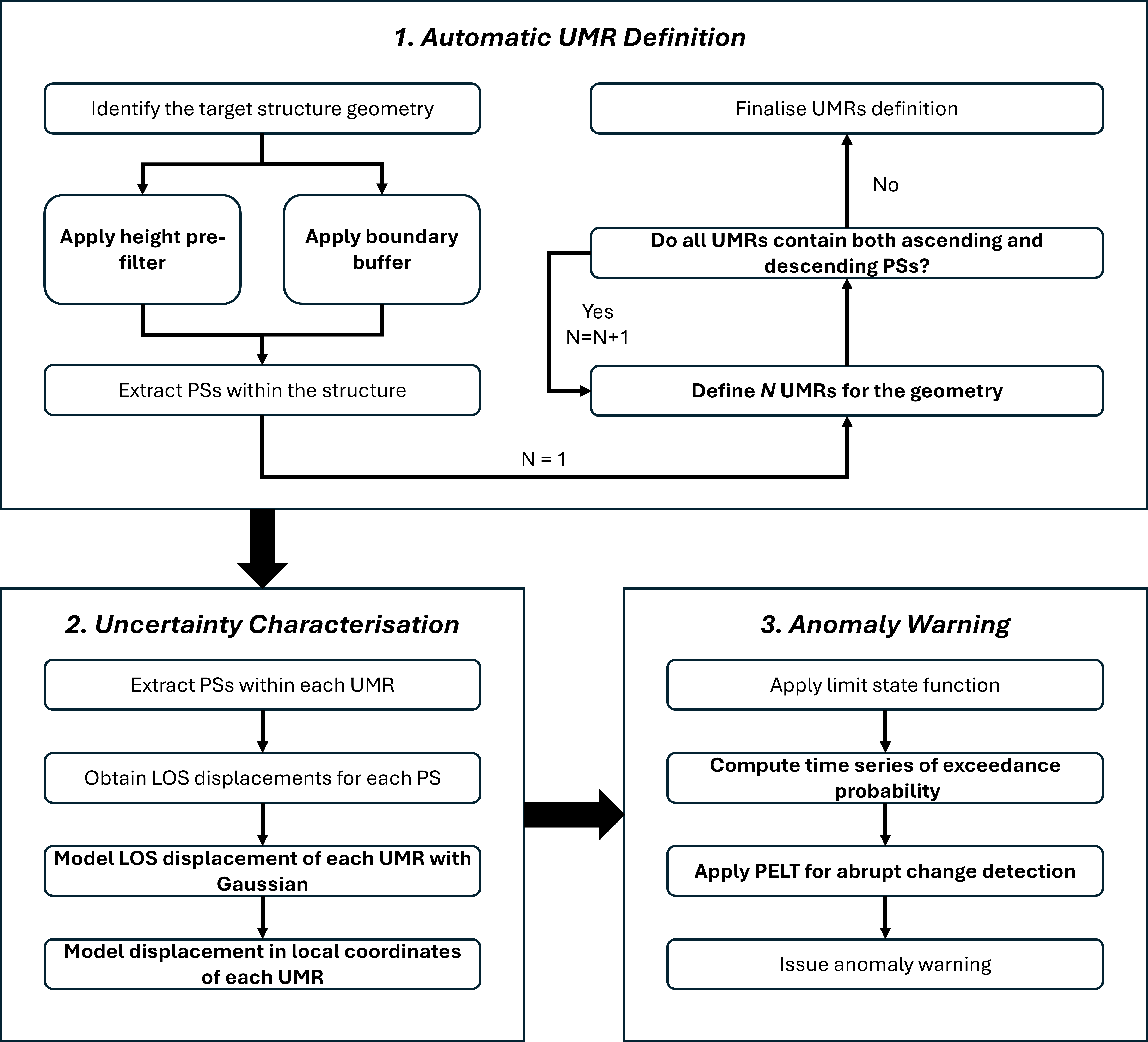}
    \caption{\revisefour{Workflow of the proposed methodology.}}
    \label{fig:flow}
\end{figure}

\revisetwo{The proposed probabilistic anomaly warning framework follows a three-stage workflow, as shown in Figure \ref{fig:flow}.} First, the geometry of the target structure and the PSs aligned with the structure were used to automatically define UMRs. Second, the LOS displacements within each UMR were modelled probabilistically, and the corresponding displacements in the local structural coordinate system were estimated. \revisetwo{Finally, after applying the limit state function, a time series of exceedance probability was obtained, and an abrupt increase detection method was applied to issue an anomaly warning.} The details of each stage are described below.

\subsection{Automated UMR definition}
\label{sec.2.1}

MT-InSAR provides LOS displacement measurements from PSs. \revise{The PS data used in this study were derived from TerraSAR-X imagery and processed by SkyGeo using their proprietary PyAntares algorithm \cite{f.j.vanleijenPersistentScattererInterferometry2014}, where PSs with a coherence value greater than 0.5 were retained. However, the proposed framework is not limited to the PS data derived from this specific MT-InSAR processing algorithm and can be applied to PS datasets produced by other processing approaches.} Initially, only the 3D geographic locations of PSs were available, without information on their alignment with specific target structures. To ensure that the selected PSs correspond to the target infrastructure, we identified the structure's geometry in the geographic information system (GIS). This geometry was defined as a boundary polygon representing the structure's outline and interior area. By overlaying the PS data with this polygon in the GIS, only the PSs located within the structural boundary were identified and retained for analysis.

Although PSs have 3D geographic locations, their positions are subject to uncertainties up to several meters \citep{yangValueCornerReflectors2019, kuaiIdentificationMechanismspecificPersistent2026}. Consequently, if the boundary polygon of the structure is used directly, some PSs that actually belong to the structure may be excluded due to these positional errors. 

\revise{To mitigate this issue, the vertical and horizontal dimensions were treated separately. For the vertical dimension, a pre-filter was applied to the PS dataset to reduce the impact of height uncertainty. For TerraSAR-X data, this uncertainty can reach approximately 8.48 meters \citep{yangValueCornerReflectors2019}. In the study area, which focuses on quay walls in the historical center of Amsterdam, such uncertainty may lead to confusion between PSs located on quay structures and those on adjacent buildings. Given that the surrounding buildings were typically at least three storey high, a conservative height range of 0 to 5 meters was adopted.}

\revise{For the horizontal dimensions, a buffer was applied to the boundary polygon to account for spatial uncertainties \citep{macchiaruloMultiTemporalInSARTransport2023, kuaiMTInSAROptimisationStructural2024}.} A buffer was defined by expanding the boundary polygon outward by a specified distance, \revise{which was determined based on the spatial resolution of the SAR images used \citep{macchiaruloMultiTemporalInSARTransport2023}. This image resolution depends on various factors, including the microwave band used and the specific SAR acquisition mode \citep{moreiraTutorialSyntheticAperture2013}.} Additionally, the structural type was also considered. For quay walls, the buffer can be further extended because the adjacent backfill soil is likely to display displacement patterns similar to those of the quay wall itself. Conversely, if the target is a bridge, further extension must be approached with caution, as the surrounding structure may exhibit entirely different displacement patterns. \revise{In this case, the buffer width $\delta_B$ was defined as the maximum value between the image resolution $\delta_I$ and twice the road width $2 \delta_R$:}

\begin{equation}
    \delta_B = \max (\delta_I, 2\delta_R)
\label{eq:buffer}
\end{equation}
\revise{The factor of two was introduced because the road is located only on one side of the quay wall, while the buffer is applied symmetrically on both sides.}

Once the buffer was defined, the corresponding PSs were extracted. These PSs provide displacement measurements in the LOS direction, as shown in Figure \ref{fig:displ_direct}. However, for SHM purposes, displacements in the local structural coordinate system (vertical (V), longitudinal (L), and transversal (T)) are more meaningful, as they relate directly to structural responses such as settlement, bending, or shear deformation. Additionally, since each structure has a unique geographical location and orientation, the LOS of a given satellite track forms a different angle with the local axes. Consequently, LOS measurements represent different combinations of displacement components, which limits their direct use in standardised and automated analyses.

\begin{figure}
    \centering
    \includegraphics[width=0.5\linewidth]{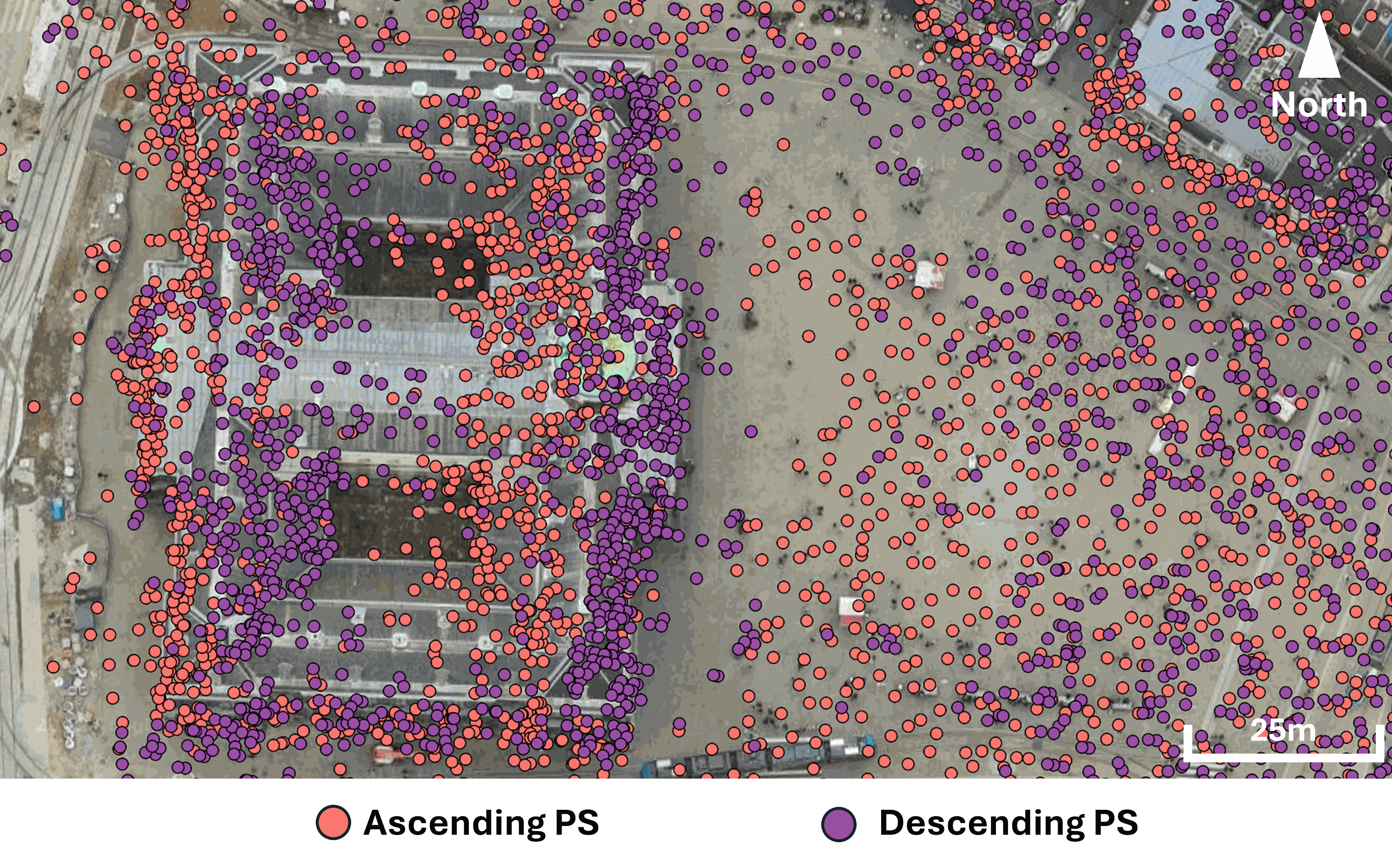}
    \caption{\revise{Spatial distribution of PSs for Dam Square and the Royal Palace in Amsterdam obtained from TerraSAR-X images.}}
    \label{fig:dedam}
\end{figure}

The projection of LOS displacement onto the local coordinate system can be achieved by combining PSs from both ascending and descending tracks, as they provide complementary viewing geometries (Figure \ref{fig:displ_direct}). However, this requires that the ascending and descending PSs represent the same physical point. In practice, such spatial coincidence rarely occurs. For example, Figure \ref{fig:dedam} shows the PSs distribution over Dam Square and the Royal Palace in Amsterdam obtained from TerraSAR-X. Due to the right-looking geometry of radar satellites, ascending tracks (moving south to north and looking east) primarily capture PSs on west-facing facades, while descending tracks (moving north to south and looking west) detect PSs on east-facing facades. Additionally, structural obstruction creates shadowed areas where no PSs are observed. Therefore, ascending and descending PSs are generally spatially separated, and directly pairing these non-coincident points for projection into the local coordinate system is not feasible. 

As introduced previously, this issue was resolved by grouping PSs into \revise{UMRs} \citep{brouwerAnalysisInSARDisplacement2021}, within which all PSs were assumed to exhibit the same displacement pattern. \revise{UMRs resolved the spatial mismatch between ascending and descending PSs and enabled their integration within a common measurement region.}
To enable efficient application of this concept to large-scale monitoring without requiring prior knowledge, such as the locations of the most severely damaged areas, we proposed an iterative and adaptive subdivision approach for defining the UMRs. This approach was governed by two criteria:

\begin{enumerate}
    \item Maximise Information: Create as many UMRs as possible to capture structural detail and improve accuracy.
    \item Ensure PSs Coverage: Guarantee the presence of both ascending and descending PSs within each UMR to allow reliable projection into the local coordinate system.
\end{enumerate}

\revise{The infrastructure polygon was initially divided into $N = 2$ UMRs, evenly along the main axis of the structure. Each UMR was then checked to ensure that it contained both ascending and descending PSs. If this criterion was satisfied for all UMRs, the entire polygon was subsequently re-divided along the same axis into $N+1$ evenly spaced UMRs to maximise the structural detail along the main axis of the structure, and the coverage test was repeated. This iterative subdivision continued until the PS coverage requirement failed in any of the resulting UMRs. The final set of UMRs was defined by the highest subdivision level that satisfied both criteria, resulting in UMRs with comparable areas.} This procedure provides an automated UMR definition suitable for large-scale SHM. 

\subsection{Uncertainty characterisation}
\label{sec:uq}
Once the UMRs were defined, LOS displacements derived from PSs were used to represent each UMR. \revise{Since PS data from ascending and descending geometries are generally acquired at different times, only the overlapping observation period was considered. Temporal interpolation was performed for both ascending and descending displacement time series to estimate displacement values at the acquisition dates of the other geometry, enabling the comparison and subsequent combination of LOS displacements from the two geometries.} However, it should be noted that the interpolated displacements, as well as the original LOS displacements, are subject to several sources of uncertainty, which need to be carefully considered in the subsequent analysis.

For a given $j$-th PS, the LOS displacement $d_{LOS,t_i}^j$ at time $t_i$ was derived from the interferometric phase difference $\Delta \phi$ between time $t_i$ and the reference time $t_0$ \citep{ferrettiPermanentScatterersSAR2001}:

\begin{equation}
    d_{LOS,t_i}^j = \frac{\lambda}{4\pi} \Delta\phi
    \label{eq:phi2los}
\end{equation}
where $\lambda$ denotes the radar wavelength.
The observed interferometric phase comprises both the displacement of structure and several non-displacement components, including atmospheric phase screens, residual orbital errors, and measurement noise \citep{ferrettiPermanentScatterersSAR2001}. 



\revisethree{Assessing the contribution of each uncertainty source is beyond the scope of this study. Instead, the variation in LOS displacements from PSs within a single UMR was used to describe the distribution of LOS displacement for that region. It is important to note that this approach describes epistemic uncertainty, reflecting a lack of knowledge about a fixed underlying displacement value which could either represent a settlement or an uplift of each UMR, rather than aleatory uncertainty caused by inherent variability of the LOS displacement. Aleatory sources of uncertainties caused by other effects, such as highly localised failure mechanism are not investigated in this study. The unknown error sources that could cause the variation from a fixed unknown value include (but are not limited to) decorrelation of the radar signal, co-registration inaccuracies, residual atmospheric phase, residual topographic phase due to DEM errors, orbital errors, other uncompensated non-displacement phase components  \citep{macchiaruloMultiTemporalInSARTransport2023}. During MT-InSAR processing, sources of errors caused by correlated components of non-displacement phase contributions, such as atmospheric delays, orbital effects, and topographic residuals, are assumed to be estimated and compensated \citep{macchiaruloMultiTemporalInSARTransport2023}. In principle, sources of epistemic uncertainty can be potentially reduced by gathering more information or data, while sources of aleatoric uncertainty are irreducible since they are related to an inherent variability. However, in this specific application, these source of uncertainties are coupled, difficult to be quantified, correctly estimated or reduced. } 

\revisethree{Consequently, the only meaningful information available is about the LOS displacements obtained at all the PSs within one UMR displaying a variability about a fixed but unknown displacement value that is caused by these (and potentially more) uncertainty sources. The shape and the parameters of the distribution representing the LOS displacements of one UMR is not known. According to the Maximum Entropy principle \citep{jaynes2003probability}, the probability density function should be chosen such that it maximises the entropy (uncertainty) without adding any unjustified structure beyond the available constraints. From the PS displacement data within a single UMR, it would be possible to evaluate the mean and the variance of the LOS displacements with a satisfactory accuracy. Consequently, the maximum entropy distribution chosen is the Gaussian distribution.} 
The validity and limitations of this approximation were further evaluated using real PS displacement data and are provided in \ref{sec:app_b}. 

The parameters of the distribution were evaluated as follows:
\begin{enumerate}
    \item For each UMR, the LOS displacements from all PSs located within it were extracted, with ascending and descending PSs treated separately due to their different LOS directions. 
    \item At each acquisition time $t_i$, the LOS displacement $d_{LOS,t_i}^j$ from the $j$-th PS were obtained. The mean (\( \mu_{t_i} \)) and variance (\( \sigma_{t_i}^2 \)), were estimated using the sample mean and unbiased sample variance \citep{casellaStatisticalInference2002}:
    \begin{equation}
        \hat{\mu}_{t_i} = \frac{1}{n} \sum_{j=1}^n d_{LOS, t_i}^j
    \label{eq:unbiased_mean}
    \end{equation}

    \begin{equation}
        \hat{\sigma}_{t_i}^2 = \frac{1}{n-1} \sum_{j=1}^n \left(d_{LOS, t_i}^j - \hat{\mu}_{t_i}\right)^2
    \label{eq:unbiased_var}
    \end{equation}
    where $n$ denotes the total number of PSs within a UMR for either the ascending or descending track, and $d_{LOS, t_i}^j$ represents the LOS displacement of the $j$-th PS at acquisition time $t_i$.
\end{enumerate}

Consequently, using Equation \ref{eq:unbiased_mean} and \ref{eq:unbiased_var}
, the characterised LOS displacement distribution at time $t_i$ for each UMR was given by the following Gaussian models: \( \mathcal{N}(\mu_\text{$d_{LOS, t_i}$, asc}, \sigma_\text{$d_{LOS, t_i}$, asc}^2) \) for ascending track and \( \mathcal{N}(\mu_\text{$d_{LOS, t_i}$, des}, \sigma_\text{$d_{LOS, t_i}$, des}^2) \) for descending track.

\begin{figure}
    \centering
    \includegraphics[width=0.6\linewidth]{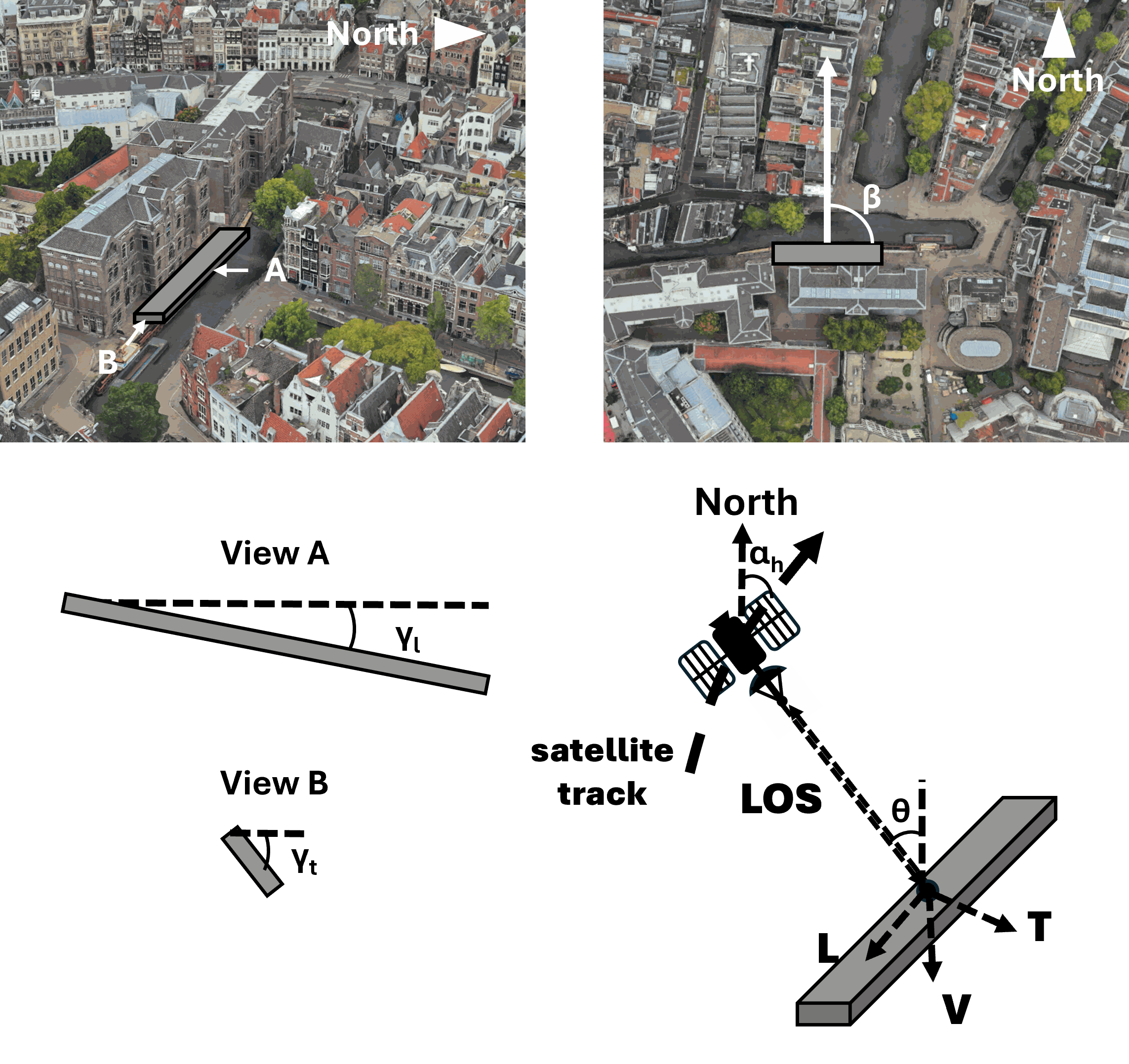}
    \caption{\revisefour{Geometric definition of angles used in the displacement projection model (Equation \ref{eq:displ_local}).}}
    \label{fig:angle}
\end{figure}

These distributions, however, describe displacements only along the LOS direction. In the context of SHM, damage mechanisms and serviceability limits are typically defined along specific principal axes, such as vertical settlement or longitudinal thermal expansion. It is therefore necessary to decompose the LOS displacements into these local components. The geometric relationship between the LOS displacement $d_{LOS,t_i}$ and its components in the transversal $d_{T,t_i}$, longitudinal $d_{L,t_i}$, and vertical $d_{V,t_i}$ directions for each UMR at time $t_i$ is expressed as follows \citep{brouwerAnalysisInSARDisplacement2021, brouwerAnalysisInsarDisplacement2021a}: 

\begin{equation}
    d_{LOS,t_i} =
    \begin{bmatrix}
        \sin\theta \, \sin\alpha_d \\ \sin\theta \, \cos\alpha_d \\ \cos\theta
    \end{bmatrix}^{T}
    R_1 R_2 R_3
    \begin{bmatrix}
        d_{T,t_i} \\
        d_{L,t_i} \\
        d_{V,t_i}
    \end{bmatrix}
\label{eq:displ_local}
\end{equation}
where $R_1$, $R_2$, $R_3$ are the rotation matrices defined as:

\begin{equation}
    R_1 = 
    \begin{bmatrix}
        \cos \beta & \sin \beta & 0 \\
        -\sin \beta & \cos \beta & 0 \\
        0 & 0 & 1 \\
    \end{bmatrix}
\end{equation}

\begin{equation}
    R_2 = 
    \begin{bmatrix}
        1 & 0 & 0 \\
        0 & \cos \gamma_l & -\sin \gamma_l \\
        0 & \sin \gamma_l & \cos \gamma_l
    \end{bmatrix}
\end{equation}

\begin{equation}
    R_3 = 
    \begin{bmatrix}
        \cos \gamma_t & 0 & \sin \gamma_t \\
        0 & 1 & 0 \\
        -\sin \gamma_t & 0 & \cos \gamma_t
    \end{bmatrix}
\end{equation}
where, $\theta$ is the satellite incidence angle; $\alpha_d$ is the azimuth of the zero-Doppler plane at the earth surface, which is approximately $90^\circ$ from heading angle $\alpha_h$; $\beta$ is the azimuth of the structure's longitudinal direction relative to the north; $\gamma_l$ is the slope in the longitudinal direction; $\gamma_t$ is the slope in the transversal direction (Figure \ref{fig:angle}).

By combining displacements derived from both ascending and descending satellite tracks, a system of two equations with three unknown displacement components was obtained:

\begin{equation}
    \begin{cases}
        d_{LOS, t_i, asc} =
    \begin{bmatrix}
        \sin\theta_{asc} \, \sin\alpha_{d, asc} \\
        \sin\theta_{asc} \, \cos\alpha_{d, asc} \\
        \cos\theta_{asc}
    \end{bmatrix}^{T}
        R_1 R_2 R_3
    \begin{bmatrix}
        d_{T,t_i} \\
        d_{L,t_i} \\
        d_{V,t_i}
    \end{bmatrix}
    \\
    \\
        d_{LOS, t_i, des} =
    \begin{bmatrix}
        \sin\theta_{des} \, \sin\alpha_{d, des} \\
        \sin\theta_{des} \, \cos\alpha_{d, des} \\
        \cos\theta_{des}
    \end{bmatrix}^{T}
        R_1 R_2 R_3
    \begin{bmatrix}
        d_{T,t_i} \\
        d_{L,t_i} \\
        d_{V,t_i}
    \end{bmatrix}
    \end{cases}
    \label{eq:func_los}
\end{equation}

\begin{figure}
    \centering \includegraphics[width=0.8\linewidth]{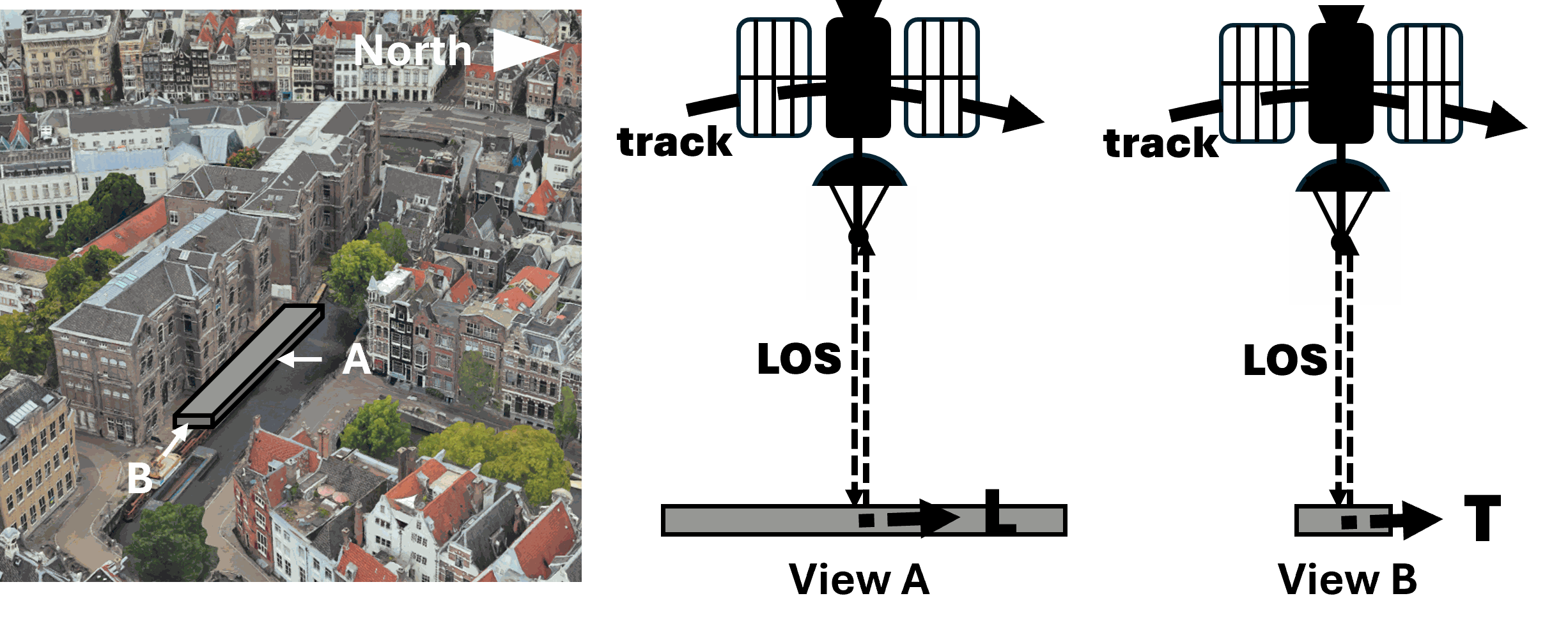}
    \caption{\revise{Relative orientation between the satellite azimuth direction and the structural axes. In view A, when the azimuth direction is aligned with the longitudinal axis of the structure, the longitudinal displacement has no component in the LOS measurement. In view B, when the azimuth direction is aligned with the transversal axis of the structure, the transversal displacement has no component in the LOS measurement.}}
    \label{fig:azi-dire}
\end{figure}

\revise{With only two equations and three unknowns, this system is mathematically underdetermined. The LOS displacement is the projection of the displacement along the structural axes onto the LOS direction. Because it is a projection, a simplifying assumption can be introduced based on the geometric relationship between the SAR viewing geometry and the structural axes to obtain a unique solution. Since SAR satellites typically operate in near-polar orbits and the LOS direction is approximately perpendicular to the satellite track, if a structure is oriented in the north–south direction, its longitudinal displacement is nearly parallel to the satellite track and thus has little contribution to the LOS displacement. Conversely, if a structure is oriented in the east–west direction, its transversal displacement aligns with the north–south direction and similarly contributes minimally to the LOS measurement \citep{farnetiMethodStructuralMonitoring2023, macchiaruloStructuralEvaluationUrban2024}. This geometric relationship is illustrated in Figure \ref{fig:azi-dire}.}

Based on this, it was assumed that the displacement component along the structural axis (longitudinal or transversal) more closely aligned with the satellite track was negligible. The specific constraint was applied as follows:

\begin{enumerate}
    \item If the angle between the structure's longitudinal axis and the satellite track was less than $45^\circ$, the longitudinal displacement was considered negligible ($d_{L,t_i} \approx 0$) in the LOS displacement.
    \item If the angle between the structure's transversal axis and the satellite track was less than $45^\circ$, the transversal displacement was considered negligible ($d_{T,t_i} \approx 0$) in the LOS displacement.
\end{enumerate}

By imposing this constraint, the system of equations is reduced to a solvable two-by-two linear system, allowing for the unique determination of the vertical displacement $d_{V,t_i}$ and one horizontal component.

Equation \ref{eq:func_los} can be reformulated in a scalar form by defining the projection coefficients $A$, $B$, and $C$ as follows:

\begin{equation}
    \begin{bmatrix}
        d_{LOS,t_i,asc} \\
        d_{LOS,t_i,des}
    \end{bmatrix}
        =
    \begin{bmatrix}
        A_a & B_a & C_a \\
        A_d & B_d & C_d
    \end{bmatrix}
    \begin{bmatrix}
        d_{T,t_i} \\
        d_{L,t_i} \\
        d_{V,t_i}
    \end{bmatrix}
\label{eq:func_los_scalar}
\end{equation}
where the detailed expressions for the coefficients are provided in \ref{sec:app_a}.

By solving Equation \ref{eq:func_los_scalar}, if \revise{the contribution of longitudinal displacement to the LOS measurement} is considered negligible ($d_{L,t_i} \approx 0$), the vertical $d_{V,t_i}$ and transversal displacements $d_{T,t_i}$ can be obtained:
\begin{equation}
    \begin{cases}
        d_{V,t_i} = \frac{A_a~d_{LOS,t_i,des}-A_d~d_{LOS,t_i,asc}}{C_d~A_a-C_a~A_d} \\
        d_{T,t_i} = \frac{C_a~d_{LOS,t_i,des}-C_d~d_{LOS,t_i,asc}}{C_a~A_d-C_d~A_a} 
    \end{cases}
    \label{eq:sl_dVdT}
\end{equation}

Since the derived expressions in Equation \ref{eq:sl_dVdT} are linear combinations of the LOS measurements, and the LOS displacements are modelled as independent Gaussian distributed variables, the resulting vertical $d_{V,t_i}$ and transversal components $d_{T,t_i}$ also follow Gaussian distributions:

\begin{equation}
    \begin{split}
    d_{V,t_i} \sim \mathcal{N}\left( 
        \frac{A_a \mu_\text{$d_{LOS, t_i,des}$} - A_d \mu_\text{$d_{LOS, t_i,asc}$}}{C_d A_a - C_a A_d}, 
        \frac{A_a^2 \sigma^2_\text{$d_{LOS, t_i,des}$} + A_d^2 \sigma^2_\text{$d_{LOS, t_i,asc}$}}{\left( C_d A_a - C_a A_d \right)^2} 
    \right) \\
    d_{T,t_i} \sim \mathcal{N}\left( 
        \frac{C_a \mu_\text{$d_{LOS, t_i,des}$} - C_d \mu_\text{$d_{LOS, t_i,asc}$}}{C_a A_d - C_d A_a}, 
        \frac{C_a^2 \sigma^2_\text{$d_{LOS, t_i,des}$} + C_d^2 \sigma^2_\text{$d_{LOS, t_i,asc}$}}{\left( C_a A_d - C_d A_a \right)^2} 
    \right)
    \end{split}
\label{eq:dVdT}
\end{equation}

Analogously, if the \revise{the contribution of transversal displacement to the LOS measurement} is considered negligible ($d_{T,t_i} \approx 0$), solving Equation \ref{eq:func_los_scalar} for vertical $d_{V,t_i}$ and longitudinal components $d_{L,t_i}$ yields:

\begin{equation}
    \begin{cases}
        d_{V,t_i} = \frac{B_a~d_{LOS,t_i,des}-B_d~d_{LOS,t_i,asc}}{C_d~B_a-C_a~B_d} \\
        d_{L,t_i} = \frac{C_a~d_{LOS,t_i,des}-C_d~d_{LOS,t_i,asc}}{C_a~B_d-C_d~B_a} 
    \end{cases}
    \label{eq:sl_dVdL}
\end{equation}

Consequently, their corresponding Gaussian distributions are determined by:

\begin{equation}
    \begin{split}
    d_{V,t_i} \sim \mathcal{N}\left( 
        \frac{B_a \mu_\text{$d_{LOS, t_i,des}$} - B_d \mu_\text{$d_{LOS, t_i,asc}$}}{C_d B_a - C_a B_d}, 
        \frac{B_a^2 \sigma^2_\text{$d_{LOS, t_i,des}$} + B_d^2 \sigma^2_\text{$d_{LOS, t_i,asc}$}}{\left( C_d B_a - C_a B_d \right)^2} 
    \right) \\
    d_{L,t_i} \sim \mathcal{N}\left( 
        \frac{C_a \mu_\text{$d_{LOS, t_i,des}$} - C_d \mu_\text{$d_{LOS, t_i,asc}$}}{C_a B_d - C_d B_a}, 
        \frac{C_a^2 \sigma^2_\text{$d_{LOS, t_i,des}$} + C_d^2 \sigma^2_\text{$d_{LOS, t_i,asc}$}}{\left( C_a B_d - C_d B_a \right)^2} 
    \right)
    \end{split}
\label{eq:dVdL}
\end{equation}

\subsection{\revisetwo{Anomaly warning}}

\revisetwo{After obtaining the displacements of the target structure in the local coordinate, a limit state function $Z$ can be introduced to describe the structural condition \citep{melchersStructuralReliabilityAnalysis2017}. }
Various limit state functions can be defined for different damage mechanisms, such as longitudinal expansion or settlement. Here, the focus was placed on vertical displacement. This choice was motivated by MT-InSAR's measurement capability and the deformation characteristics of the investigated structures. Due to the orientation between the structure's axis and the satellite track, one horizontal component could not be reliably detected by MT-InSAR, whereas the vertical deformation remains consistently observable. \revise{Beyond these measurement constraints, vertical displacement is also physically highly relevant for the quay walls investigated in this study. These quay walls are located in the Netherlands, where subsidence is an important challenge for infrastructure management. \revisetwo{Moreover, various typical failure mechanisms of quay walls, including degradation of foundation and supporting elements, can induce progressive settlement, making vertical displacement a relevant indicator for the assessment of the investigated structures \citep{ARKAmsterdamseRisicobeoordeling2024, sharmaAnalysisProceduresAccounting2024}.}}

Accordingly, within this framework, the limit state function was defined as:

\revisethree{
\begin{equation}
    Z_{t_i} = d_{V, max} - s_{t_i}~d_{V, t_i}
    \label{eq:LSF}
\end{equation}
}
where $d_{V, max}$ is the maximum vertical allowable displacement, specified by safety standards \citep{curRapport16620082008}, and $d_{V, t_i}$ is the actual time-dependent vertical displacements referenced to the first acquisition, derived from the LOS measurements. \revisethree{The parameter $s_{t_i}$ indicates the governing deformation direction, defined as:}

\revisethree{\begin{equation}
    s_{t_i} = 
    \begin{cases}
        +1, &\mu_{V,t_i} \geq 0\\
        -1, &\mu_{V,t_i} <0
    \end{cases}
\end{equation}}
\revisethree{where $\mu_{V,t_i}$ is the mean vertical displacement of the UMR at time $t_i$, calculated from Equation \ref{eq:dVdT} and \ref{eq:dVdL}.}

A sign convention was adopted in which upward vertical displacement $d_{V, t_i}$ was taken as positive. \revisethree{Under the assumption of uniform motion within each UMR, the governing deformation direction was determined by the sign of the mean vertical displacement. Uplift and subsidence were not considered distinct failure mechanisms, but rather opposite directions of vertical displacement. Accordingly, each UMR can be characterised by a single governing deformation direction at a given time, and the limit state function is evaluated exclusively for that direction. It is worth mentioning that this uniform motion assumption would not hold for highly localised failure mechanisms. For example, uneven degradation of pile foundations may cause rotation of a quay wall, resulting in uplift on one side and settlement on the other. In such cases, the mean displacement may approach zero due to the coexistence of opposite deformation directions within the same UMR, potentially masking the actual deformation pattern. In this case, the UMR definition could be refined to identify separate regions exhibiting uniform deformation directions. However, refining the number of UMR might not be possible for failure mechanisms with significant deformation gradients within a single region where very few PSs might be available.}


\revisetwo{The exceedance probability can be expressed as a bounded integral:}

\begin{equation}
    P_e(t_i) = \iint_{Z_{t_i} \leq 0} f_{d_{LOS,t_i,asc}}(s_a) f_{d_{LOS,t_i,des}}(s_d) ds_a ds_d 
    \label{eq:pf}
\end{equation}
where $f_{d_{LOS,t_i,asc}}(s_a)$ and $f_{d_{LOS,t_i,des}}(s_d)$ are the probability density functions for the ascending and descending LOS displacement at $t_i$.


\revisetwo{The functional form of Equation \ref{eq:pf} is common in reliability analysis, and its solutions has been studied extensively in the literature \citep{hasoferExactInvariantSecondMoment1974, melchersStructuralReliabilityAnalysis2017}. In particular, Equation \ref{eq:pf}, can be solved using exact analytical expressions only in low-dimensional cases with explicit limit state functions. For cases involving non-linear or implicit limit state functions or high-dimensional variable spaces, analytical solutions become intractable. In such cases, the integral can be evaluated using approximate methods like the First-Order Reliability Method or numerical approaches such as Monte Carlo simulation \citep{hasoferExactInvariantSecondMoment1974, melchersStructuralReliabilityAnalysis2017}.}


\revise{In this study, the LOS displacements for each PSs from ascending or descending track were assumed to be statistically independent and Gaussian distributed. Under this assumption, the vertical displacement $d_{V,t_i}$ also follows a Gaussian distribution (Equation \ref{eq:dVdT} or \ref{eq:dVdL}). \revisetwo{The probability of exceedance can be evaluated analytically by introducing the standard normal variable:}}

\begin{equation}
    U_{t_i} = \frac{d_{V,t_i} - \mu_{V,t_i}}{\sigma_{V,t_i}}
\end{equation}
\revise{where $\mu_{V,t_i}$ and $\sigma_{V,t_i}$ denote the mean and standard deviation of the vertical displacement at time $t_i$, respectively.}

\revisetwo{Consequently, the probability of exceedance can be analytically evaluated using the standard normal cumulative distribution function, $\Phi(\cdot)$ \citep{melchersStructuralReliabilityAnalysis2017}:}

\begin{equation}
    \begin{aligned}
        P_e(t_i) &= \max \left\{ P_e \left( d_{V,t_i} \ge d_{V, \max} \right), P_e \left( d_{V,t_i} \le -d_{V, \max} \right) \right \}\\
                 &= \max \left\{ P_e \left(U_{t_i} \ge \left(\frac{d_{V, \max}-\mu_{V, t_i}}{\sigma_{V, t_i}} \right) \right),  P_e \left(U_{t_i} \le \left(\frac{-d_{V, \max}-\mu_{V, t_i}}{\sigma_{V, t_i}} \right) \right) \right \} \\
                 &= \max \left\{ 1 - \Phi \left(\frac{d_{V, \max}-\mu_{V, t_i}}{\sigma_{V, t_i}} \right), \Phi\left(\frac{-d_{V, \max}-\mu_{V, t_i}}{\sigma_{V, t_i}} \right) \right\}
    \end{aligned}
    \label{eq:analytical_pf}
\end{equation}

\revisetwo{By computing the probability of exceedance at different times of the UMR, we constructed its temporal evolution, that is, a curve displaying how the exceedance probability evolves in time. }

\revisetwo{Traditionally, probabilistic assessments such as reliability-based analysis have adopted a fixed threshold on the reliability index or the probability of exceedance. For instance, de Gijt and Broeken \citep{degijtQuayWallsSecond2013} suggested a target reliability index of 3.3, corresponding to an exceedance probability of 0.0005, as an anomaly warning criterion for historical quay walls in Amsterdam.} However, since the acquisition period of MT-InSAR is generally much shorter than the full service life of infrastructures, MT-InSAR cannot capture the entire history of vertical displacements. Therefore, fixed-threshold approaches are less suitable when assessments rely solely on the limited temporal coverage provided by MT-InSAR.

\begin{figure}
    \centering
    \begin{subfigure}[t]{0.4\linewidth}
        \centering
        \includegraphics[width=\linewidth]{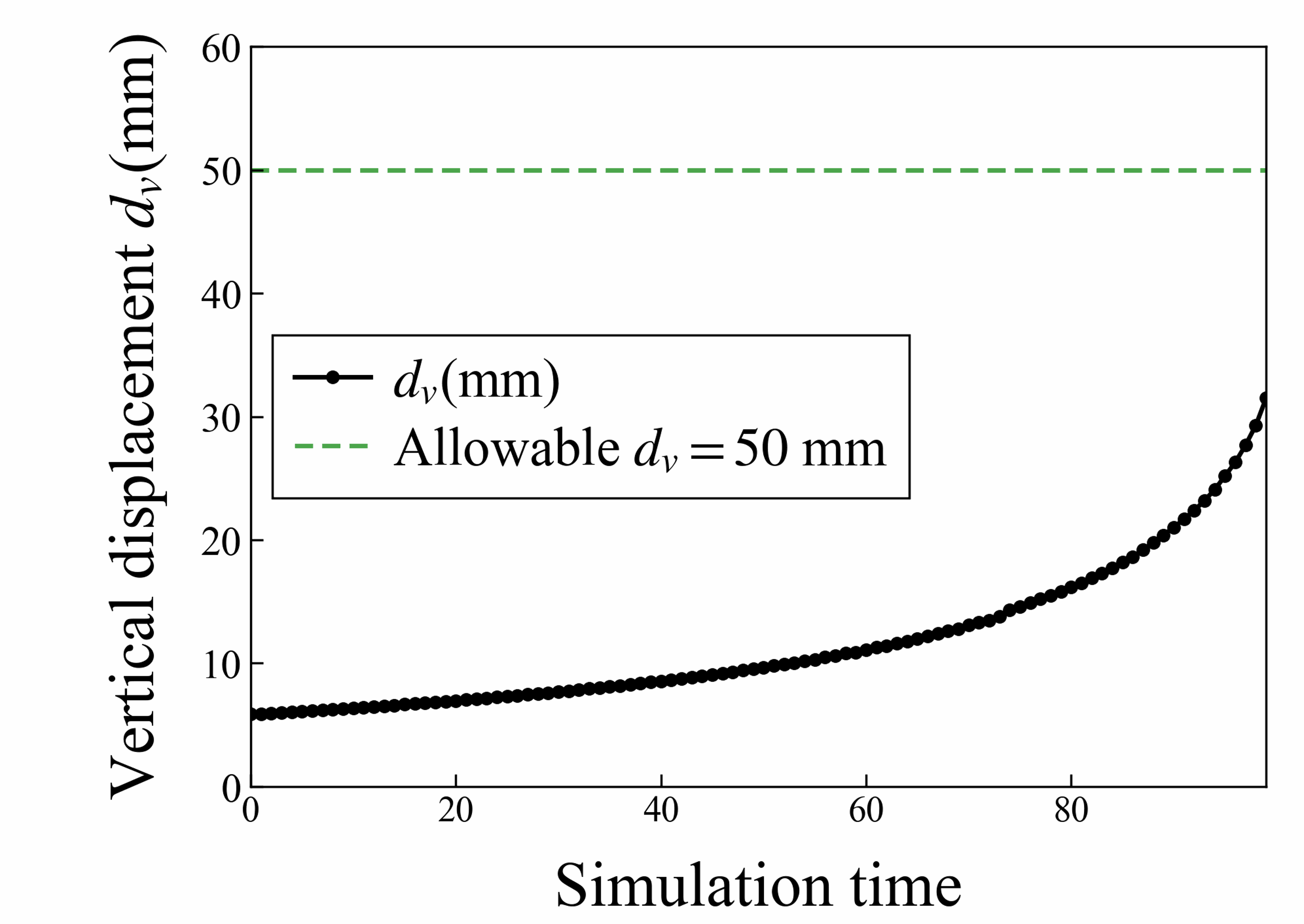}
        \caption{}
        \label{fig:marnix_a}
    \end{subfigure}
    \hfill
    \begin{subfigure}[t]{0.4\linewidth}
        \centering
        \includegraphics[width=\linewidth]{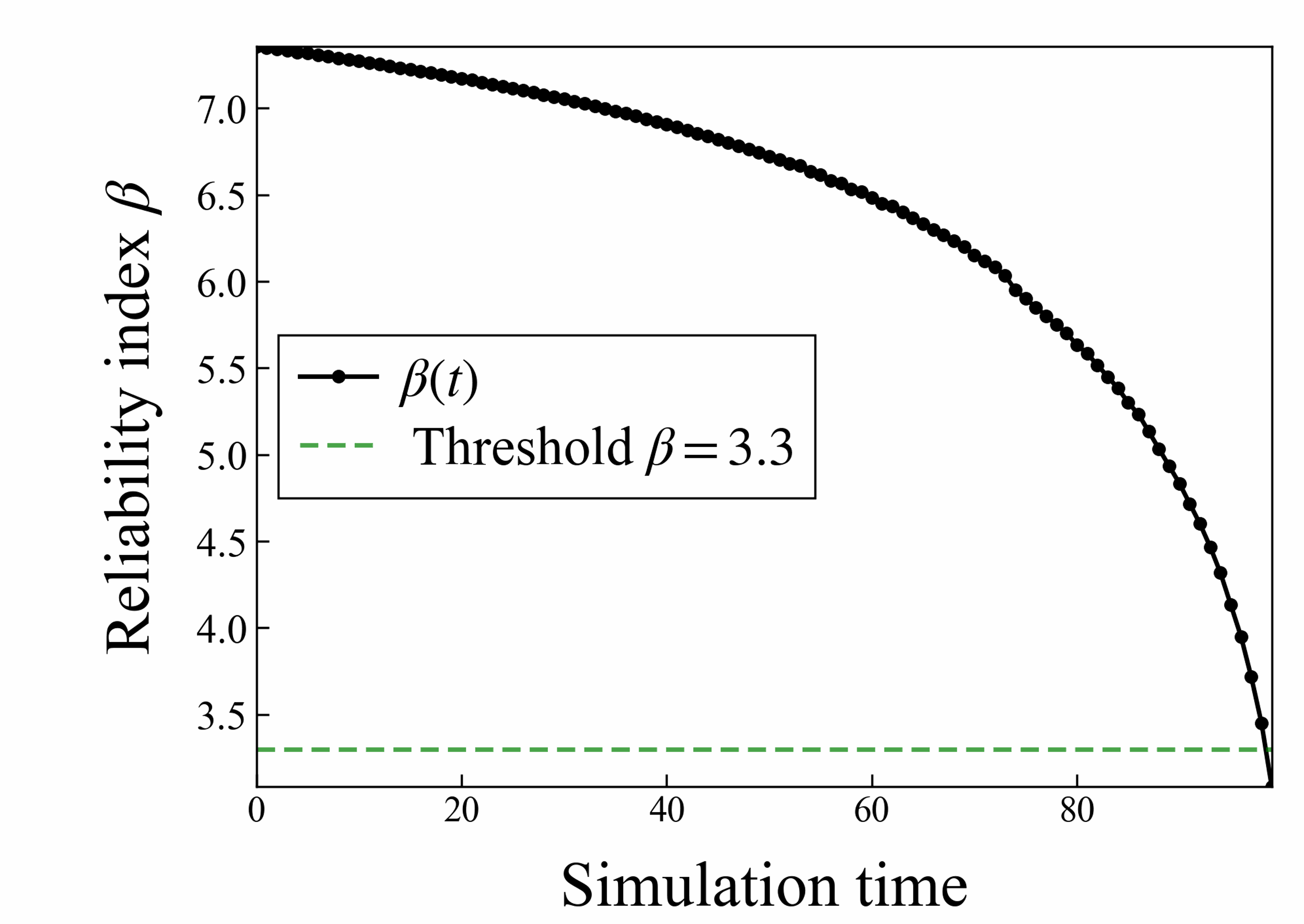}
        \caption{}
        \label{fig:marnix_b}
    \end{subfigure}
    \vspace{1em}
    \begin{subfigure}[t]{0.4\linewidth}
        \centering
        \includegraphics[width=\linewidth]{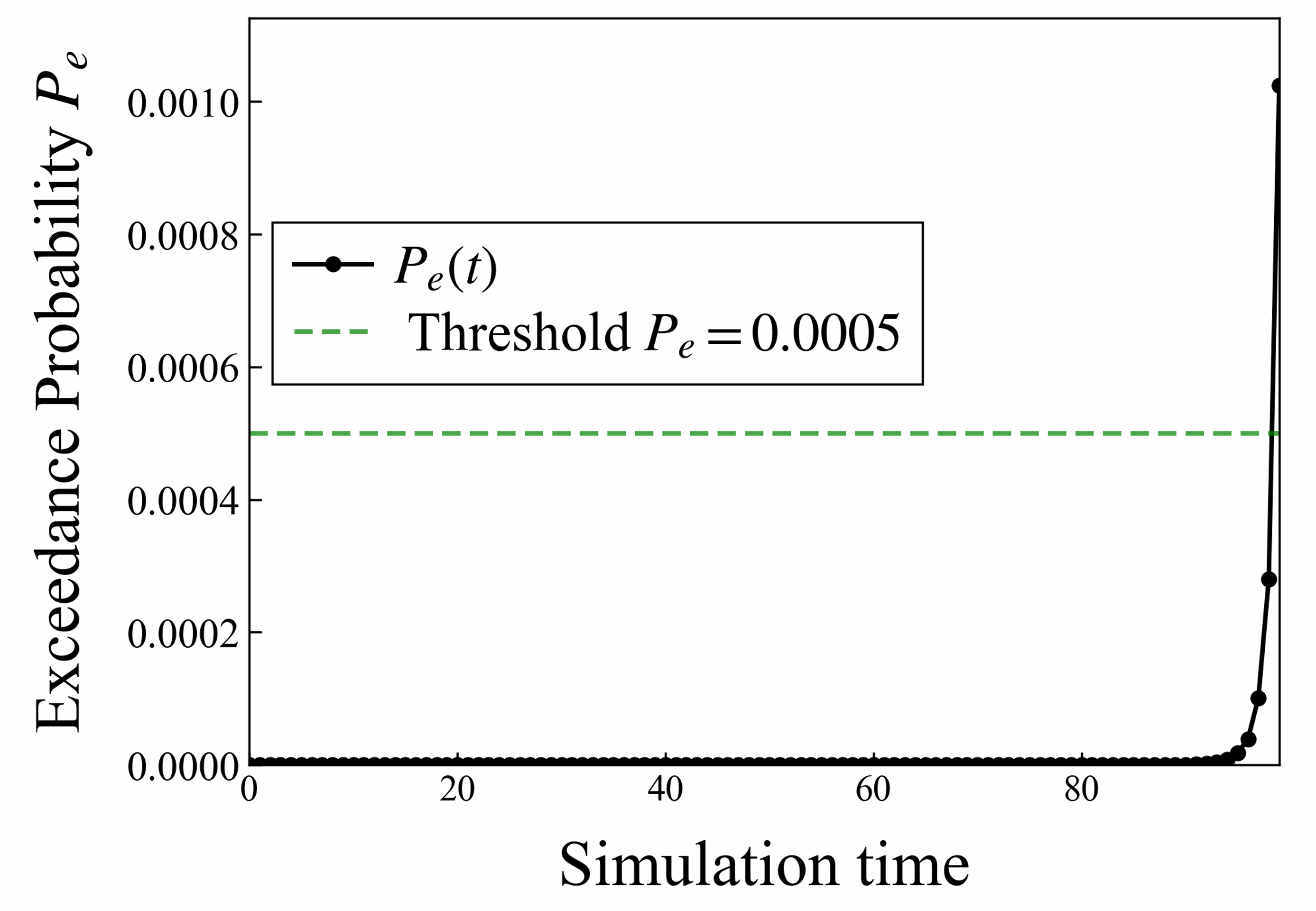}
        \caption{}
        \label{fig:marnix_c}
    \end{subfigure}
    \hfill
    \begin{subfigure}[t]{0.4\linewidth}
        \centering
        \includegraphics[width=\linewidth]{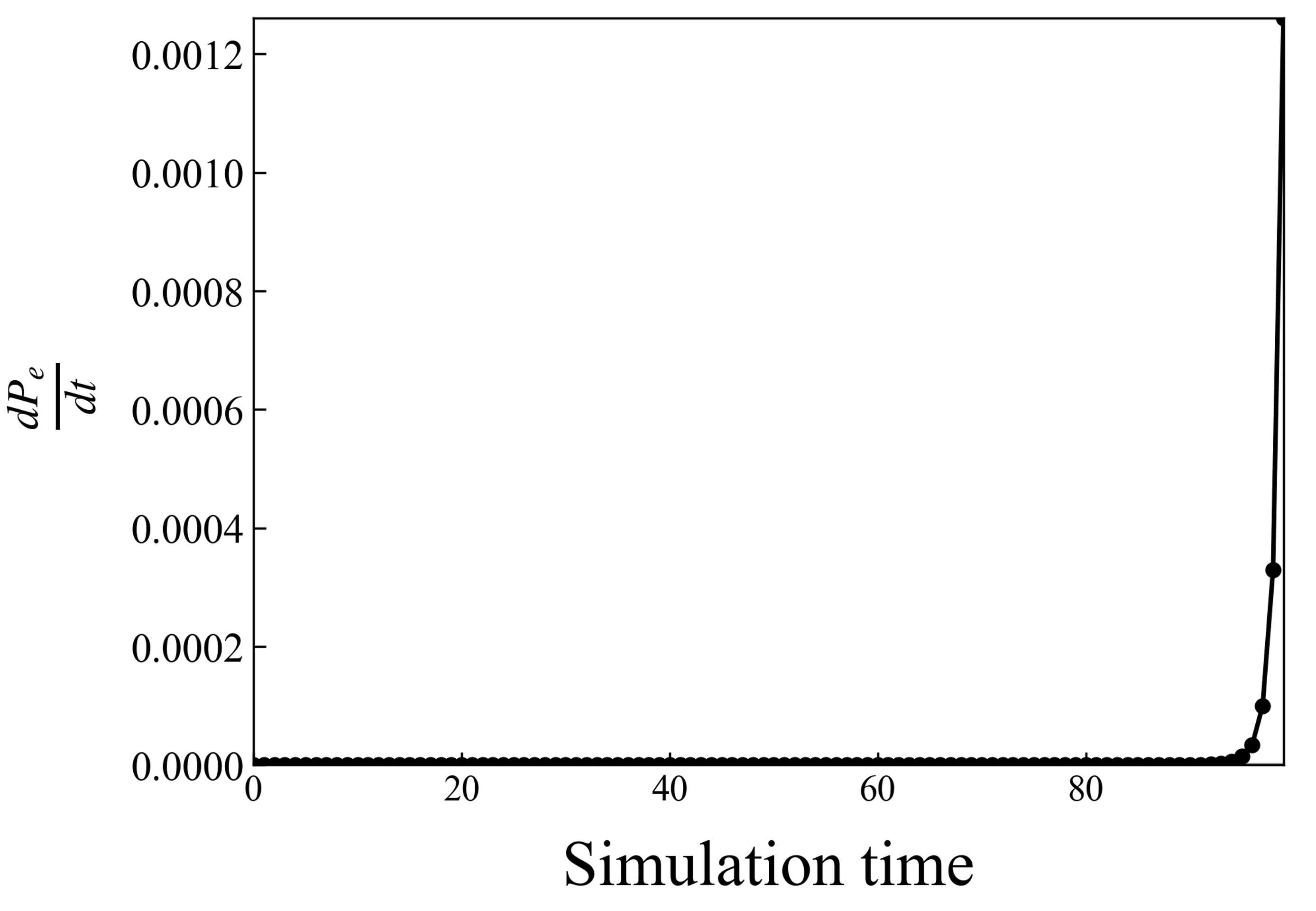}
        \caption{}
        \label{fig:marnix_d}
    \end{subfigure}
    \caption{\revisetwo{Numerical simulation results of the quay wall from damage initiation to total collapse at a surface point \citep{sharmaAnalysisProceduresAccounting2024} (maximum allowable vertical displacement 50 mm): (a) vertical displacement, (b) reliability index, (c) exceedance probability and (d) first derivative of the exceedance probability. }}
    \label{fig:marnix}
\end{figure}

\revisetwo{In this study, an abrupt-increase point detection method was proposed to determine the warning timing for deformation anomaly, for two main reasons.} 
\begin{enumerate}
    \item 
    Although MT-InSAR provides relative rather than absolute displacements, this limitation is not critical because infrastructure displacement typically evolves in a nonlinear manner. Sharma et al. \citep{sharmaAnalysisProceduresAccounting2024} investigated the progressive failure of piles, the most common failure mechanism of quay walls in Amsterdam. As illustrated in Figure \ref{fig:marnix_a}, the simulated displacement at the quay wall surface remained almost unchanged during the early stage, and only increased gradually as the piles approached critical damage. Notably, their model assumed a uniform degradation rate of the piles over time. In reality, deterioration of materials in infrastructure often accelerates significantly in the later stages of its service life. Therefore, even though MT-InSAR captures only relative displacements after the first acquisition, the resulting nonlinear displacement trend still contains essential information for identifying potential anomalies.
    \item     
    \revisetwo{More importantly, $P_e(t_i)$ was calculated based on the vertical displacement $d_{V, t_i}$ at time step $t_i$ via Equation \ref{eq:analytical_pf}. Due to the shape of the cumulative distribution function of a normally distributed random variable with zero mean and unit variance, $P_e(t_i)$ becomes highly sensitive as the observed displacement $d_{V, t_i}$ approaches the maximum allowable displacement $d_{V, max}$. This behaviour is illustrated in Figure \ref{fig:marnix_c}, where the probability of exceedance exhibits an abrupt nonlinear increase immediately before collapse. Correspondingly, the slope of $P_e$ (Figure \ref{fig:marnix_d}) rises sharply near the critical point. This pronounced increase in the probability of exceedance thus provides a more sensitive and practical anomaly warning indicator using MT-InSAR data.}
\end{enumerate}

\revisetwo{In this paper, we adopted the pruned exact linear time (PELT) algorithm \citep{killickOptimalDetectionChangepoints2012} to detect the abrupt-increase point in the probability of exceedance.} PELT is a change-point detection method that identifies time points where the statistical properties of a time series shift. It was selected for its computationally efficient and its ability to robustly capture abrupt changes without manual threshold tuning. 

Although all case studies in this paper were conducted retrospectively using recorded MT-InSAR data, the change-point detection was implemented in an online monitoring manner to mimic real-time operation. Specifically, the time series was processed sequentially, with one new data point incorporated at each time step. At each step, PELT was applied only to the data available up to that time, rather than to the complete time series. \revisethree{Note that the MT-InSAR displacement time series were derived from the full image stack. Furthermore, linear interpolation was applied prior to the sequential analysis to temporally align ascending and descending observations, potentially incorporating observations acquired after the interpolated dates. Consequently, although the PELT-based change-point detection was implemented in an online manner, the overall framework did not operate in a fully real-time setting.}

\revisethree{The PELT algorithm idenfities the change points by dividing the exceedance probability time series into segments, such that the observations within each segment remain as close as possible to their respective mean, while penalising the introduction of too many segments.}
\revisethree{At time $t_m$, the available exceedance probability time series obtained is $[P_{e,t_0},P_{e,t_1},\ldots,P_{e,t_m}]$. The optimal segmentation is determined by minimising the following cost function:}

\revisethree{
\begin{equation}
    \hat{\mathcal{B}}_m = \underset{\mathcal{B}_m}{\arg\min} \left[ \sum_{k=0}^{K_m} C\left(\mathbf{P}_{e,b_k:b_{k+1}-1}\right) + \zeta K_m \right]
\end{equation}
where $\mathcal{B}_m=\{b_1,\ldots,b_{K_m}\}$ represents the set of change-point indices, with $b_0=0$
and $b_{K_m+1}=m+1$. $K_m$ is the number of change points. $\zeta$ is the penalty parameter
controlling the number of change points, set to $1\times 10^{-5}$ in this study considering the relatively small magnitude of the cost associated with variations in the exceedance probability. The influence of this parameter on the warning dates was further examined through a sensitivity analysis (\ref{app:appe}). $C$ is the cost function. Here the least squares cost function was adopted, which measures the sum of squared deviations from the mean exceedance probability within each segment:}

\revisethree{
\begin{equation}
    C\left(\mathbf{P}_{e,b_k:b_{k+1}-1}\right) = \sum_{j=b_k}^{b_{k+1}} \left(P_{e,j}-\bar{P}_{e,b_k:b_{k+1}-1}\right)^2
\end{equation}
where $\bar{P}_{e,b_k:b_{k+1}-1}$ denotes the mean exceedance probability over the corresponding segment.}

\revisethree{The minimum segment was set to $1$, requiring each segment to contain at least one exceedance probability value. Change point detection was initiated only when at least $10$ observations were available. A warning was triggered at the first time when PELT identified at least one change point. Importantly, the estimated change point date represents the inferred onset of the change, whereas the first detection date corresponds to the earliest time when the change was identified.}

\revisetwo{The resulting warning represents a proactive indicator rather than a direct prediction of imminent structural collapse. It allows stakeholders to undertake further investigation and, where necessary, preventive maintenance.}


\section{Case studies}
\revise{To illustrate the applicability of the proposed framework, we applied it to three quay walls in Amsterdam's historical centre: Grimburgwal, Entrepotdok and Houtmankade East. Grimburgwal and Entrepotdok had recently collapsed after years of service. \revisetwo{By analysing the historical InSAR data, the aim was to assess whether the framework could have facilitated a timely warning of deformation anomaly.} \revisethree{The third case, Houtmankade East, presents a dual scenario: a damaged part (subsequently replaced) was used to test the framework's warning capabilities, while the remaining undamaged part served as a control to provide preliminary evidence of its robustness against false alarms.}}

\subsection{Grimburgwal}

\begin{figure}
    \centering
    \begin{subfigure}[t]{0.48\linewidth}
        \centering
        \includegraphics[width=\linewidth]{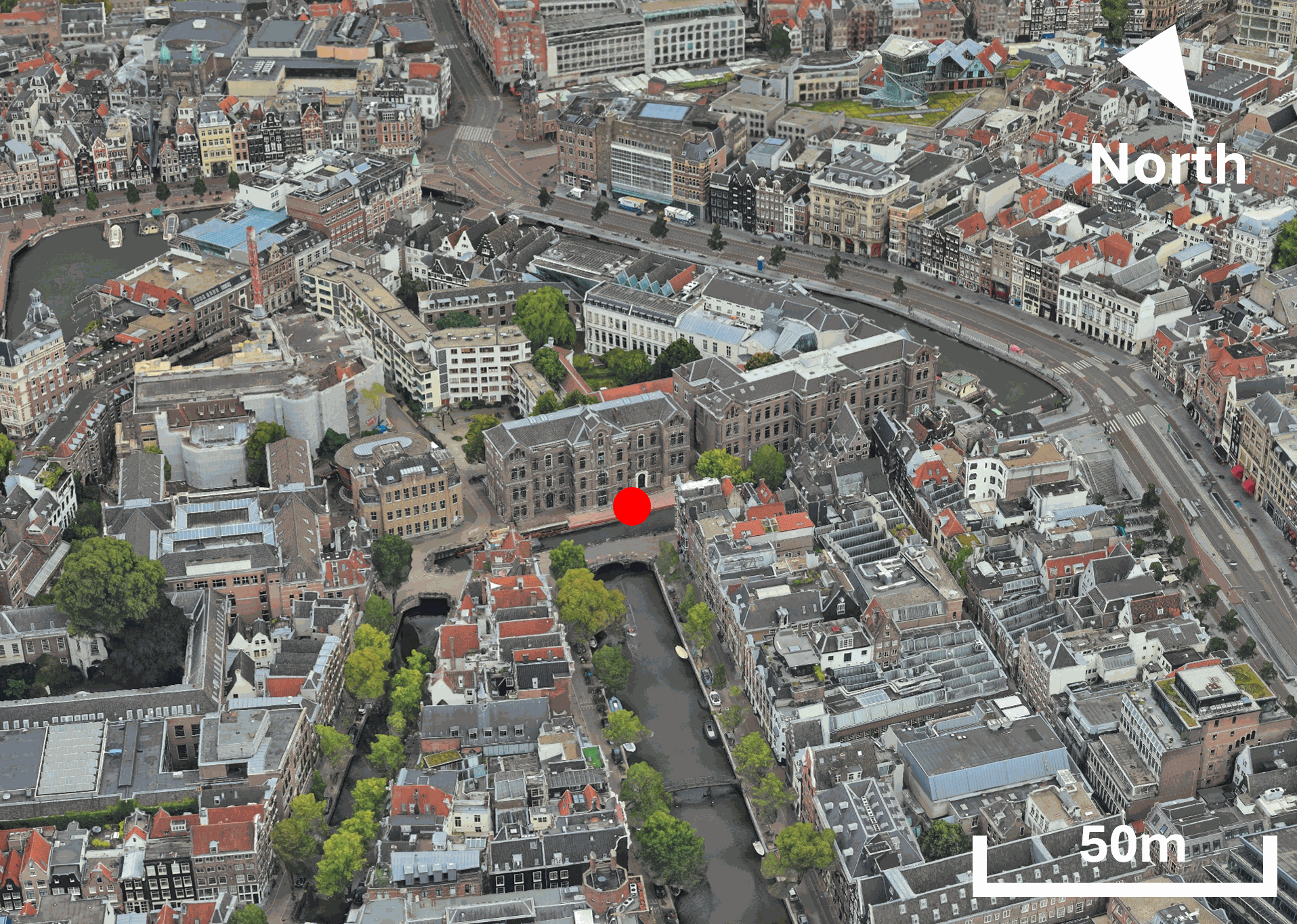}
        \caption{}
        \label{fig:grim_a}
    \end{subfigure}
    \hfill
    \begin{subfigure}[t]{0.48\linewidth}
        \centering
        \includegraphics[width=\linewidth]{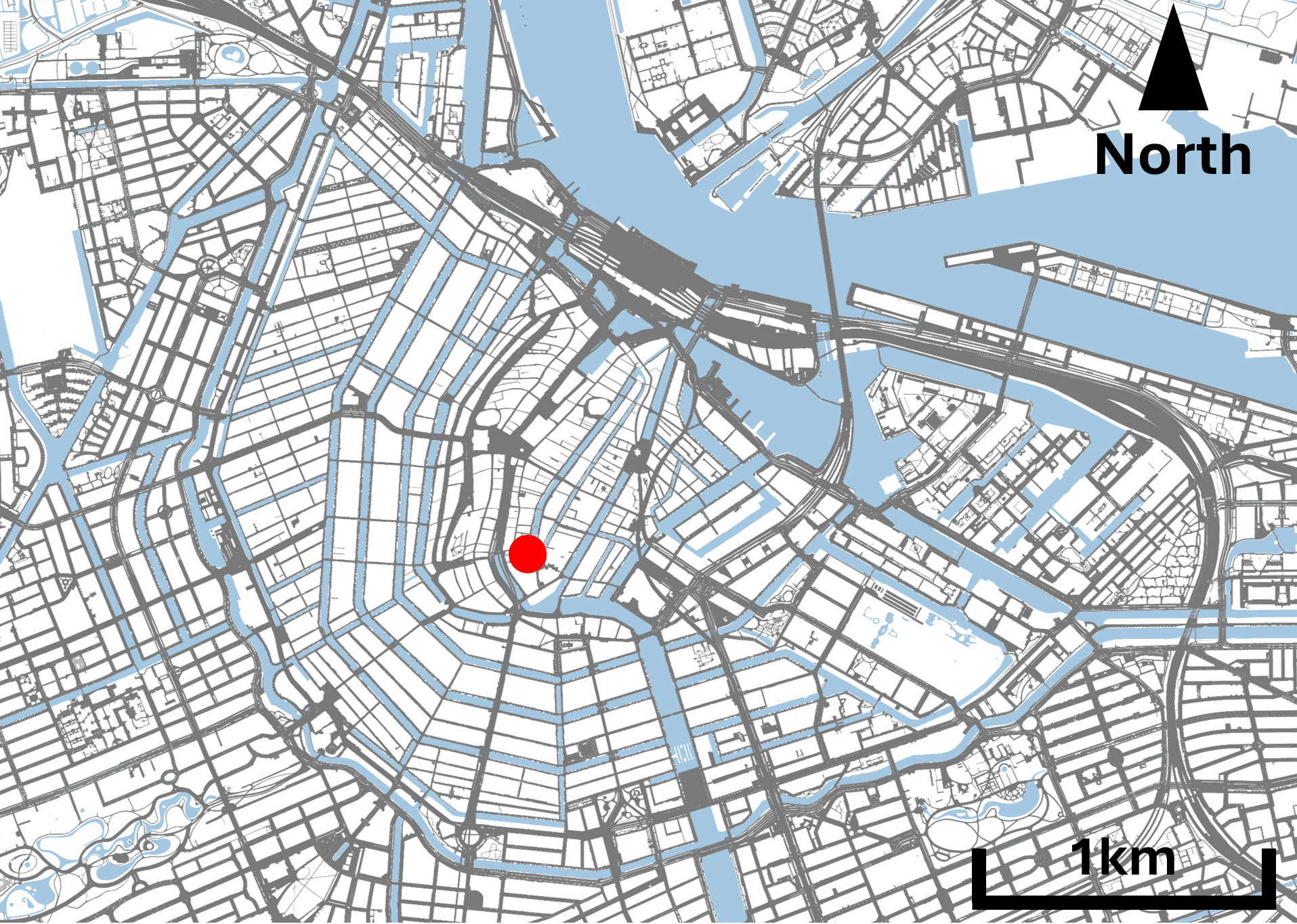}
        \caption{}
        \label{fig:grim_b}
    \end{subfigure}
    \vspace{1em}
    \begin{subfigure}[t]{0.48\linewidth}
        \centering
        \includegraphics[width=\linewidth]{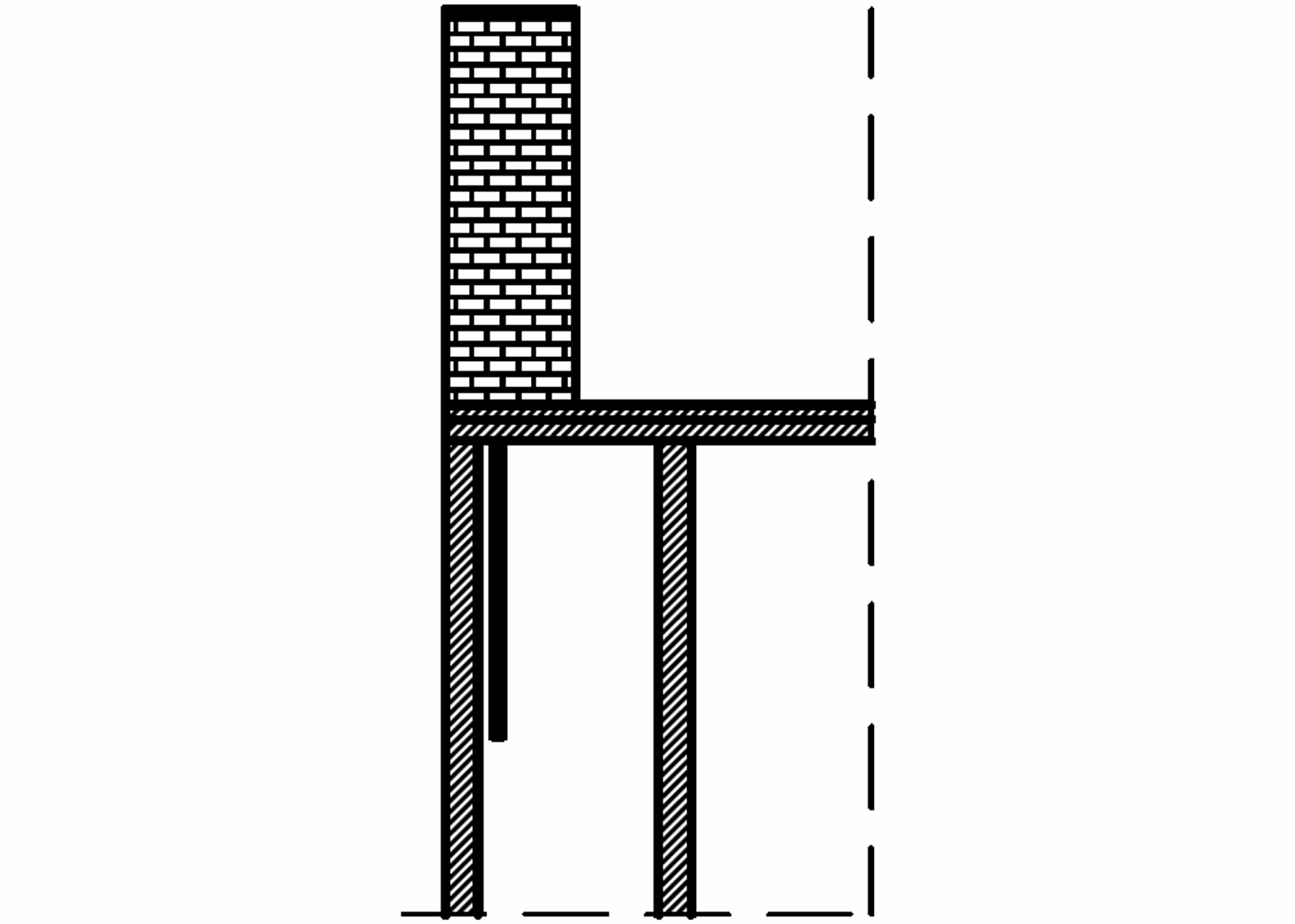}
        \caption{}
        \label{fig:grim_c}
    \end{subfigure}
    \hfill
    \begin{subfigure}[t]{0.48\linewidth}
        \centering
        \includegraphics[width=\linewidth]{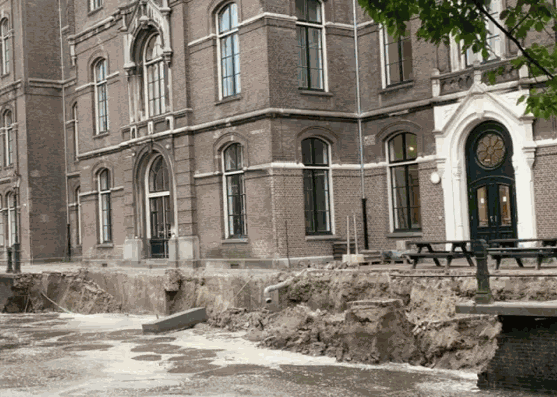}
        \caption{}
        \label{fig:grim_d}
    \end{subfigure}
    \caption{\revise{Grimburgwal quay: (a) aerial view, where the red marker denotes its location (image \copyright\ Google); (b) location in Amsterdam, where the red marker denotes its location (basemap using data from \citep{dutchnationalgeo-informationpublicserviceplatformpdokBasisregistratieGrootschaligeTopografie2025}); (c) cross-section sketch of the structure \citep{korffBezwijkenGrimburgwalLeerpunten2021}; (d) photograph taken after the collapse, showing the collapsed area \citep{fabriqueGrimburgwalProvidesLessons2021}.}}
    \label{fig:grim}
\end{figure}

Grimburgwal quay, situated in Amsterdam's historic centre (Figure \ref{fig:grim_a} and \ref{fig:grim_b}), was constructed around 1875 \citep{korffBezwijkenGrimburgwalLeerpunten2021}. During its construction, the previous quay wall was removed to a depth of approximately 1.5 meters below ground level; however, some older structures, built before 1875, remained beneath the current quay. Figure \ref{fig:grim_c} shows a cross-section sketch based on available information \citep{korffBezwijkenGrimburgwalLeerpunten2021}. The total length of the analysed quay section is around 70 meters. In general, Grimburgwal quay had a retaining height of approximately 4 meters.

\begin{figure}
    \centering
    \begin{subfigure}{0.48\linewidth}
        \centering
        \includegraphics[width=1\linewidth]{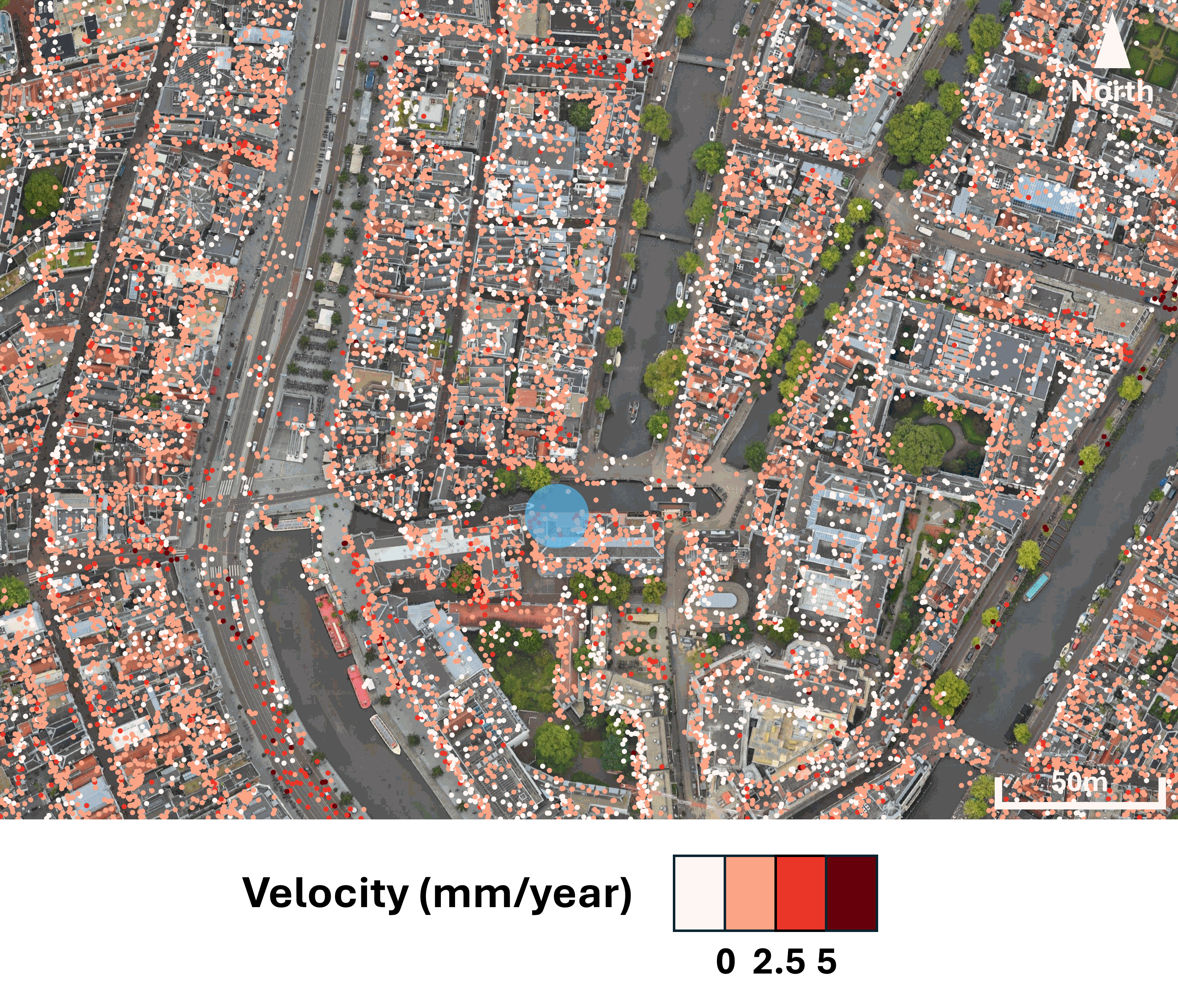}
        \caption{}
        \label{fig:ps_grim_asc}
    \end{subfigure}
    \hfill
    \begin{subfigure}{0.48\linewidth}
        \centering
        \includegraphics[width=1\linewidth]{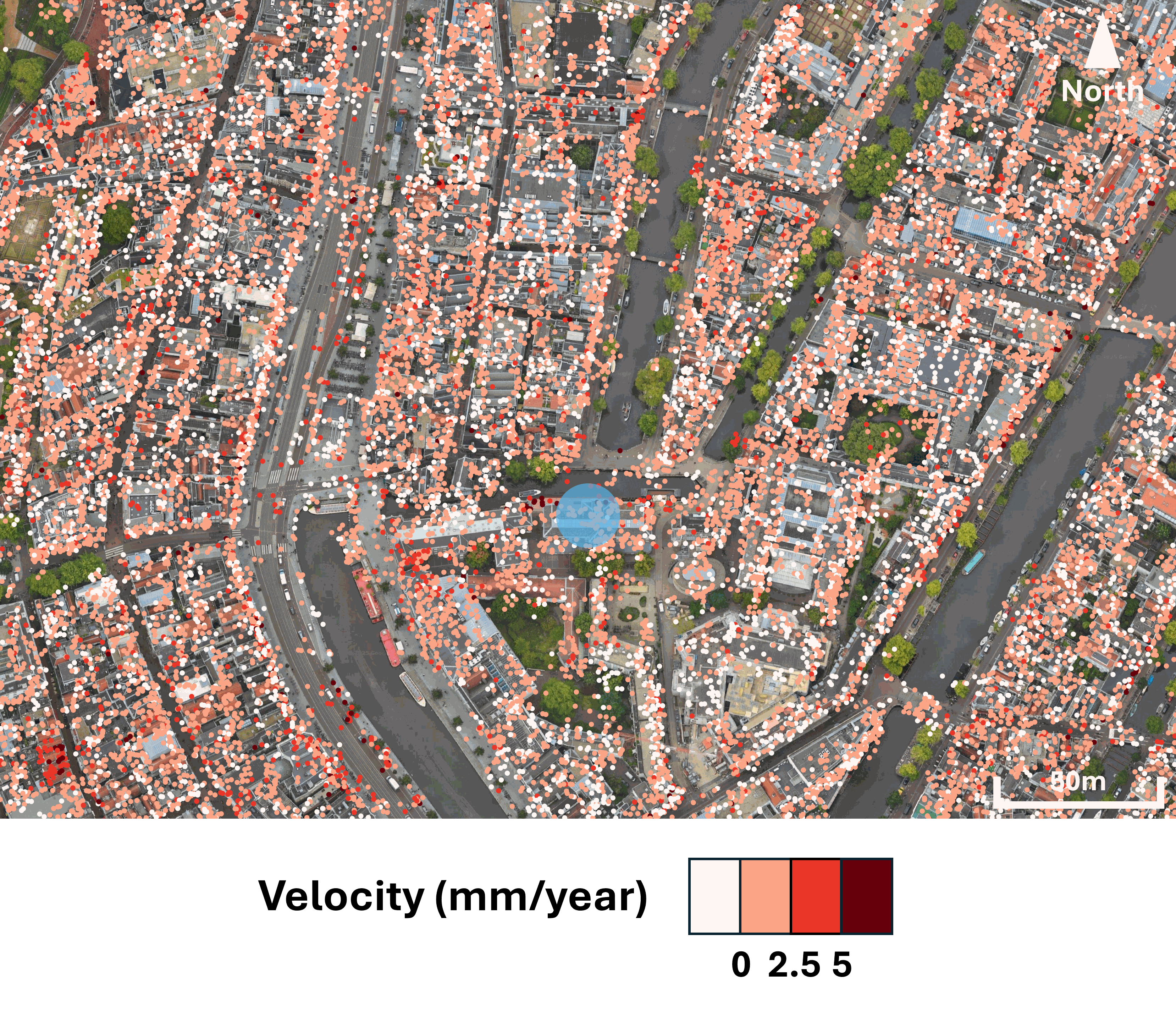}
        \caption{}
        \label{fig:ps_grim_des}
    \end{subfigure}
    \caption{\revisefour{Spatial distribution of PSs in the area around Grimburgwal quay (Basemap \copyright\ Google) from: (a) ascending and (b) descending track.}}
    \label{fig:ps_grim}
\end{figure}

On September 1, 2020, the Girmburgwal quay collapsed as shown in Figure \ref{fig:grim_d}. Footage from a security camera recorded during the collapse reveals that the quay inclined toward the canal, causing a section of approximately 25 meters to collapse into it \citep{korffCollapseGrimburgwalHistoric2022}.
Figure \ref{fig:ps_grim} presents the distribution of PSs derived from the ascending (a) and descending (b) tracks in the Grimburgwal area. \revisefour{The coloured points represent PS locations, and the colour scale indicates the estimated LOS velocity in mm/year. The blue marker denotes the location of the Grimburgwal quay.} The high density of PSs provided a robust basis for reliable time-series analysis of ground displacements. The PSs and their LOS displacement time series were obtained from TerraSAR-X images with a spatial resolution of 3 meters, acquired between March 2016 and December 2020. \revise{A total of 138 ascending and 145 descending images were used in the analysis.} These images were processed by SkyGeo using their proprietary PyAntares algorithm, which is based on the method proposed by Van Leijen \cite{f.j.vanleijenPersistentScattererInterferometry2014}.

\subsection{Entrepotdok}
\begin{figure}
    \centering
    \begin{subfigure}[t]{0.48\linewidth}
        \centering
        \includegraphics[width=\linewidth]{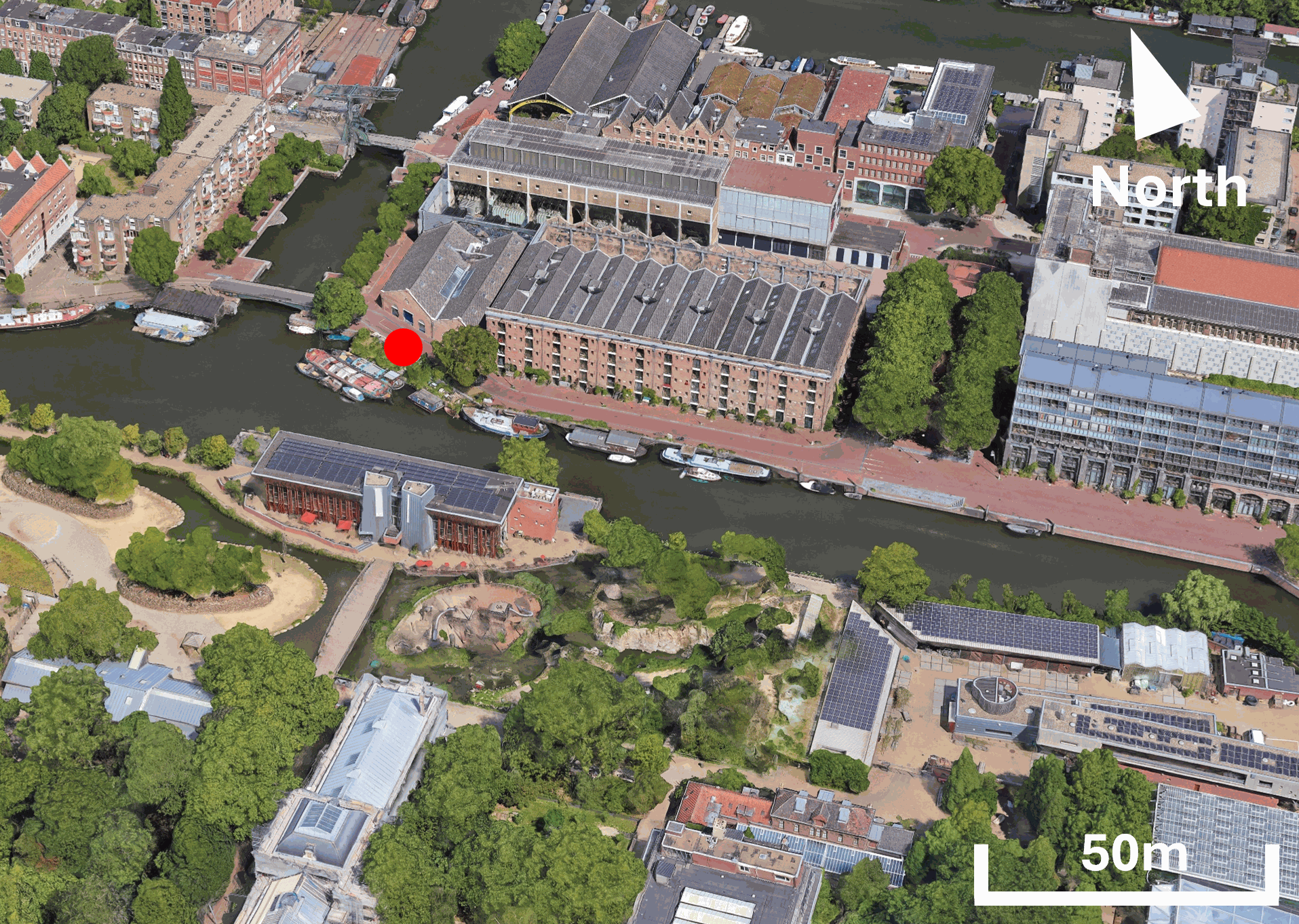}
        \caption{}
        \label{fig:entre_a}
    \end{subfigure}
    \hfill
    \begin{subfigure}[t]{0.48\linewidth}
        \centering
        \includegraphics[width=\linewidth]{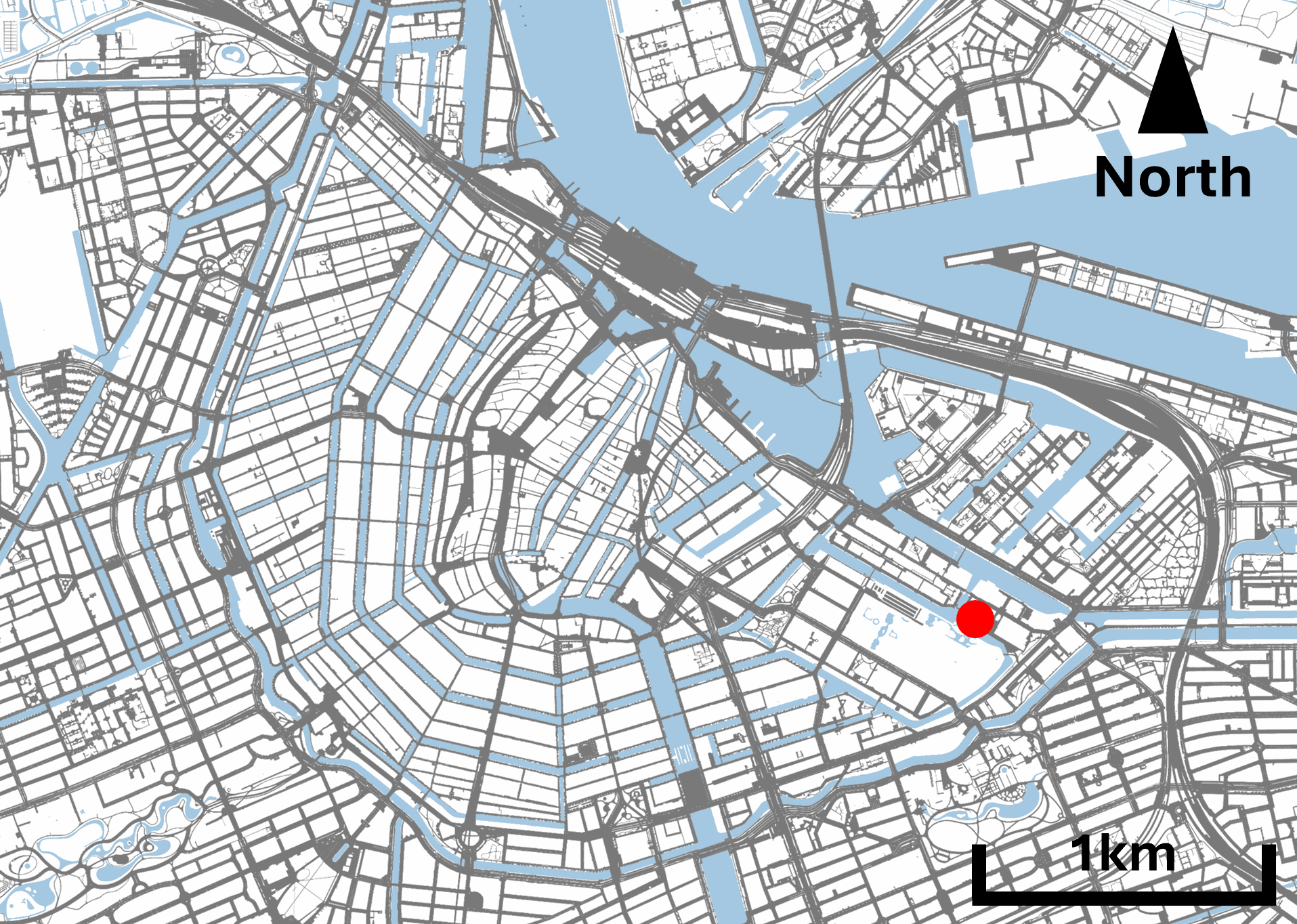}
        \caption{}
        \label{fig:entre_b}
    \end{subfigure}
    \vspace{1em}
    \begin{subfigure}[t]{0.48\linewidth}
        \centering
        \includegraphics[width=\linewidth]{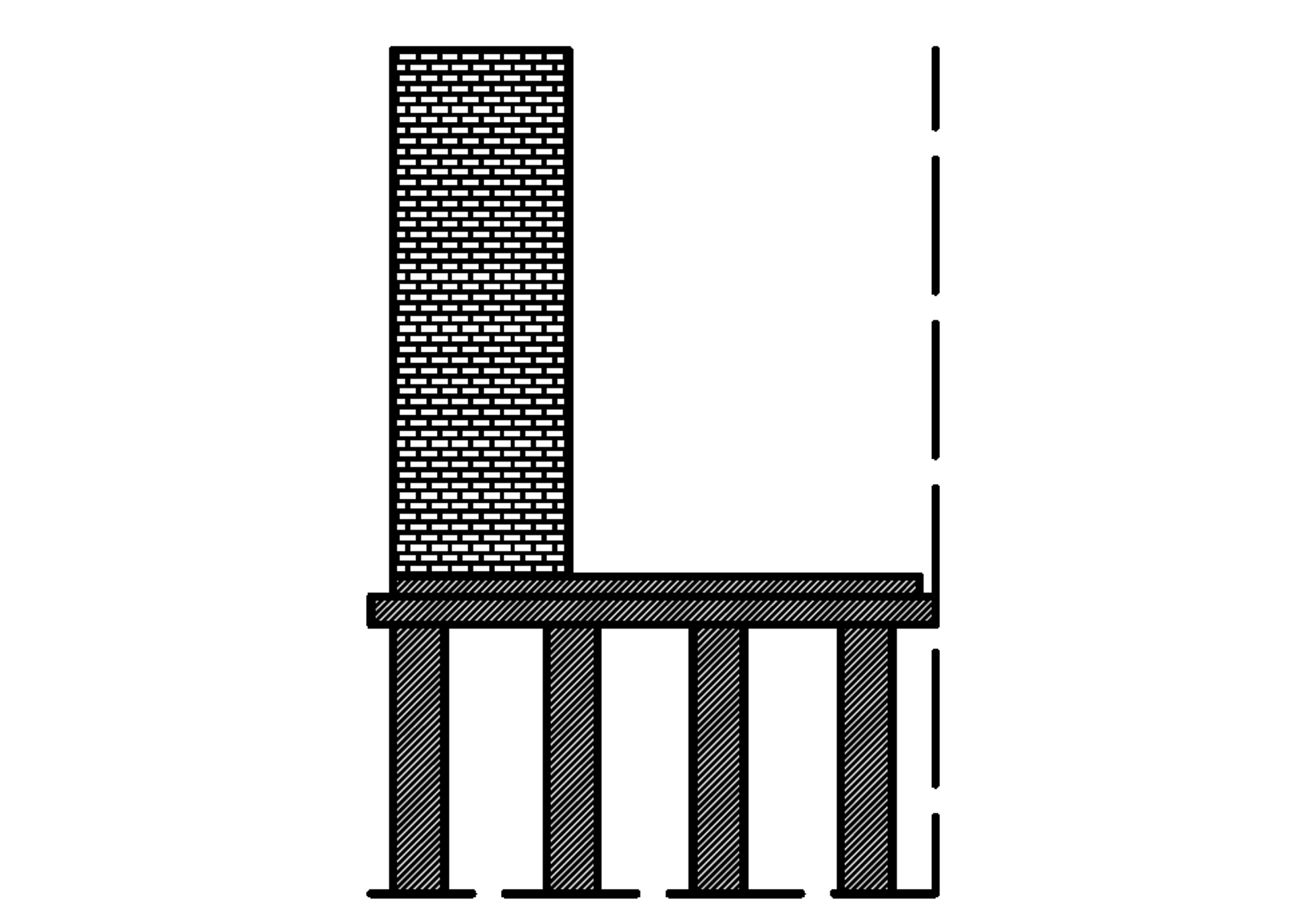}
        \caption{}
        \label{fig:entre_c}
    \end{subfigure}
    \hfill
    \begin{subfigure}[t]{0.48\linewidth}
        \centering
        \includegraphics[width=\linewidth]{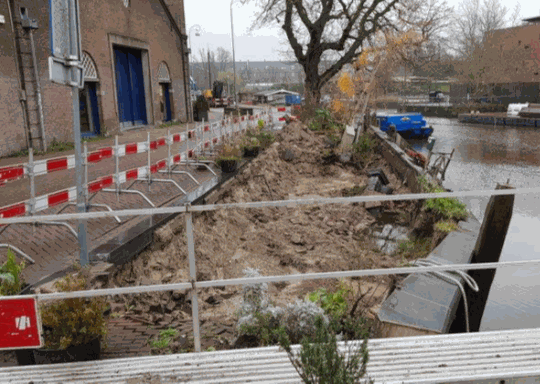}
        \caption{}
        \label{fig:entre_d}
    \end{subfigure}
    \caption{\revise{Entrepotdok quay: (a) aerial view, where the red marker denotes its location (image \copyright\ Google); (b) location in Amsterdam, where the red marker denotes its location (basemap using data from \citep{dutchnationalgeo-informationpublicserviceplatformpdokBasisregistratieGrootschaligeTopografie2025}); (c) cross-section sketch of the structure \citep{AmsterdamInspectiePortaal}; (d) photograph taken after the collapse, showing the collapsed area \citep{buurtorganisatie1018KadeEntrepotdokIngestort2017}.}}
    \label{fig:entre}
\end{figure}

Entrepotdok quay, situated northeast of Amsterdam’s historic centre (Figure \ref{fig:entre_a} and \ref{fig:entre_b}), was constructed as part of the installation of a large lock between Entrepotdok 85 and 86. \revisefour{According to the records provided by Municipality of Amsterdam, the quay was renovated in 1995 \citep{AmsterdamInspectiePortaal}.} During the renovation, the masonry section from the capstone to the waterline was replaced, while the pile foundation remained unchanged. \revisefour{Figure \ref{fig:entre_c} represents a cross-section sketch based on the technical drawing provided by the Municipality of Amsterdam after the renovation \citep{AmsterdamInspectiePortaal}}. The total length of the analysed quay section is around 50 meters. In general, the quay wall has a retaining height of approximately 2 meters.

\begin{figure}
    \centering
    \begin{subfigure}{0.48\linewidth}
        \centering
        \includegraphics[width=1\linewidth]{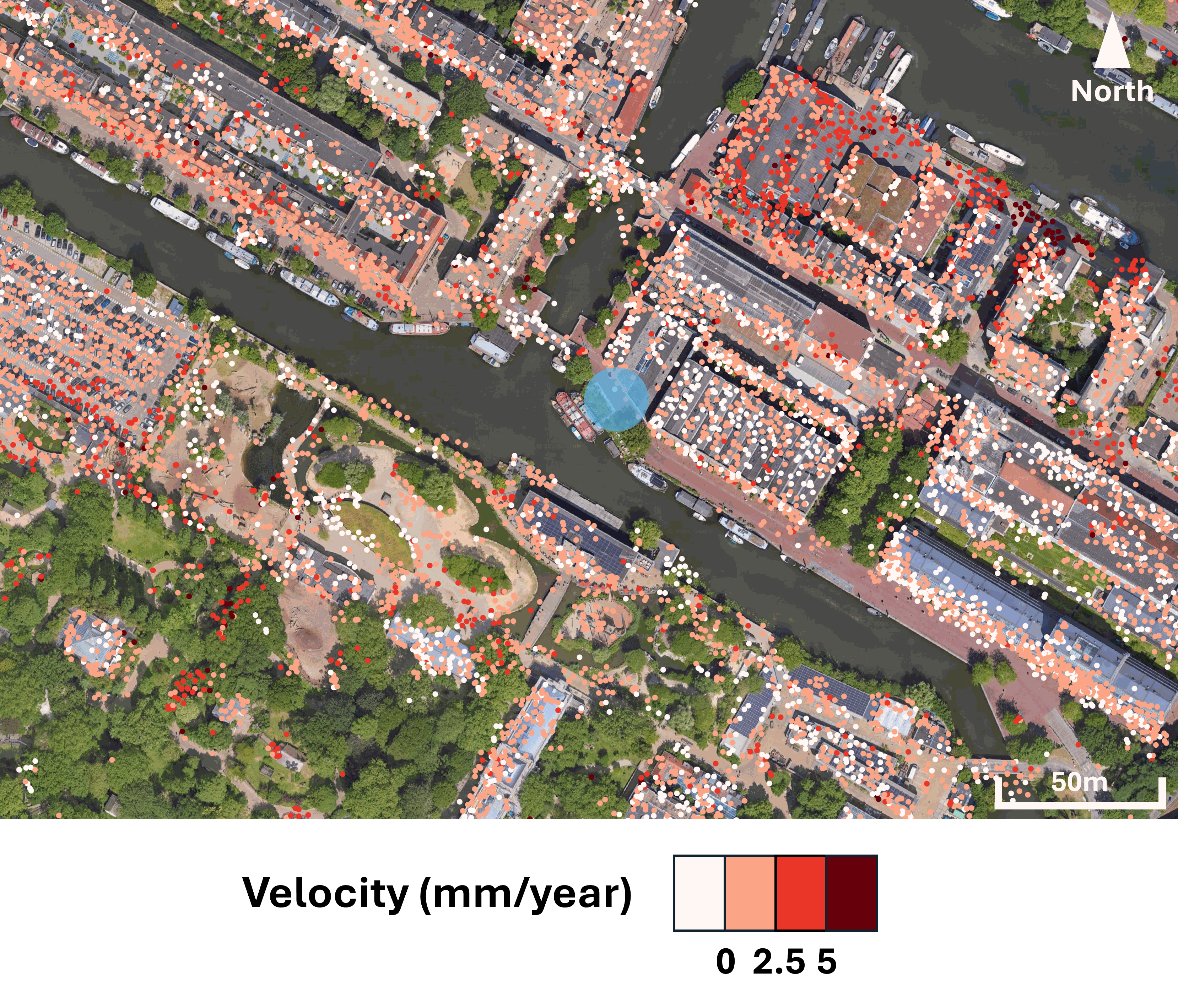}
        \caption{}
        \label{fig:ps_entre_asc}
    \end{subfigure}
    \hfill
    \begin{subfigure}{0.48\linewidth}
        \centering
        \includegraphics[width=1\linewidth]{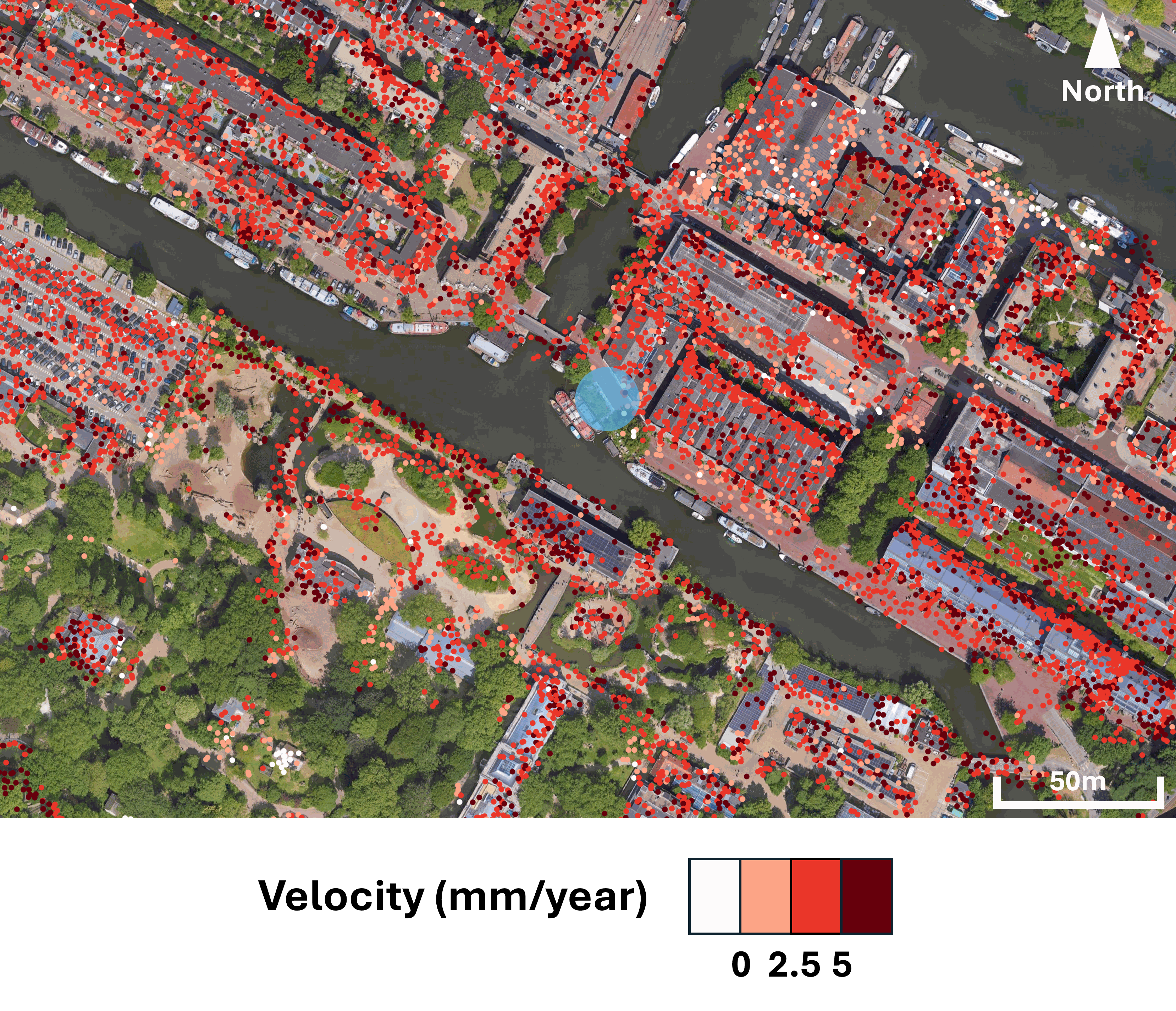}
        \caption{}
        \label{fig:ps_entre_des}
    \end{subfigure}
    \caption{\revisefour{Spatial distribution of PSs in the area around Entrepotdok quay (Basemap \copyright\ Google) from: (a) ascending and (b) descending track.}}
    \label{fig:ps_entre}
\end{figure}

On the night of December 11, 2017, the Entrepotdok quay collapsed, as shown in Figure \ref{fig:entre_d} \citep{buurtorganisatie1018KadeEntrepotdokIngestort2017}. On January 1, 2018, a temporary sheet pile wall was installed to prevent further damage to the quay and to protect nearby buildings \citep{gemeenteamsterdamVeiligheidsmaatregelenDashboardProgramma2025}; however, the collapsed quay wall remained in place.
Figure \ref{fig:ps_entre} shows the distribution of PSs derived from the ascending (a) and descending (b) tracks in the Entrepotdok area. \revisefour{The coloured points represent PS locations, and the colour scale indicates the estimated LOS velocity in mm/year. The blue marker denotes the location of the Entrepotdok quay.} The dense coverage of PSs provided a basis for reliable time-series analysis of ground displacements. \revise{The PSs and their LOS displacement time series were obtained from TerraSAR-X images (51 ascending-track and 59 descending-track images) with a spatial resolution of 3 meters, acquired between March 2016 and March 2018, and processed by SkyGeo using their proprietary PyAntares algorithm \cite{f.j.vanleijenPersistentScattererInterferometry2014}.}

\subsection{\revise{Houtmankade East}}

\begin{figure}
    \centering
    \begin{subfigure}[t]{0.48\linewidth}
        \centering
        \includegraphics[width=\linewidth]{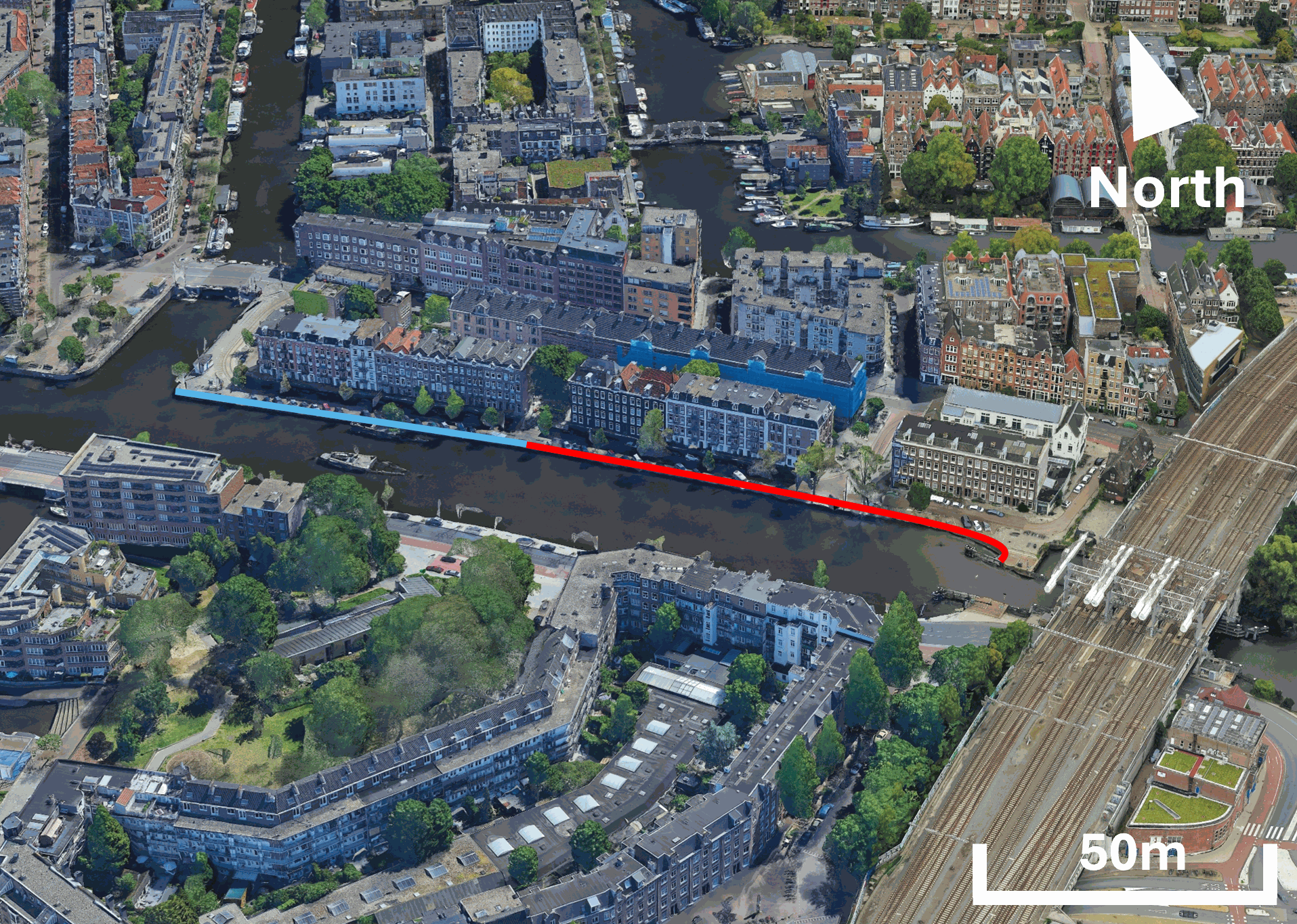}
        \caption{}
        \label{fig:hout_a}
    \end{subfigure}
    \hfill
    \begin{subfigure}[t]{0.48\linewidth}
        \centering
        \includegraphics[width=\linewidth]{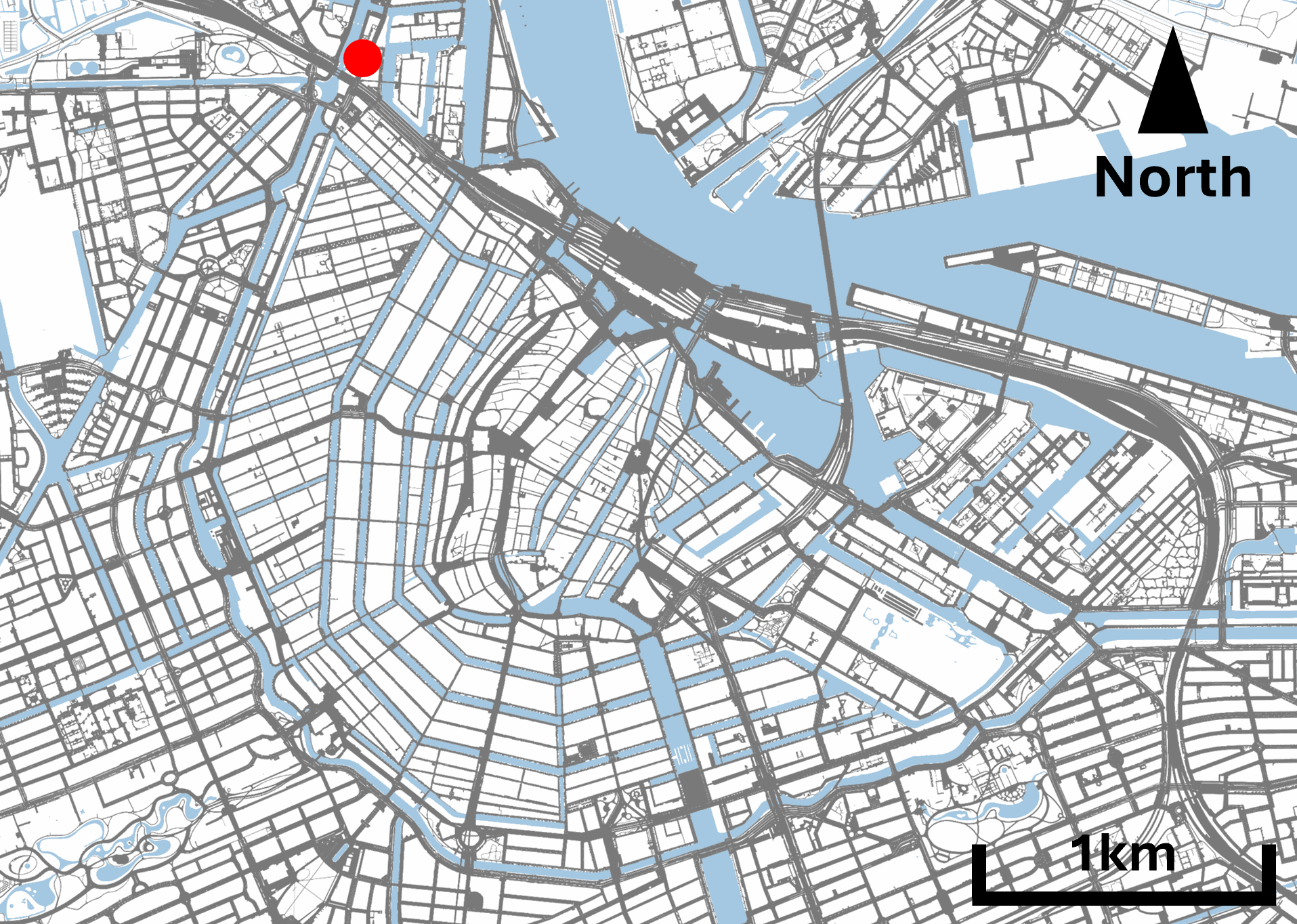}
        \caption{}
        \label{fig:hout_b}
    \end{subfigure}
    \vspace{1em}
    \begin{subfigure}[t]{0.48\linewidth}
        \centering
        \includegraphics[width=\linewidth]{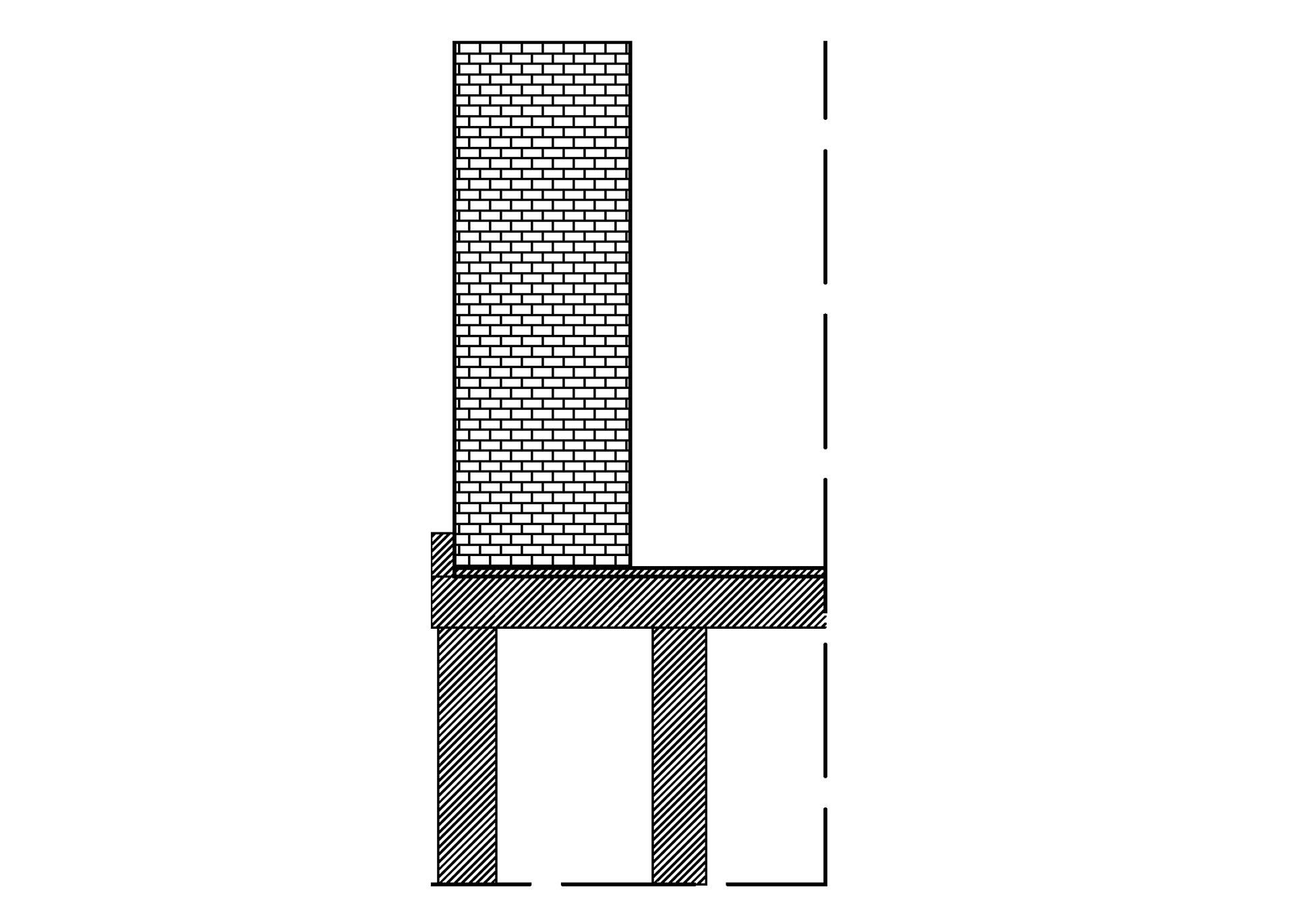}
        \caption{}
        \label{fig:hout_c}
    \end{subfigure}
    \hfill
    \begin{subfigure}[t]{0.48\linewidth}
        \centering
        \includegraphics[width=\linewidth]{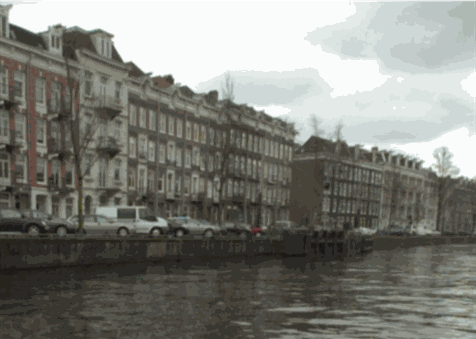}
        \caption{}
        \label{fig:hout_d}
    \end{subfigure}
    \caption{\revise{Houtmankade East quay: (a) aerial view, where the blue line denotes Sections 1 to 3 (safe sections) and the red line denotes Sections 4 and 5 (replaced sections) (image \copyright\ Google); (b) location in Amsterdam, where the red marker denotes its location (basemap using data from \citep{dutchnationalgeo-informationpublicserviceplatformpdokBasisregistratieGrootschaligeTopografie2025}); (c) cross-section sketch of the structure \citep{AmsterdamInspectiePortaal}; (d) photograph of the quay wall showing its general condition \citep{AmsterdamInspectiePortaal}.}}
    \label{fig:hout}
\end{figure}

\revisefour{Houtmankade East quay is located northwest of Amsterdam's historic centre, along the Westergracht canal (Figures \ref{fig:hout_a} and \ref{fig:hout_b}). According to the Municipality of Amsterdam, the quay has a total length of approximately 230 meters and consists of five sections \citep{AmsterdamInspectiePortaal}.} Sections 1 to 3, highlighted by the blue line in Figure \ref{fig:hout_a}, represent the newer part of the quay wall and were constructed in 1990, 2002 and 2002, respectively, with a total length of approximately 80 meters. Section 4 was probably constructed around 1960, while the construction year of Section 5 is unknown. These two older sections are highlighted by the red line in Figure \ref{fig:hout_a}. \revisefour{The Municipality of Amsterdam documented cross-section sketches for Sections 3 and 4, which are shown in Figure \ref{fig:hout_c} \citep{AmsterdamInspectiePortaal}.} As cross-section information is unavailable for the remaining sections, the available cross-section sketches of Sections 3 and 4 were used as references, and a representative retaining height of approximately 2.5 meters was adopted for the quay wall.

\revise{In June 2019, Sweco assessed the quay wall and observed local deformation at the transition between Sections 3 and 4 \citep{AmsterdamInspectiePortaal}. This observation raised concerns regarding the structural condition of this quay wall and prompted a more detailed underwater assessment by Baars-CIPRO in September 2019. This assessment concluded that Sections 4 and 5 suffered structural damage, whereas Sections 1 to 3 were considered safe \citep{AmsterdamInspectiePortaal}. Based on these inspection results, Sections 4 and 5 were selected for replacement. The replacement works started in October 2023 and were completed in 2026 by the Municipality of Amsterdam.}

\revise{This case study is therefore suitable for evaluating two aspects of the proposed method. \revisethree{Sections 1 to 3, where no structural problems were reported, serve as reference areas for assessing potential false alarms, providing preliminary evidence of the framework's false alarm performance.} Section 4 is the most relevant section for warning evaluation, because it corresponds to the location where the first documented abnormal symptoms were observed.}

\begin{figure}
    \centering
    \begin{subfigure}{0.48\linewidth}
        \centering
        \includegraphics[width=1\linewidth]{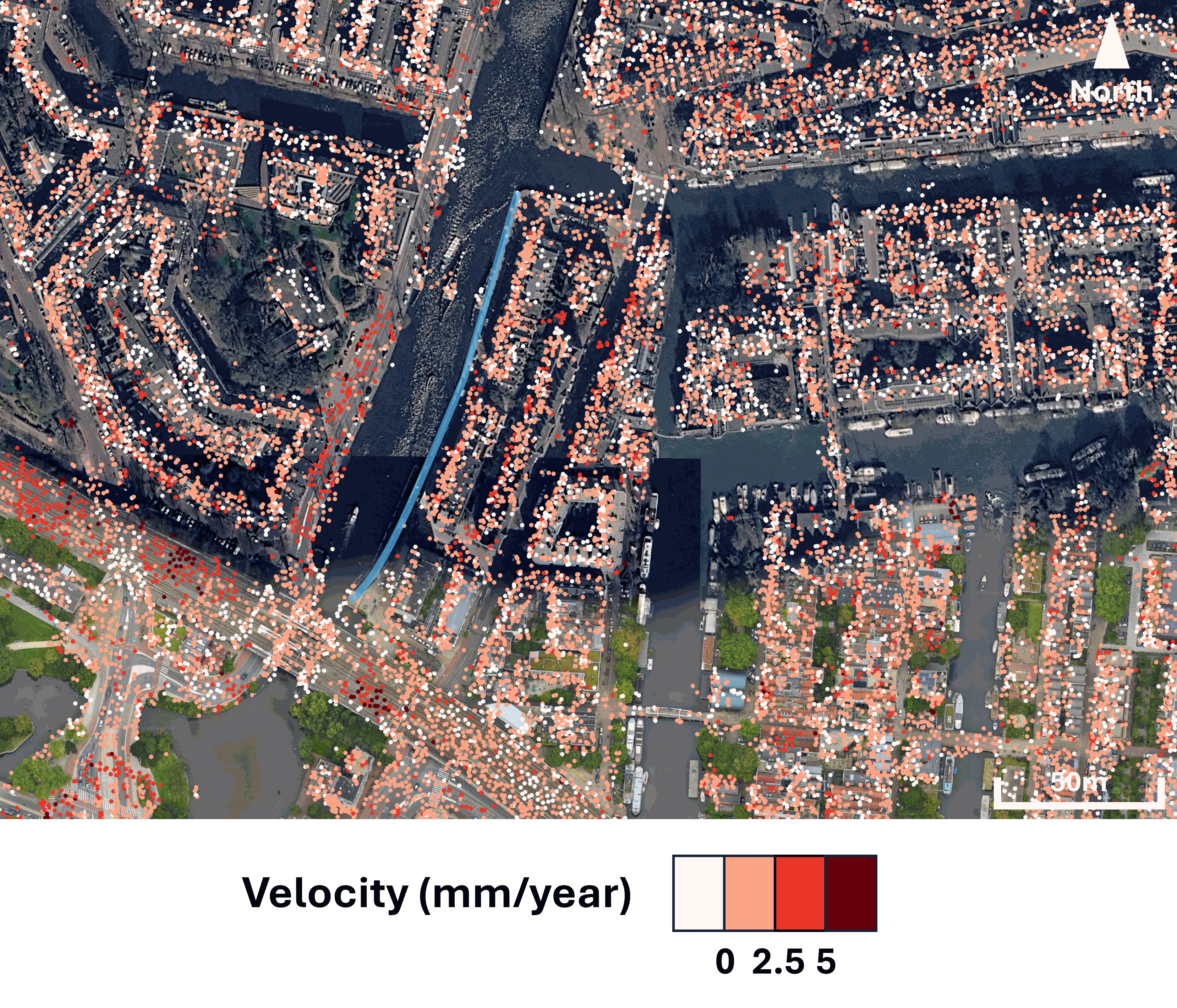}
        \caption{}
        \label{fig:ps_hout_asc}
    \end{subfigure}
    \hfill
    \begin{subfigure}{0.48\linewidth}
        \centering
        \includegraphics[width=1\linewidth]{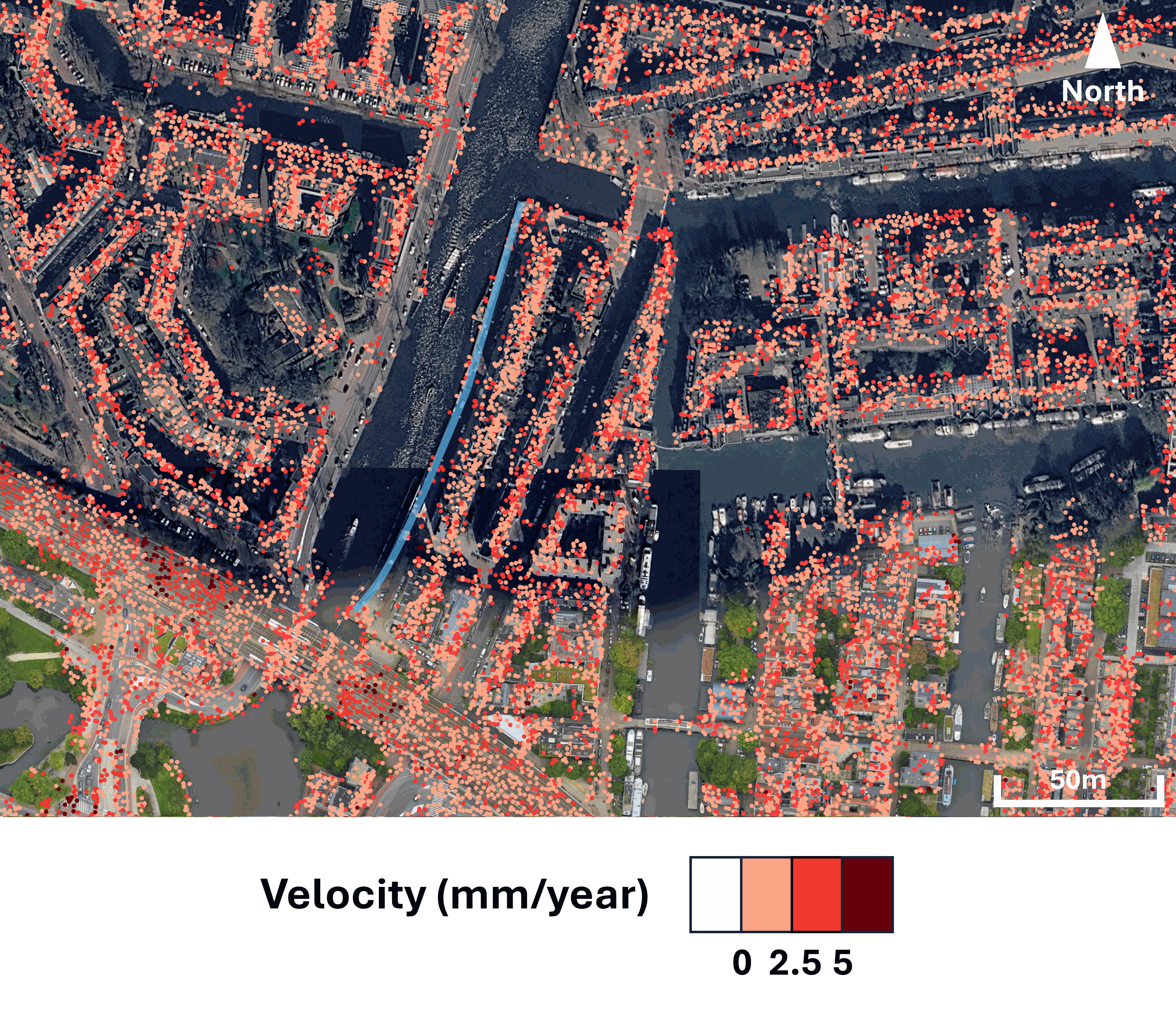}
        \caption{}
        \label{fig:ps_hout_des}
    \end{subfigure}
    \caption{\revisefour{Spatial distribution of PSs in the area around Houtmankade East quay (Basemap \copyright\ Google) from: (a) ascending and (b) descending track.}}
    \label{fig:ps_hout}
\end{figure}

\revise{Figure \ref{fig:ps_hout} presents the distribution of PSs derived from the ascending (a) and descending (b) tracks in the Houtmankade East area. \revisefour{The coloured points represent PS locations, and the colour scale indicates the estimated LOS velocity in mm/year. The blue line denotes the location of the Houtmankade East quay.} The relatively high PS density provides a suitable basis for reliable time-series analysis of ground displacement. The PSs and their LOS displacement time series were obtained from TerraSAR-X images with a spatial resolution of 3 meters, including 109 ascending-track and 114 descending-track images acquired between March 2016 and December 2019. These images were processed by SkyGeo using their proprietary PyAntares algorithm \cite{f.j.vanleijenPersistentScattererInterferometry2014}.}

\section{Results}
\revisefour{The proposed framework was applied to the selected quay wall cases. The resulting exceedance probabilities and warning detections were evaluated to assess the performance of the proposed approach.}
\subsection{Grimburgwal}
\label{sec:Grim.1}

\begin{figure*}
    \centering
    \begin{subfigure}{0.48\linewidth}
        \centering
        \includegraphics[width=1\linewidth]{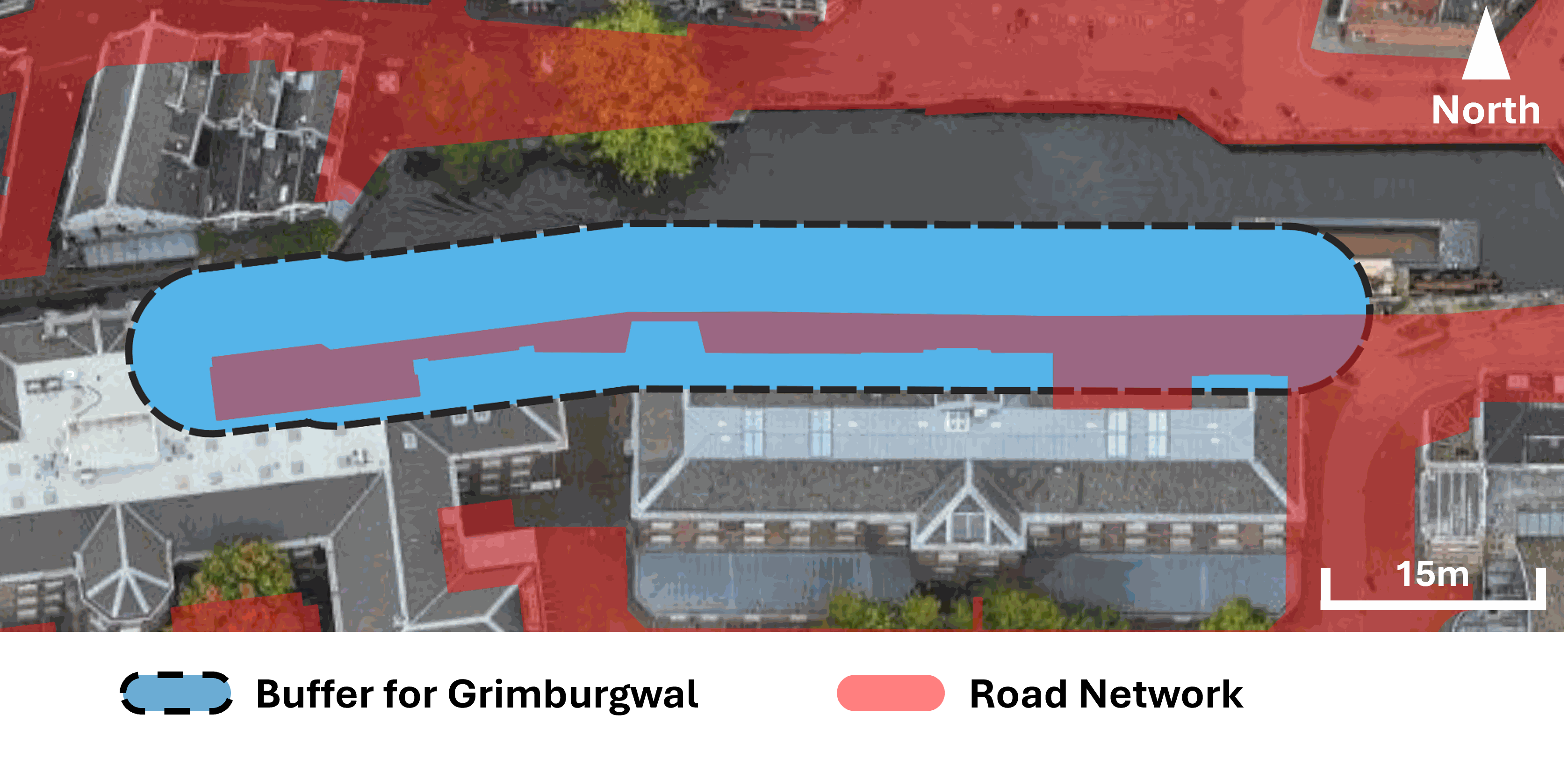}
        \caption{}
        \label{fig:grim_UMRsvsroad}
    \end{subfigure}
    \hfill
    \begin{subfigure}{0.48\linewidth}
        \centering
        \includegraphics[width=1\linewidth]{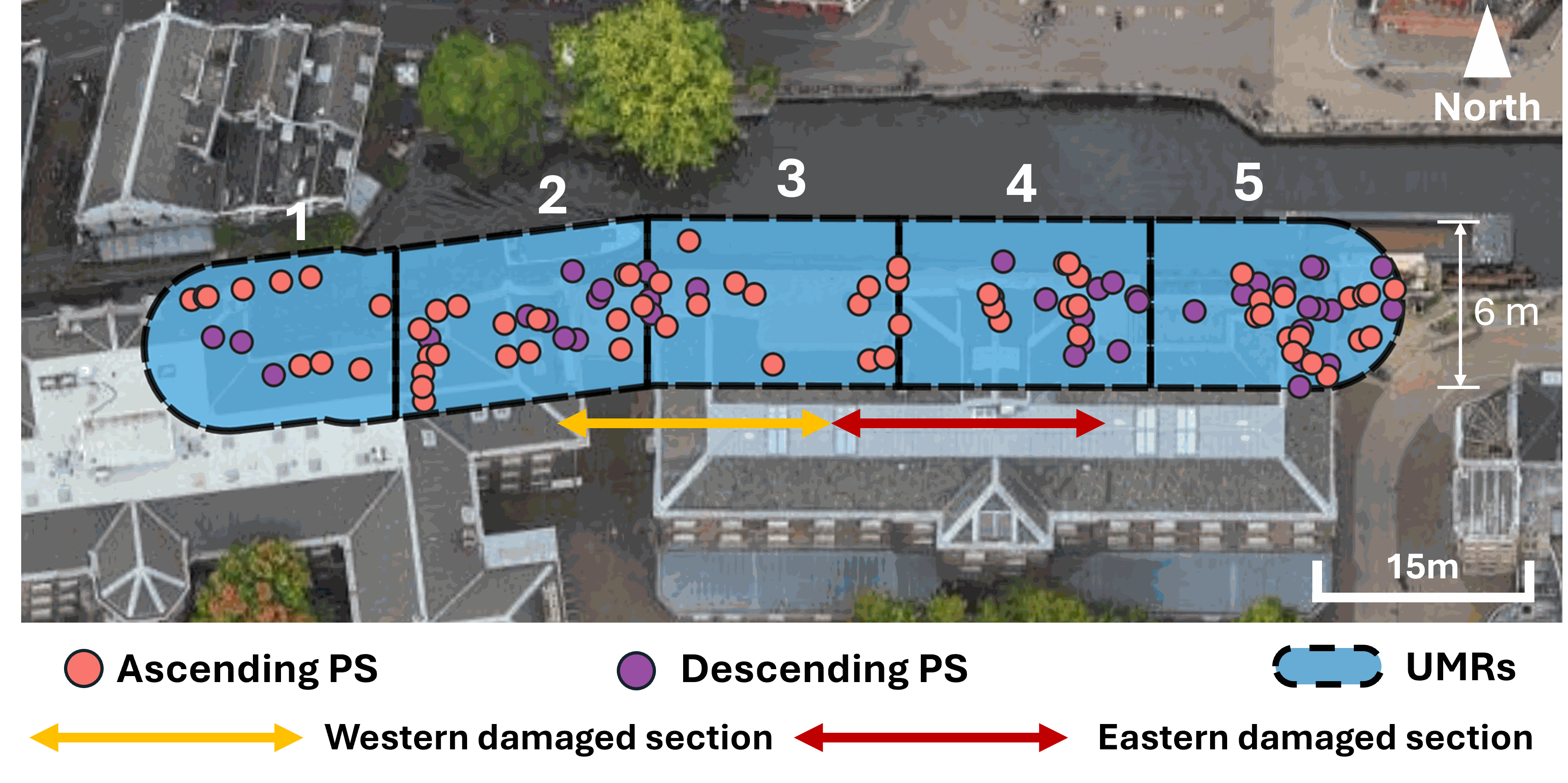}
        \caption{}
        \label{fig:grim_UMRs_regions}
    \end{subfigure}
    \begin{subfigure}{0.48\linewidth}
        \centering
        \includegraphics[width=1\linewidth]{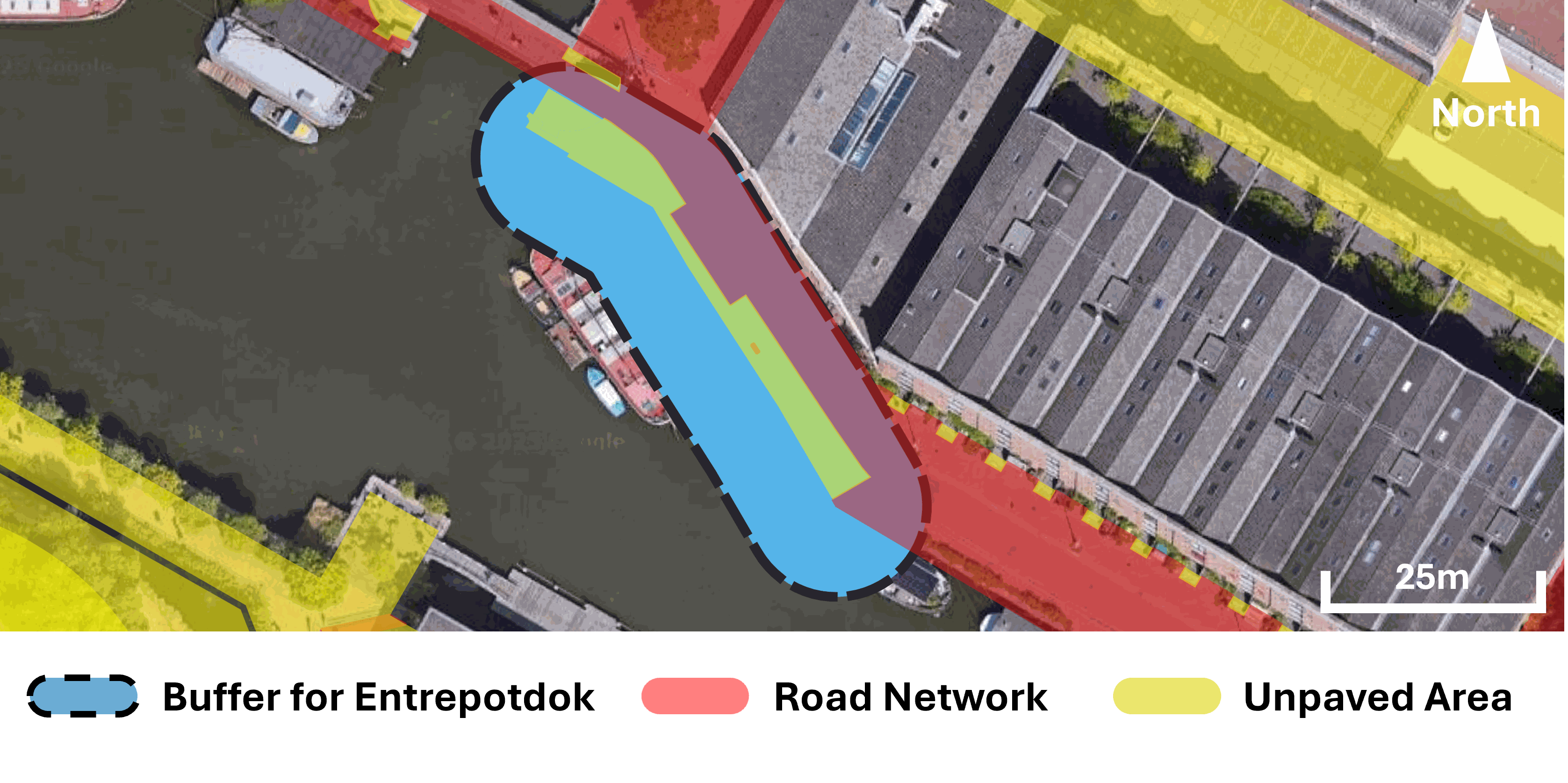}
        \caption{}
        \label{fig:entre_UMRsvsroad}
    \end{subfigure}
    \hfill
    \begin{subfigure}{0.48\linewidth}
        \centering
        \includegraphics[width=1\linewidth]{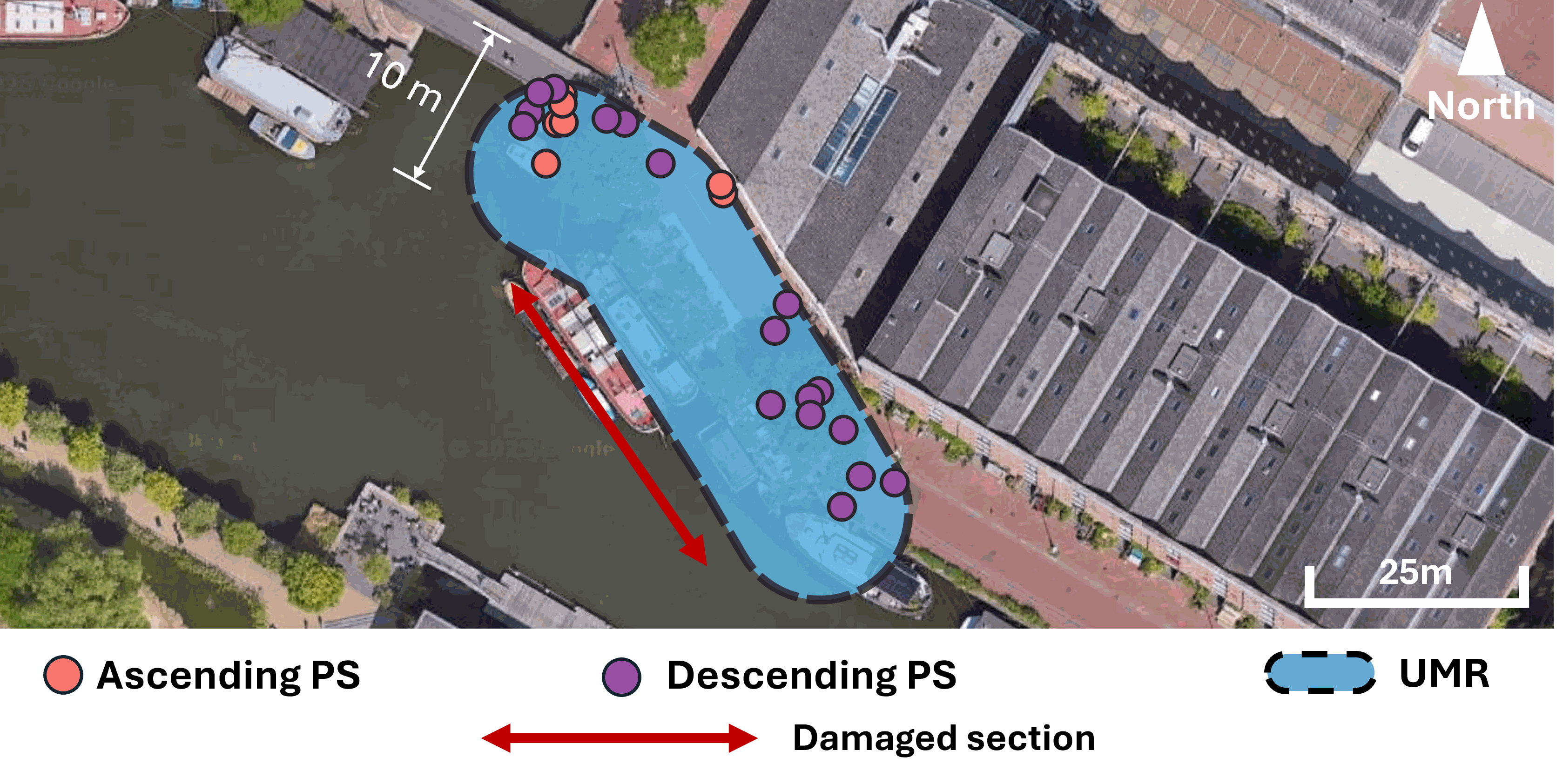}
        \caption{}
        \label{fig:entre_UMRs_regions}
    \end{subfigure}
    \begin{subfigure}{0.48\linewidth}
        \centering
        \includegraphics[width=1\linewidth]{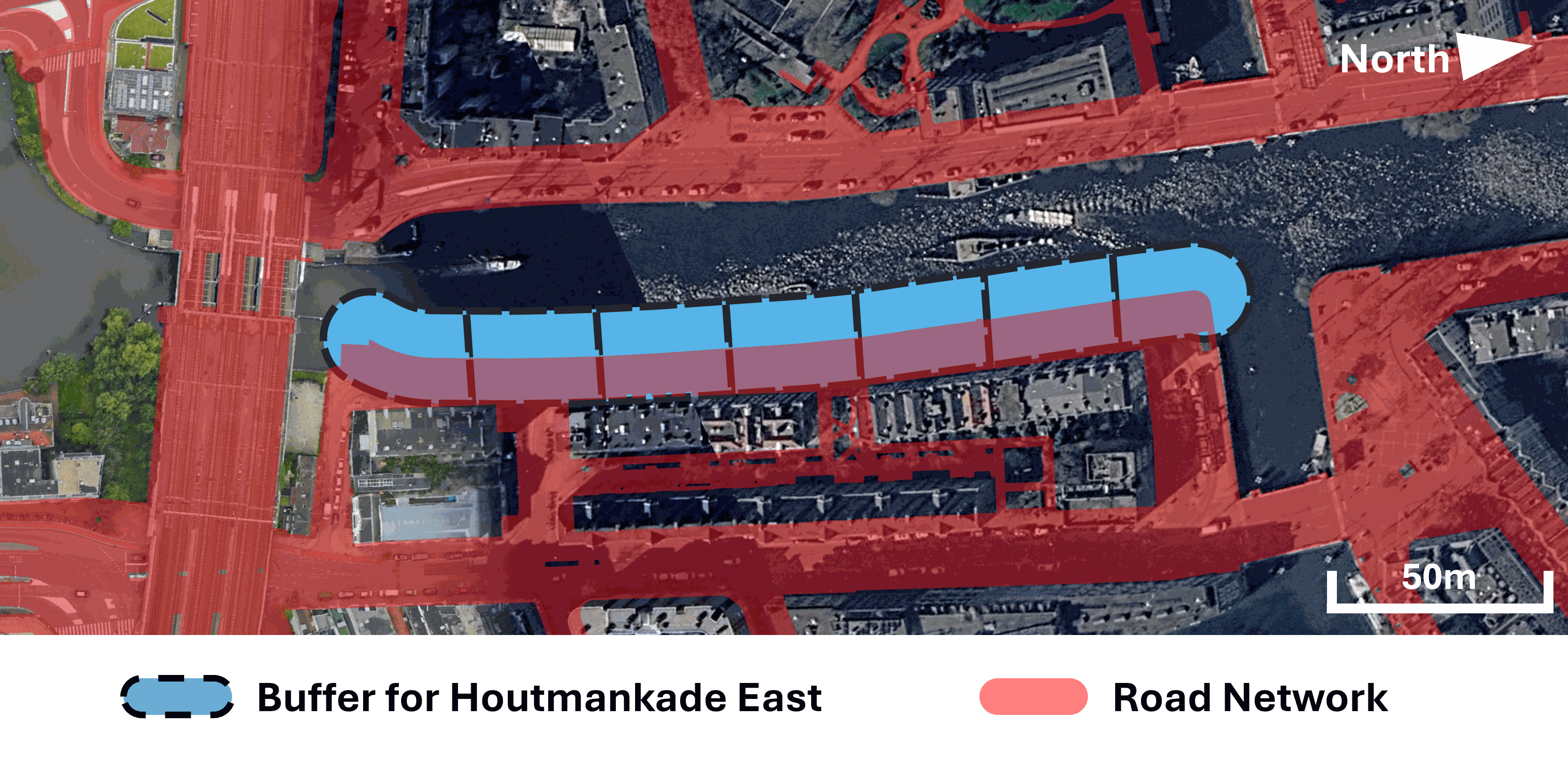}
        \caption{}
        \label{fig:hout_UMRsvsroad}
    \end{subfigure}
    \hfill
    \begin{subfigure}{0.48\linewidth}
        \centering
        \includegraphics[width=1\linewidth]{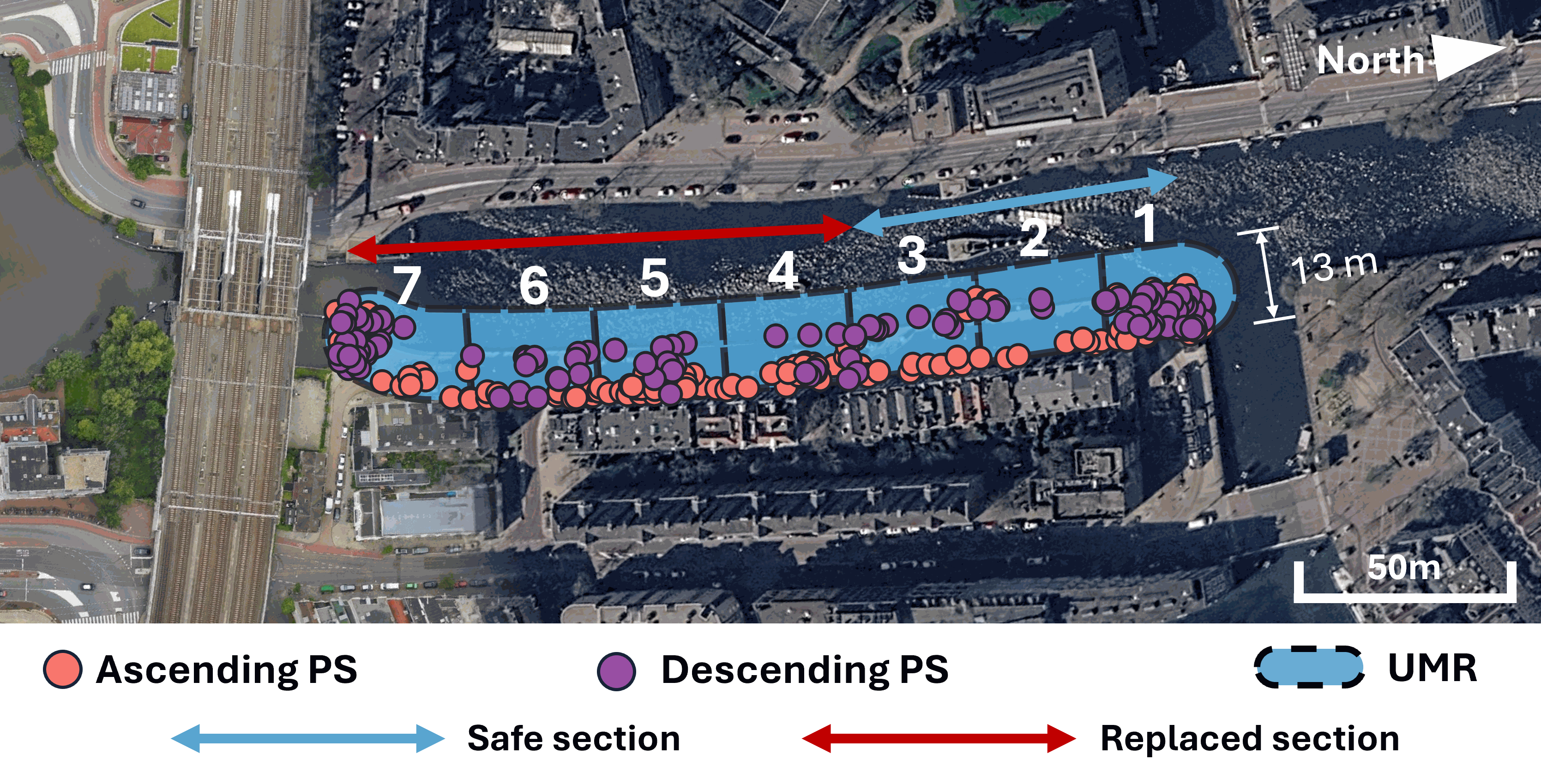}
        \caption{}
        \label{fig:hout_UMRs_regions}
    \end{subfigure}
    \caption{\revise{Defined buffer and the surrounding road network geometry for (a) Grimburgwal, (c) Entrepotdok and (e) Houtmankade East. Defined UMRs with PSs from ascending and descending tracks for (b) Grimburgwal, (d) Entrepotdok and (f) Houtmankade East  (basemap \copyright\ Google).}}
    \label{fig:UMRs}
\end{figure*}

\revise{PSs were first filtered for the Grimburgwal case. For the vertical dimension, only PSs with heights ranging from $0$ to $5$ m were retained. For the horizontal dimensions, a buffer was defined according to Section \ref{sec.2.1}. Considering the spatial resolution of TerraSAR-X data, the image-related uncertainty was set to $\delta_I = 6$ m, following the general approach described by Macchiarulo et al. \citep{macchiaruloStructuralEvaluationUrban2024}. The road width $\delta_R$ was estimated from the PDOK dataset (Dutch National Geo-Information Public Service Platform), as illustrated in Figure \ref{fig:grim_UMRsvsroad} \citep{dutchnationalgeo-informationpublicserviceplatformpdokBasisregistratieGrootschaligeTopografie2025}. The extracted road geometry indicates that the road width along Grimburgwal does not exceed 3 m. According to Equation \ref{eq:buffer}, the resulting buffer width was therefore set to 6 m for this study area. This choice ensures sufficient spatial coverage of the quay structure while limiting the inclusion of surrounding urban features.}

Figure \ref{fig:grim_UMRs_regions} shows the defined buffer together with the ascending and descending PSs located within it. UMRs were automatically defined following the method described in Section \ref{sec.2.1}. In total, five UMRs were identified, of which UMR 2, 3 and 4 approximately corresponded to the observed collapse area \citep{korffBezwijkenGrimburgwalLeerpunten2021, korffCollapseGrimburgwalHistoric2022}.

\begin{figure}
    \centering
    \begin{subfigure}[t]{0.48\linewidth}
        \centering
        \includegraphics[width=\linewidth]{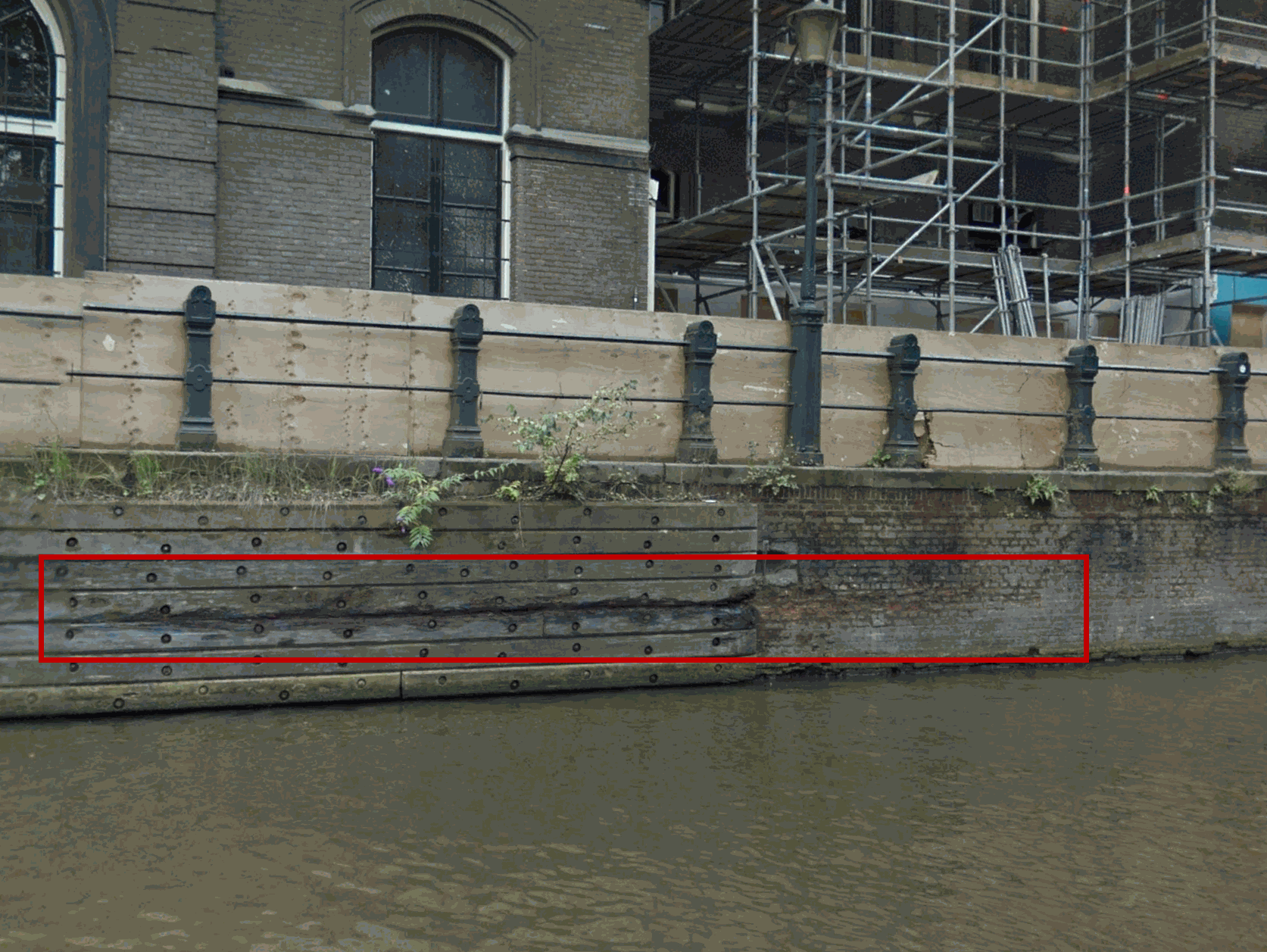}
        \caption{}
        \label{fig:grimburgwal_symbol_a}
    \end{subfigure}
    \hfill
    \begin{subfigure}[t]{0.48\linewidth}
        \centering
        \includegraphics[width=\linewidth]{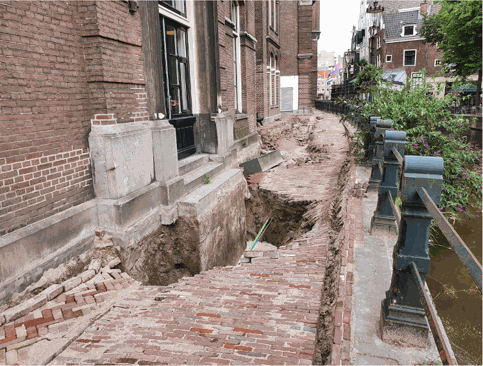}
        \caption{}
        \label{fig:grimburgwal_symbol_b}
    \end{subfigure}
    \caption{\revise{Evidence supporting the progressive collapse of Grimburgwal: (a) ship collisions, where the red box highlights the affected part (image \copyright Google, August 2014); (b) localised subsidence observed prior to the collapse \citep{korffCollapseGrimburgwalHistoric2022}.}}
    \label{fig:grimburgwal_symbol}
\end{figure}

\revise{In particular, UMR 4 and a small portion of UMR 3 correspond to the eastern damaged section, whereas the remaining part of UMR 3 and UMR 2 correspond to the western damaged section (Figure \ref{fig:grim_UMRs_regions}). Previous investigations indicated that the collapse initiated in the eastern section and subsequently propagated westward, where progressive tilting occurred prior to failure \citep{korffBezwijkenGrimburgwalLeerpunten2021, korffCollapseGrimburgwalHistoric2022}. Supporting evidence is presented in Figure \ref{fig:grimburgwal_symbol}. The eastern section is located at a turning point frequently used by tour boats, where repeated collisions have been reported (Figure \ref{fig:grimburgwal_symbol_a}). Consistent with this observation, pronounced localised subsidence was observed in the eastern section shortly before collapse (Figure \ref{fig:grimburgwal_symbol_b}), suggesting more severe structural degradation compared to the western section.}

\begin{figure*}
    \centering
    \begin{subfigure}{0.4\linewidth}
        \centering
        \includegraphics[width=1\linewidth]{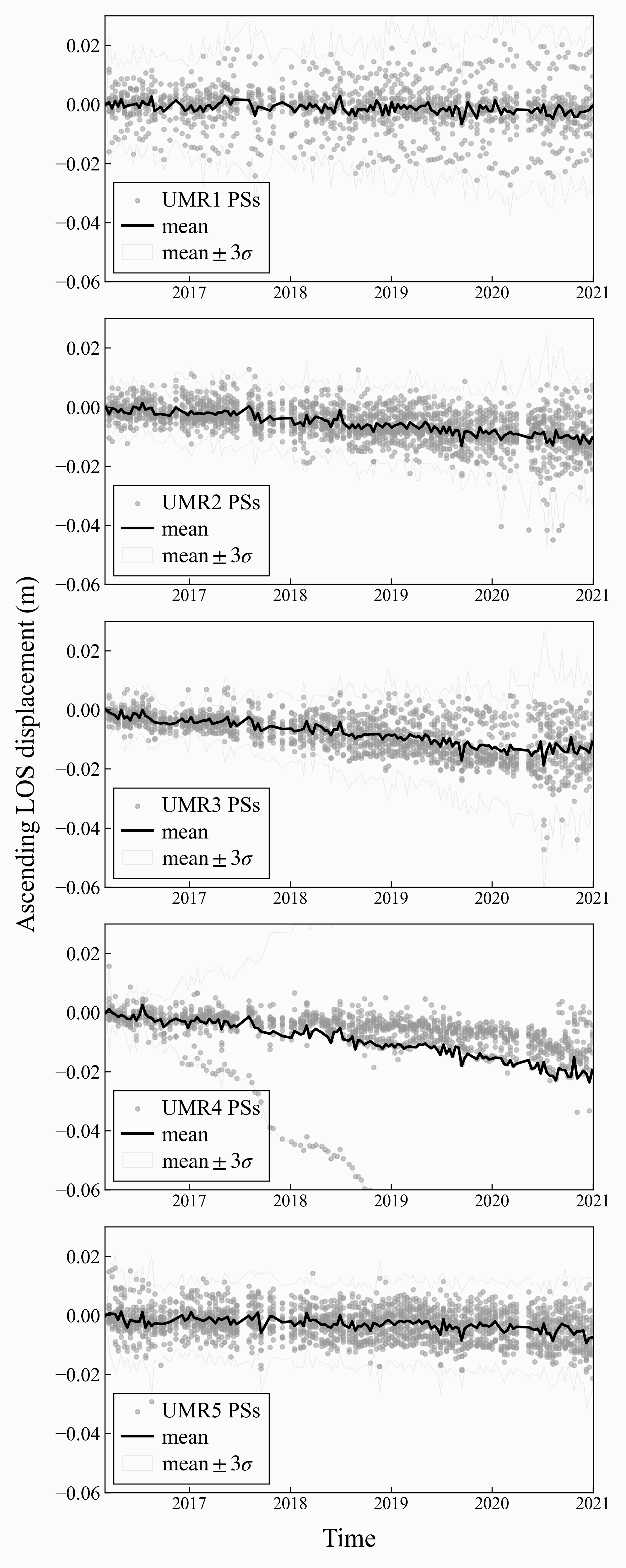}
        \caption{}
        \label{fig:los_grim_a}
    \end{subfigure}
    \hfill
    \begin{subfigure}{0.4\linewidth}
        \centering
        \includegraphics[width=1\linewidth]{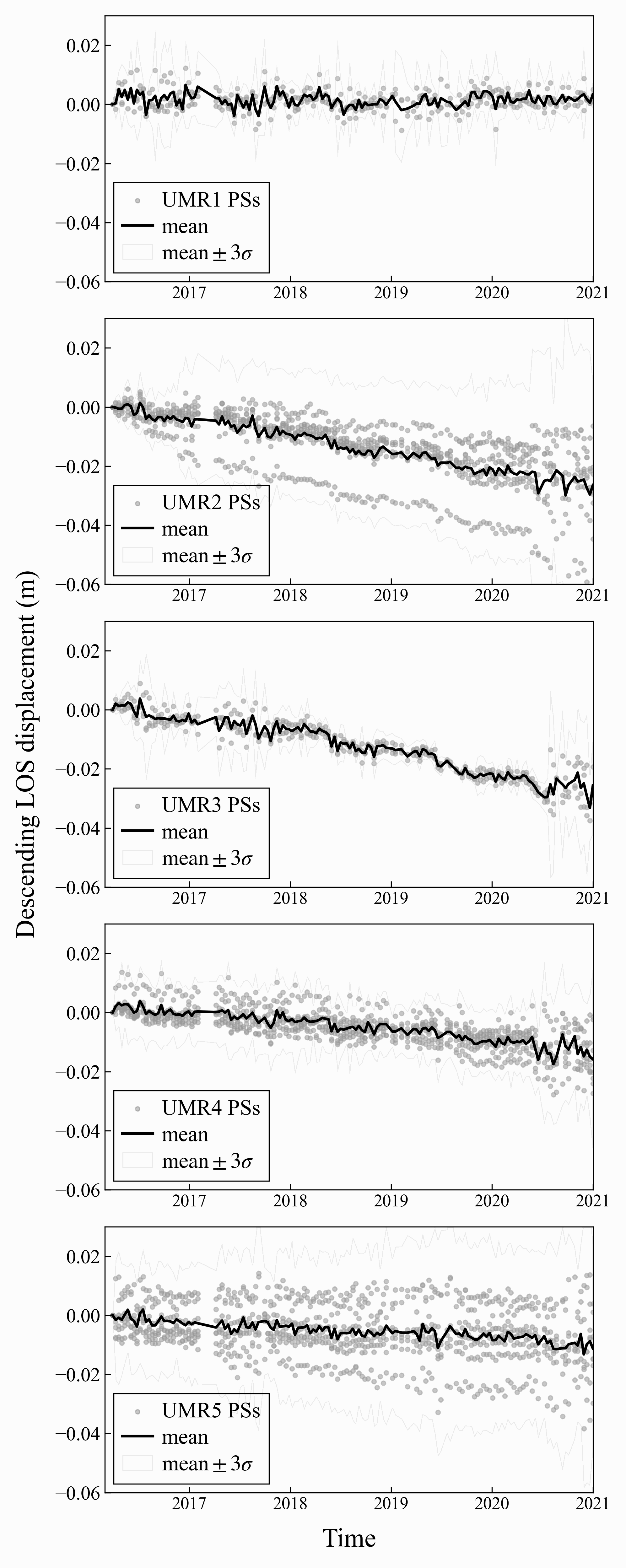}
        \caption{}
        \label{fig:los_grim_d}
    \end{subfigure}
    \caption{\revise{LOS displacements of PSs for each Grimburgwal UMR from: (a) ascending and (b) descending track.}}
    \label{fig:los_grim}
\end{figure*}

To assess the deformation pattern of the quay wall, LOS displacement time series of the PSs in each UMR were analysed. Figure \ref{fig:los_grim} presents the results for the ascending (a) and descending (b) tracks, including individual PS observations, mean displacement, and associated uncertainty range. Within each UMR, significant variability is evident among individual PSs. This is particularly notable in UMR 4 for the ascending track, where large deviations of LOS displacements from PSs result in a mean LOS displacement that may be statistically unrepresentative. Relying solely on this mean could lead to underestimation or overestimation of structural deformation, resulting in an erroneous condition assessment. These results demonstrate that mean LOS displacement alone is insufficient for reliably capturing the deformation pattern and that associated uncertainties must be taken into account.

\begin{figure*}
    \centering
    \begin{subfigure}{0.4\linewidth}
        \centering
        \includegraphics[width=1\linewidth]{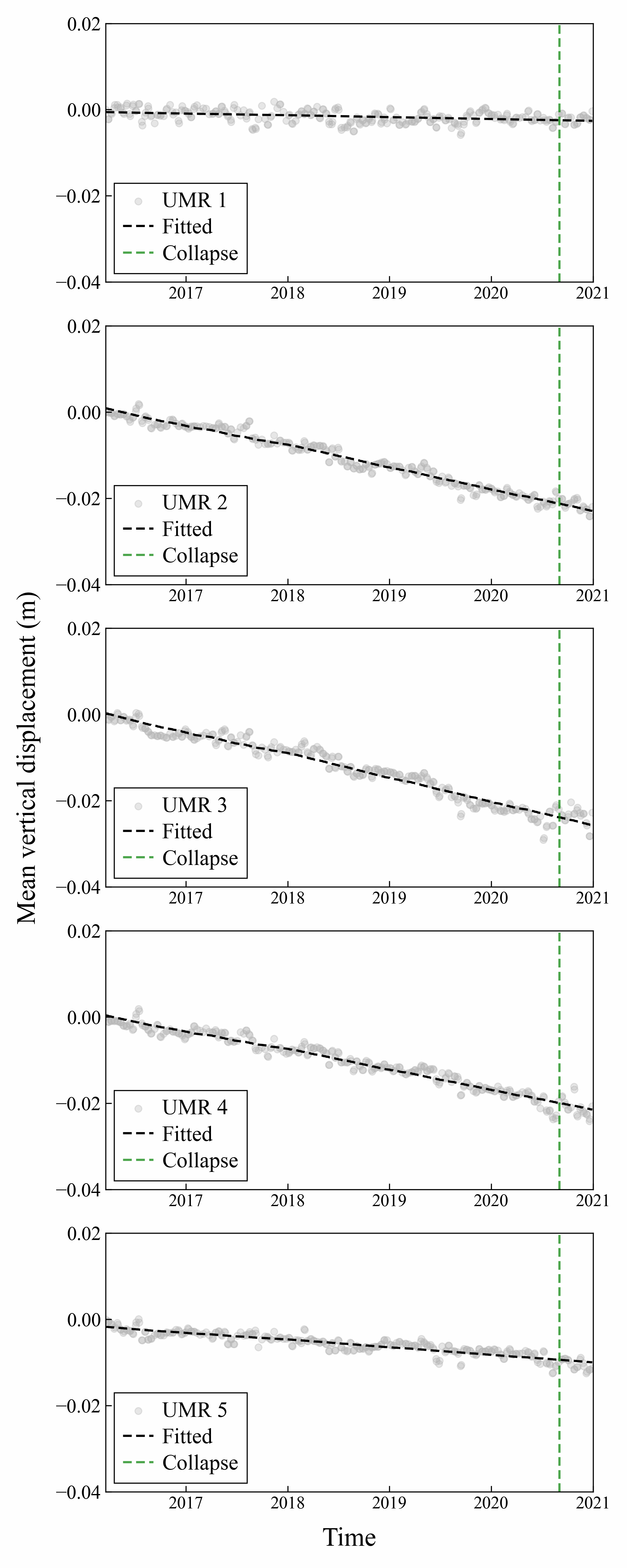}
        \caption{}
        \label{fig:grimburgwal_results_ts}
    \end{subfigure}
    \hfill
    \begin{subfigure}{0.4\linewidth}
        \centering
        \includegraphics[width=1\linewidth]{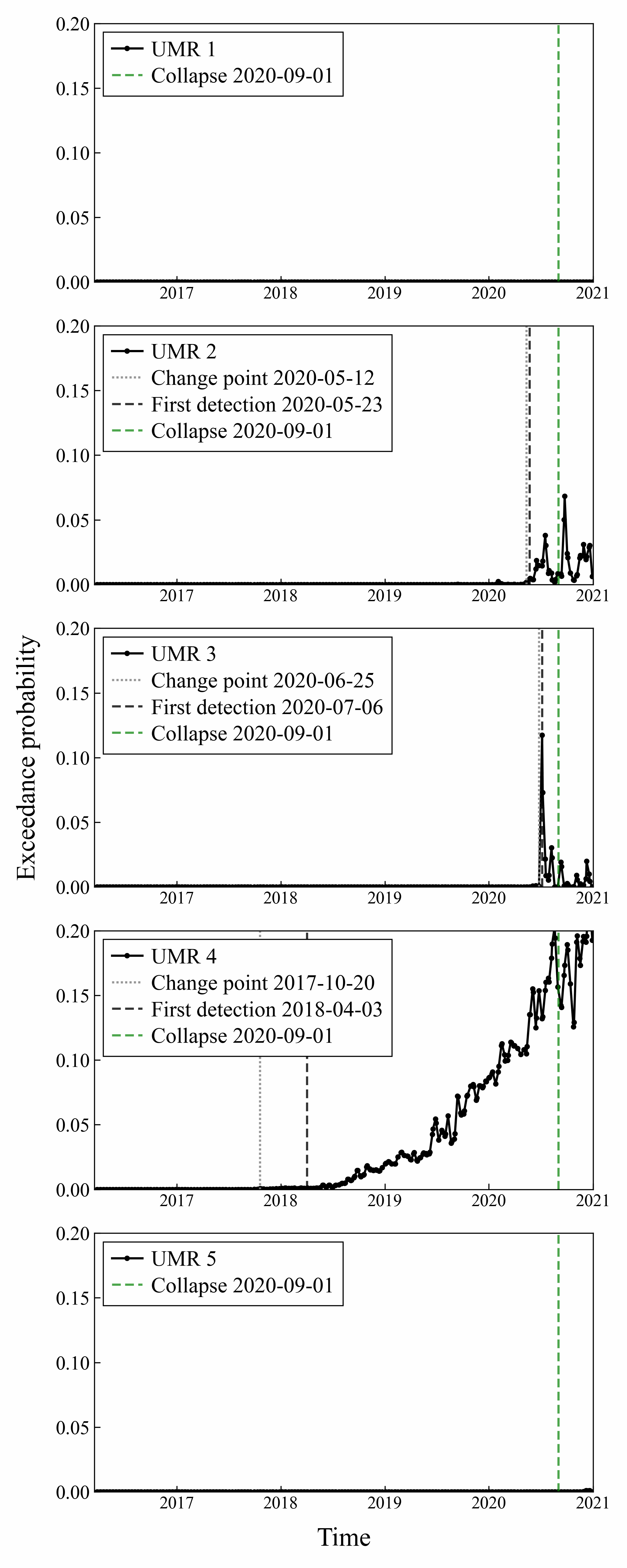}
        \caption{}
        \label{fig:grimburgwal_results_pf}
    \end{subfigure}
    \caption{\revisethree{Grimburgwal results of each UMR: (a) mean vertical displacement time series and (b) exceedance probability over time.}}
    \label{fig:grimburgwal_results}
\end{figure*}

In the proposed framework, LOS displacements were modelled as Gaussian distributions, with mean and variance estimated using Equations \ref{eq:unbiased_mean} and \ref{eq:unbiased_var}. As illustrated in Figure \ref{fig:grim_UMRs_regions}, the transversal axis of the Grimburgwal quay is aligned with the satellite track in the north-south direction, rendering LOS observations insensitive to displacements in this direction. By assuming \revise{the contribution of the transversal component to the LOS displacement to be negligible}, vertical and longitudinal displacement components were therefore resolved using Equation \ref{eq:sl_dVdL}. Figure \ref{fig:grimburgwal_results_ts} presents the resulting time series of the mean vertical displacement, along with a linear regression fitted to the entire dataset.

\revisefour{As Dutch standards do not specify a maximum allowable vertical displacement for quay walls, a limit of $1\%$ of the retaining height, commonly adopted for horizontal displacement \citep{curRapport16620082008}}, was also applied for vertical displacements, following \cite{bettsPortBotanyExpansion2011}. This yielded a threshold of $40$ mm for Grimburgwal.

Comparison of the MT-InSAR-derived vertical displacements against this threshold revealed clear differences between UMRs (Figure \ref{fig:grimburgwal_results_ts}). UMR 1 and UMR 5 appeared relatively stable, with vertical displacements of $0.3$ mm and $11$ mm, respectively, between March 2016 and December 2020. In contrast, UMRs 2, 3, and 4 exhibited significantly larger displacements, reaching $22$ mm, $23$ mm, and $21$ mm. \revise{Yet none exceeded the $40$ mm threshold, highlighting the limitations of traditional deterministic methods (i.e., fixed displacement threshold) when applied to MT-InSAR data.}

\revisetwo{The limit state function (Equation \ref{eq:LSF}) was used to calculate the probability of exceedance for each UMR using Equation \ref{eq:analytical_pf}. The results are presented in Figure \ref{fig:grimburgwal_results_pf}. The probability of exceedance for UMR 1 and 5 remained close to zero throughout the entire observation period. In contrast, UMRs 2 and 3 exhibit a sharp increase around mid-2020. UMR 4 shows a different trend, with a gradual increase starting in 2018, followed by a continuous rise until the collapse.}

\revisethree{Applying the PELT method to these exceedance probabilities enabled the automatic detection of abrupt changes. The first detection dates, at which warnings were triggered, are marked by black dashed lines, while the corresponding estimated change point dates are indicated by grey dotted lines. The actual collapse date of 2020-09-01 is indicated by a green dashed line. The earliest anomaly warning was triggered in UMR 4 on 2018-04-03, more than two years before the collapse, followed by warnings in UMR 2 on 2020-05-23 and in UMR 3 on 2020-07-06.}
Overall, the proposed framework would have provided warnings at least two months prior to collapse. The triggering UMRs (UMRs 2, 3, and 4) correspond closely to the actual collapse region. Notably, UMR 4, which corresponds to the eastern damaged section, issued a warning approximately two years earlier than UMRs 2 and 3. This is consistent with the observations that damage initiated in the eastern section and subsequently led to collapse.
\revisetwo{It should be emphasised that a warning triggered by a deformation anomaly does not necessarily indicate an imminent structural collapse. Rather, it serves as a proactive indicator, providing stakeholders with critical time to conduct detailed in-situ investigations or implement preventive maintenance.}

\subsection{Entrepotdok}

\revise{Applying the filter methods proposed in Section \ref{sec.2.1}, PSs were vertically restricted to heights between $0$ and $5$ m. For the horizontal buffer, the image-related uncertainty was set to $\delta_I = 6$ m, based on the spatial resolution of TerraSAR-X data and following \citep{macchiaruloStructuralEvaluationUrban2024}. The top surface of the Entrepotdok quay consists of a roadway and an adjacent unpaved area. Figure \ref{fig:entre_UMRsvsroad} shows these features derived from the PDOK dataset \citep{dutchnationalgeo-informationpublicserviceplatformpdokBasisregistratieGrootschaligeTopografie2025}, with a total width of approximately 5 meters. This combined width is represented by $\delta_R$, reflecting the full width of the quay-top area. According to Equation \ref{eq:buffer}, resulting in $\delta_B = 10$ m for this case.}
With this buffer, ascending PSs are concentrated to the northwest of the quay wall, preventing subdivision into multiple UMRs while still ensuring that both ascending and descending PSs are present in each segment. Therefore, the entire buffer was treated as a single UMR.

\begin{figure*}
    \centering
    \begin{subfigure}{0.4\linewidth}
        \centering
        \includegraphics[width=\linewidth]{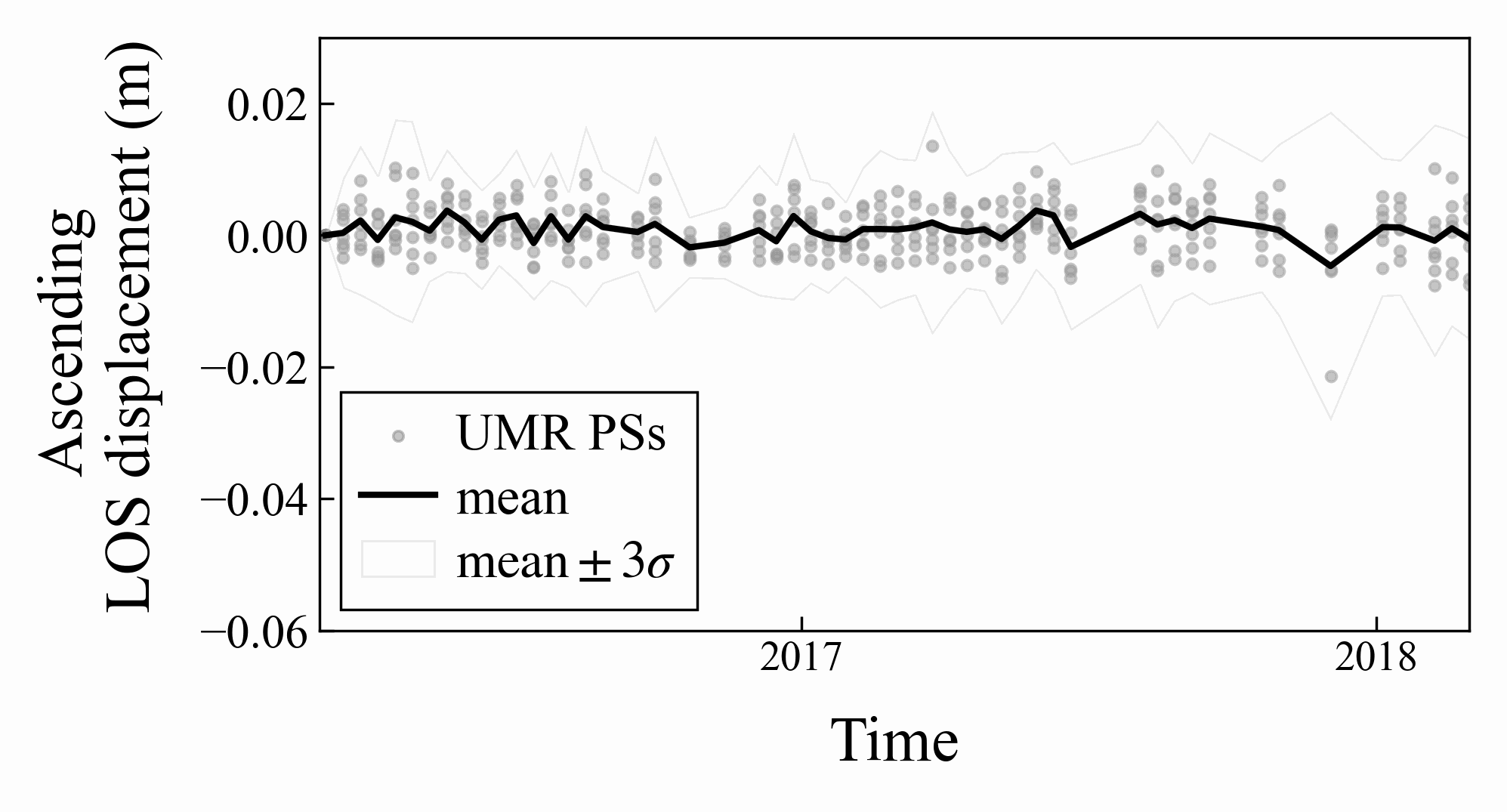}
        \caption{}
        \label{fig:los_entre_a}
    \end{subfigure}
    \hfill
    \begin{subfigure}{0.4\linewidth}
        \centering
        \includegraphics[width=\linewidth]{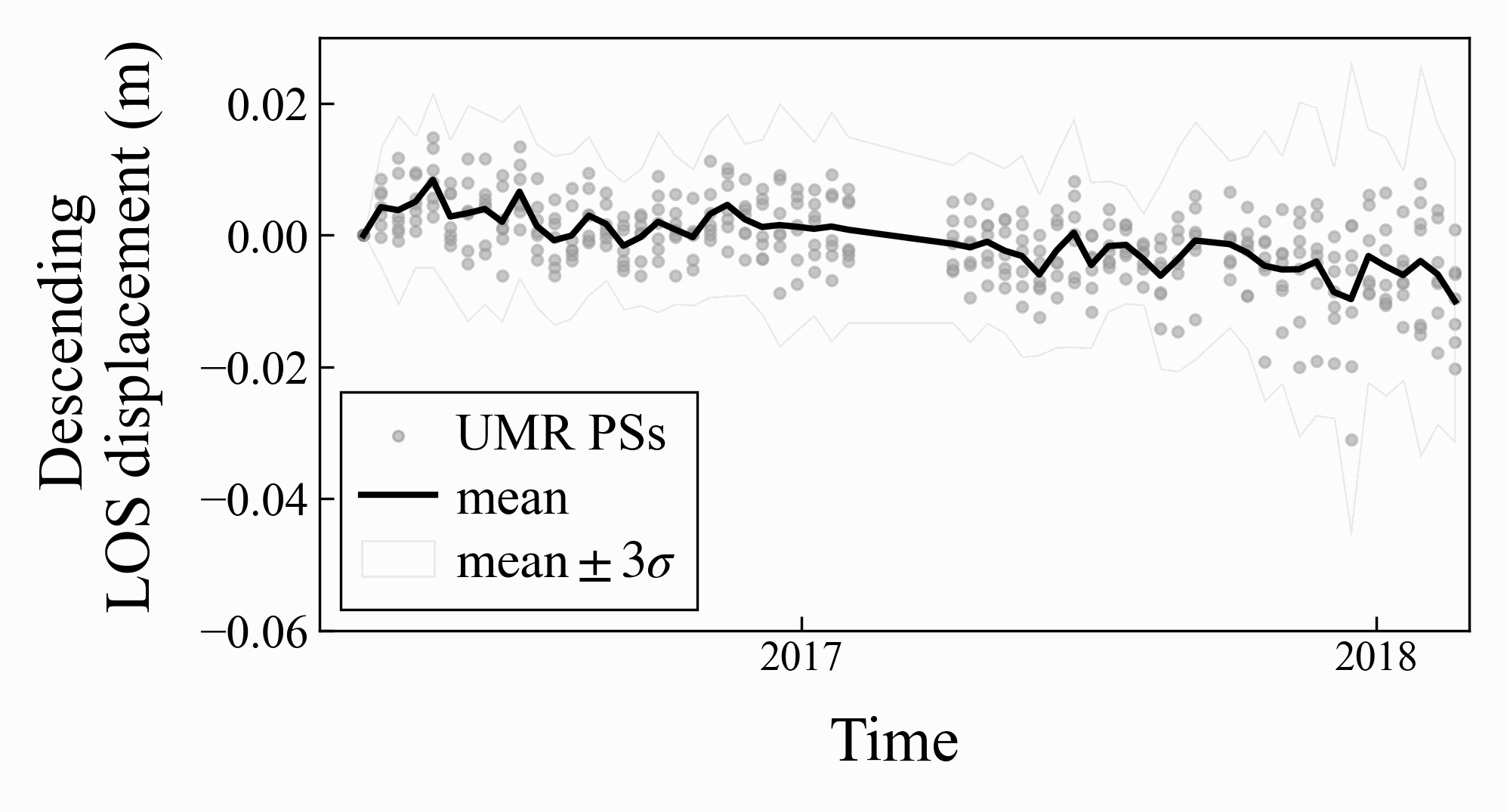}
        \caption{}
        \label{fig:los_entre_d}
    \end{subfigure}
    \caption{\revise{LOS displacements of PSs for Entrepotdok from: (a) ascending and (b) descending track.}}
    \label{fig:los_entre}
\end{figure*}

The LOS displacement time series of each PS in the UMR are shown in Figure \ref{fig:los_entre}, alongside the mean LOS displacement. Significant variability is evident among individual PSs, highlighting the importance of accounting for uncertainties. The LOS displacements were modelled as Gaussian distributions incorporating all PSs within the UMR, with the mean and variance estimated using Equations \ref{eq:unbiased_mean} and \ref{eq:unbiased_var}.

\begin{figure*}
    \centering
    \begin{subfigure}{0.4\linewidth}
        \centering
        \includegraphics[width=\linewidth]{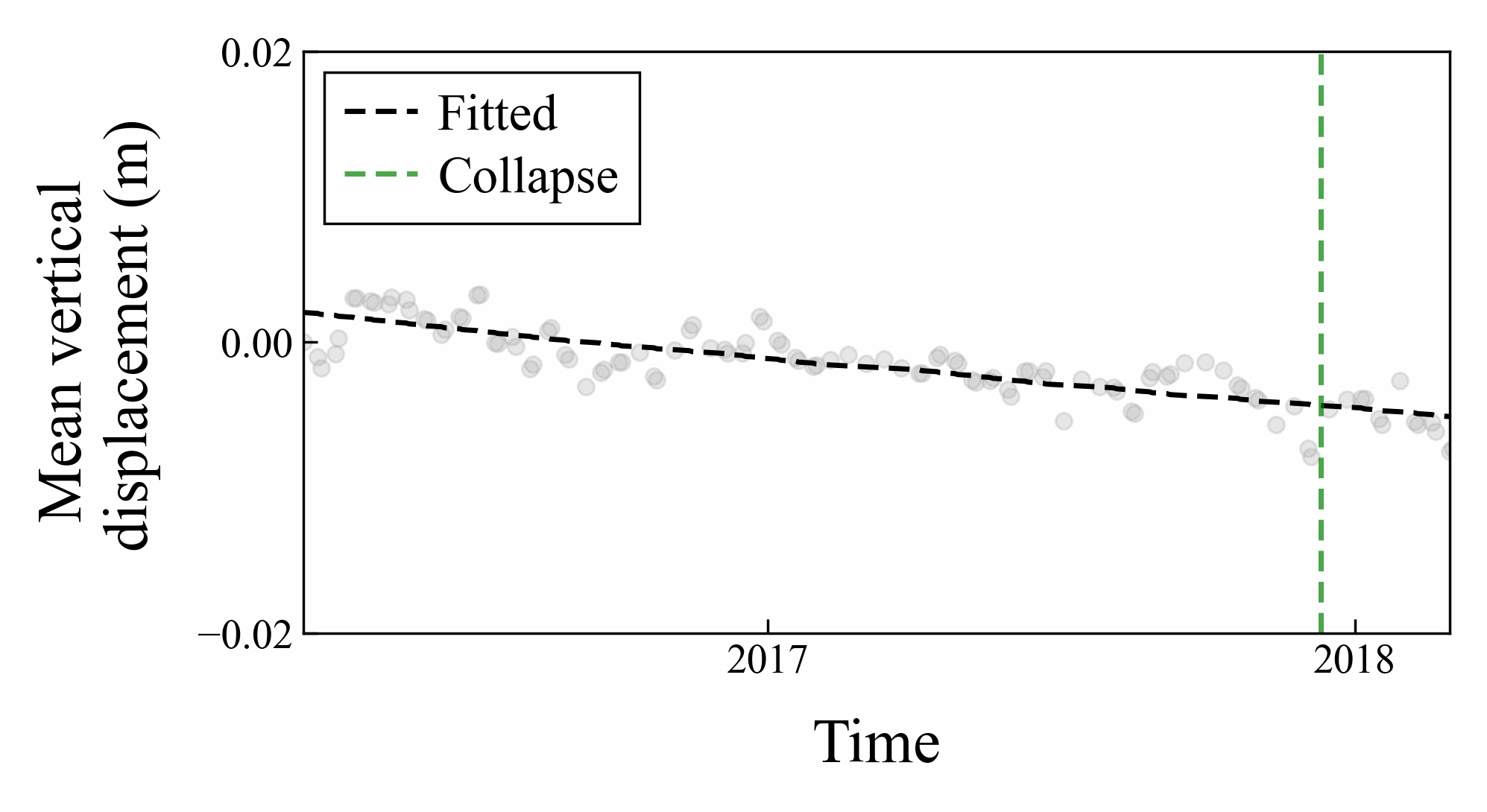}
        \caption{}
        \label{fig:entre_results_ts}
    \end{subfigure}
    \hfill
    \begin{subfigure}{0.4\linewidth}
        \centering
        \includegraphics[width=\linewidth]{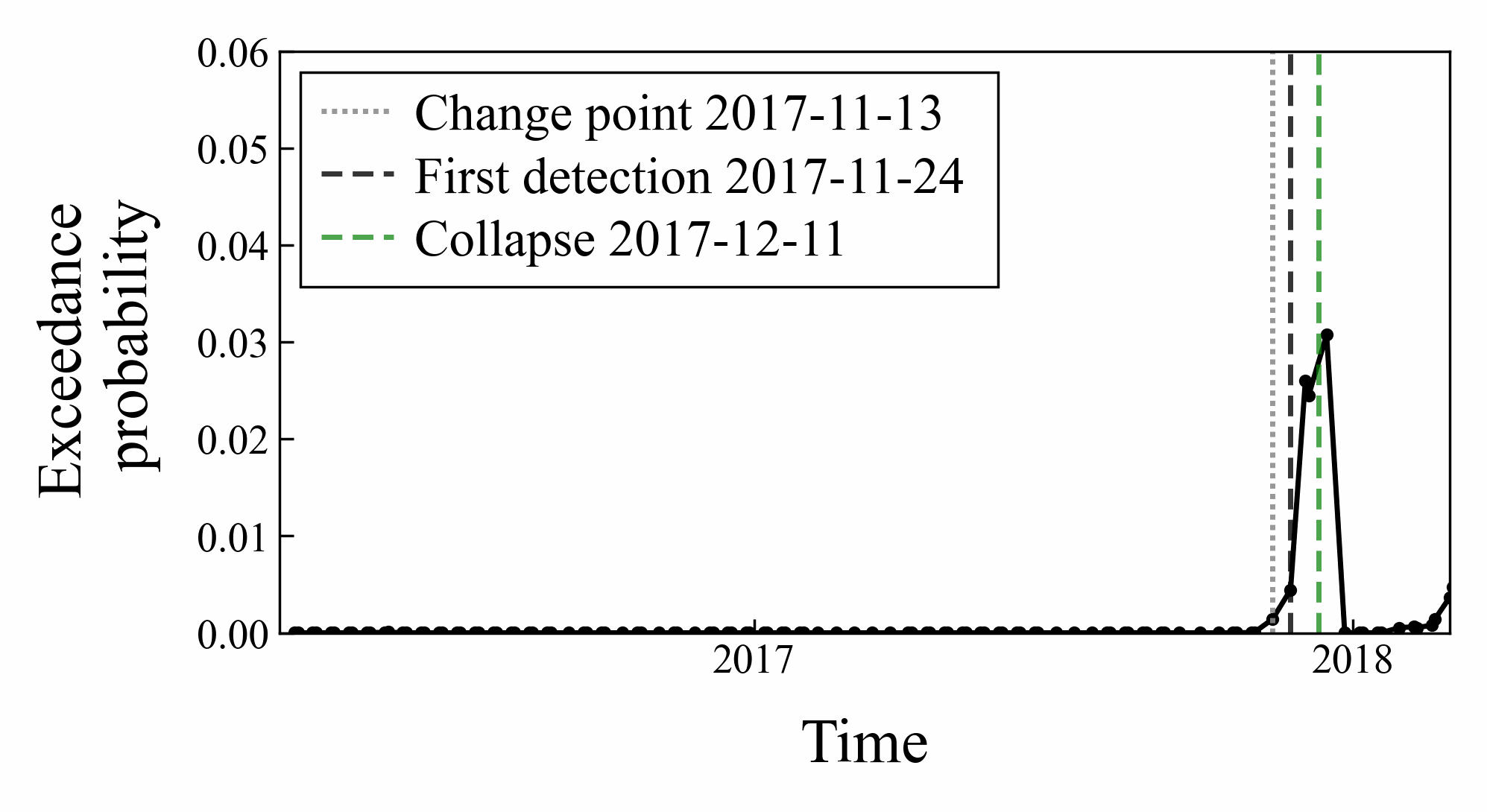}
        \caption{}
        \label{fig:entre_results_pf}
    \end{subfigure}
    \caption{\revisethree{Entrepotdok results: (a) mean vertical displacement time series and (b) exceedance probability over time.}}
    \label{fig:entre_results}
\end{figure*}

As shown in Figure \ref{fig:entre_UMRs_regions}, and in contrast to the Grimburgwal case, the longitudinal axis of the Entrepotdok forms an angle smaller than $45^\circ$ with respect to the satellite track. Consequently, \revise{the contribution of the longitudinal component to the LOS displacement} was assumed to be negligible, allowing the vertical and transversal displacements to be resolved using Equation \ref{eq:sl_dVdT}. Figure \ref{fig:entre_results_ts} presents the time series of the mean vertical displacement, along with a linear regression trend fitted to the entire dataset.

As in the previous case, the maximum allowable vertical displacement was set to 1\% of the retaining height, i.e $20$ mm. \revise{Since the observed vertical displacement did not exceed this limit, this collapse would be classified as safe under traditional threshold-based approaches.} \revisetwo{However, applying the limit state function to calculate the probability of exceedance via Equation \ref{eq:analytical_pf} again revealed critical information.} As shown in Figure \ref{fig:entre_results_pf}, a distinct temporal pattern emerges. \revisetwo{The probability of exceedance remained near zero until October 2017}, followed by a sudden increase, shortly before the collapse. After January 2018, the probability stabilised again, coinciding with the installation of the temporary sheet pile wall.
\revisethree{Using the PELT method, the abrupt increase in the probability of exceedance was automatically detected (Figure \ref{fig:entre_results_pf}). The first detection date, at which the warning was triggered, is marked by a black dash line, while the corresponding estimated change point date is indicated by a grey dotted line. The warning date was detected as 2017-11-24, approximately three weeks before the collapse.}


\subsection{\revise{Houtmankade East}}

\revise{Following the same procedure for positional uncertainties, vertical filtering was set to $0-5$ m for PSs. Considering the horizontal buffer, the image-related uncertainty was set to $\delta_I = 6$ m, based on the spatial resolution of TerraSAR-X data \citep{macchiaruloStructuralEvaluationUrban2024}. The road width along Houtmankade East was estimated from the PDOK dataset (Figure \ref{fig:hout_UMRsvsroad}) \citep{dutchnationalgeo-informationpublicserviceplatformpdokBasisregistratieGrootschaligeTopografie2025}, with $\delta_R = 6.5$ m. Using Equation \ref{eq:buffer}, the buffer width was thus determined as $\delta_B = 13$ m.}

\revise{Figure \ref{fig:hout_UMRs_regions} shows the defined buffer together with the ascending and descending PSs located within it. UMRs were automatically defined following the method described in Section \ref{sec.2.1}, resulting in a total of seven regions. UMRs 1–3 correspond to the undamaged part and were used to evaluate the potential for false alarms within the framework. UMRs 4–7 represent the replaced parts, with UMR 4 corresponding to the location where local deformation was detected and serving as the first trigger for detailed assessment and replacement.}

\begin{figure*}
    \centering
    \begin{subfigure}{0.4\linewidth}
        \centering
        \includegraphics[width=\linewidth]{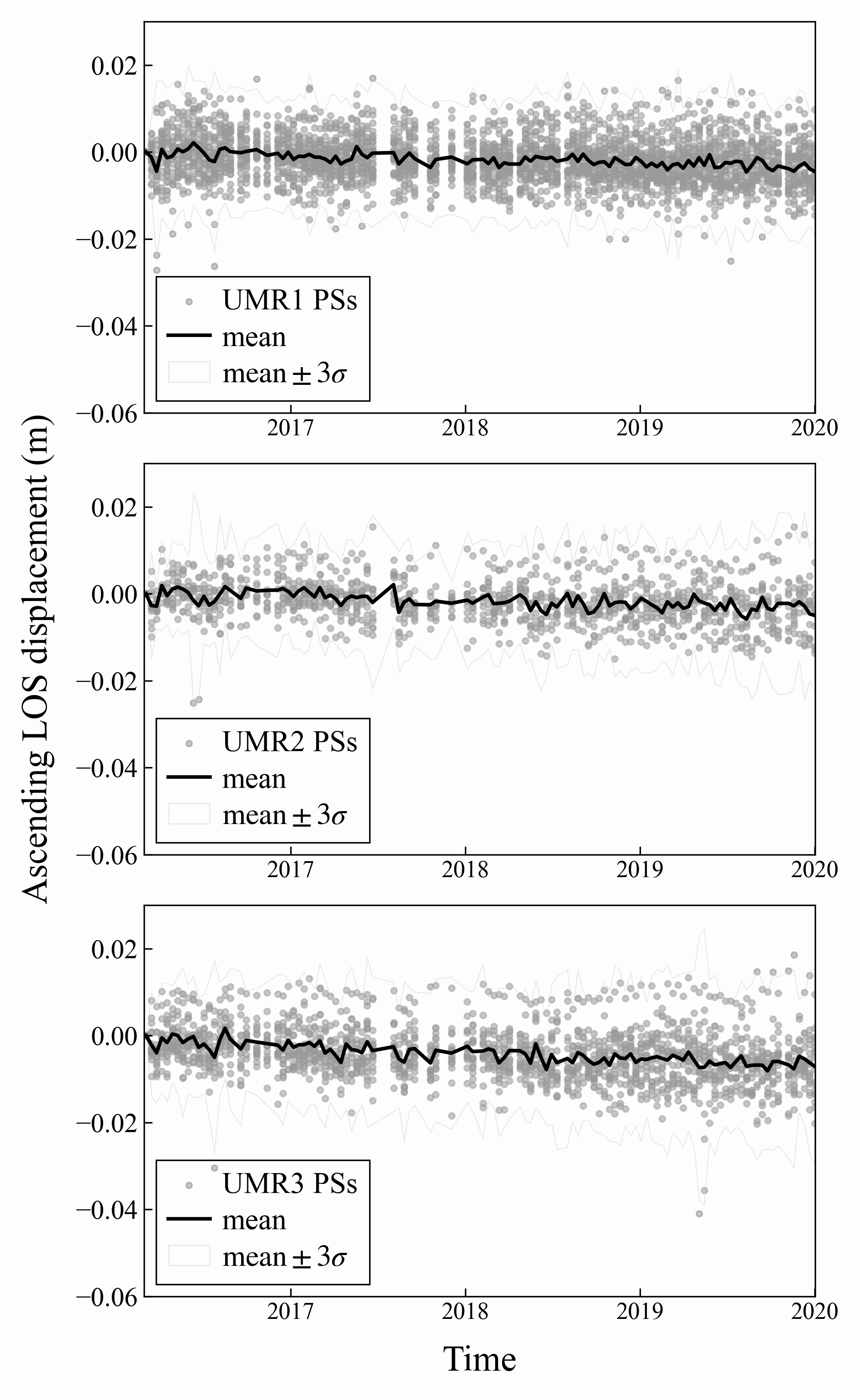}
        \caption{}
        \label{fig:los_hout_a3}
    \end{subfigure}
    \hfill
    \begin{subfigure}{0.4\linewidth}
        \centering
        \includegraphics[width=\linewidth]{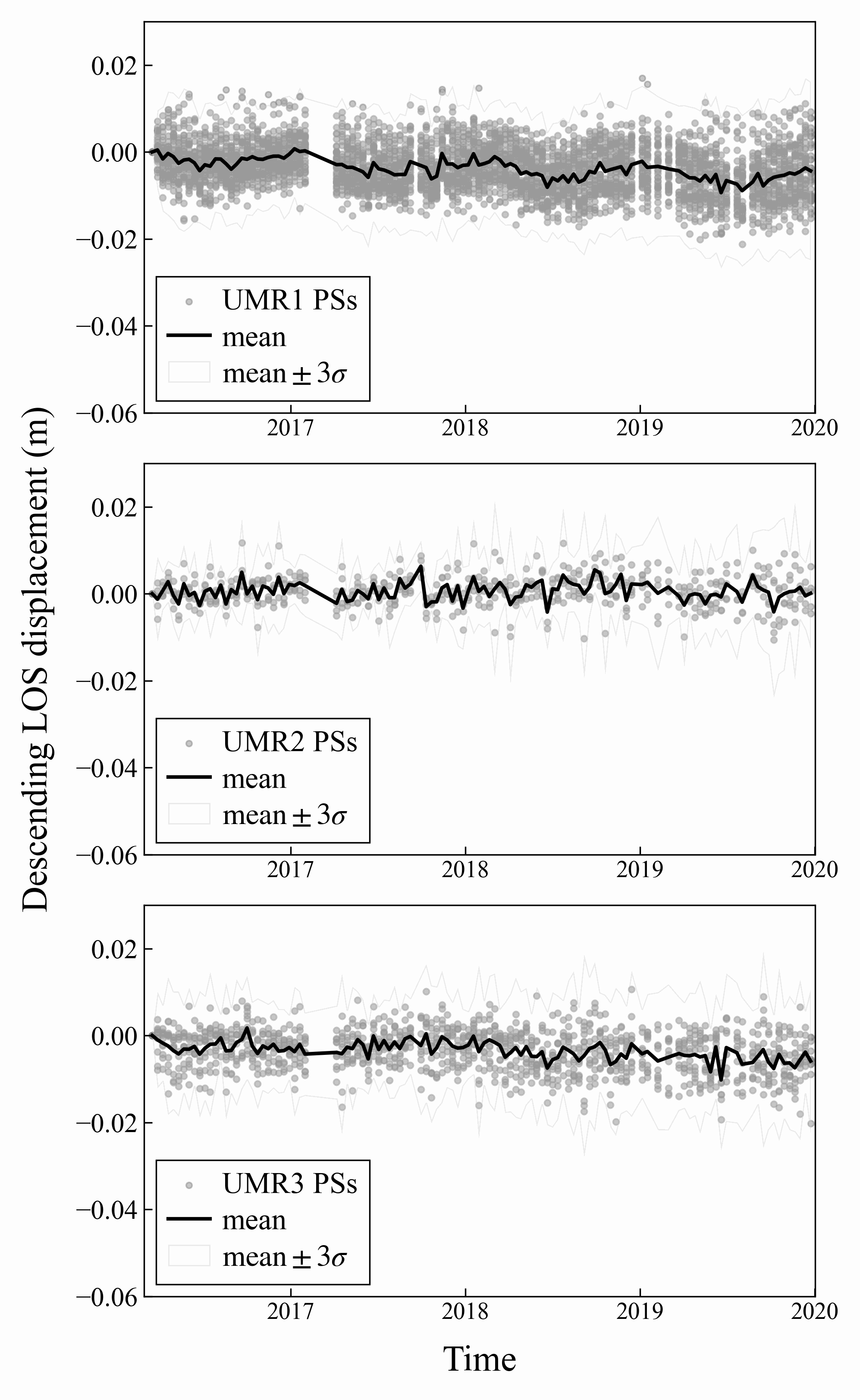}
        \caption{}
        \label{fig:los_hout_d3}
    \end{subfigure}
    \caption{\revise{LOS displacements of PSs in UMRs 1-3 for Houtmankade East, corresponding to the section that was reported to be in a safe condition from: (a) ascending and (b) descending track.}}
    \label{fig:los_hout3}
\end{figure*}

\begin{figure*}
    \centering
    \begin{subfigure}{0.4\linewidth}
        \centering
        \includegraphics[width=\linewidth]{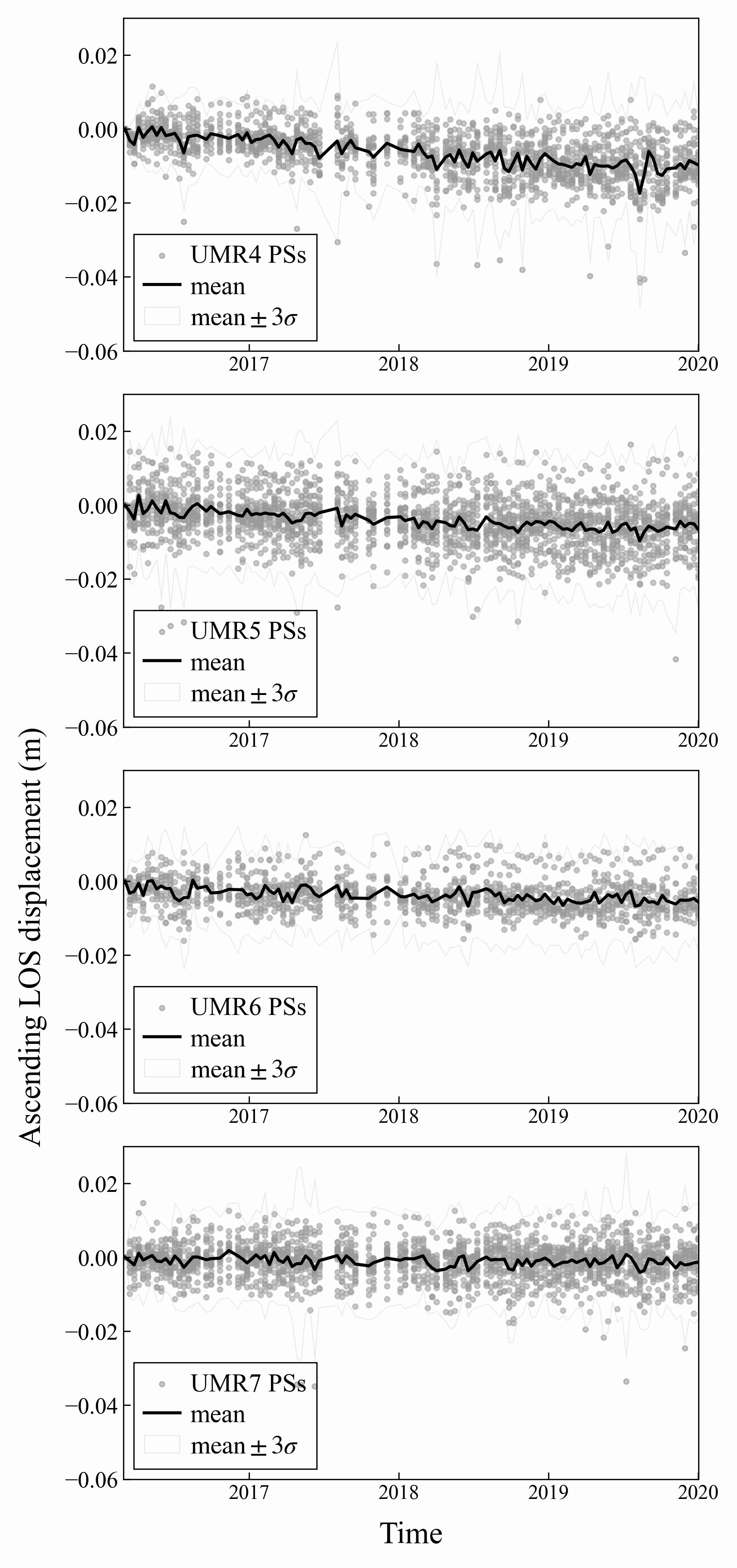}
        \caption{}
        \label{fig:los_hout_a4}
    \end{subfigure}
    \hfill
    \begin{subfigure}{0.4\linewidth}
        \centering
        \includegraphics[width=\linewidth]{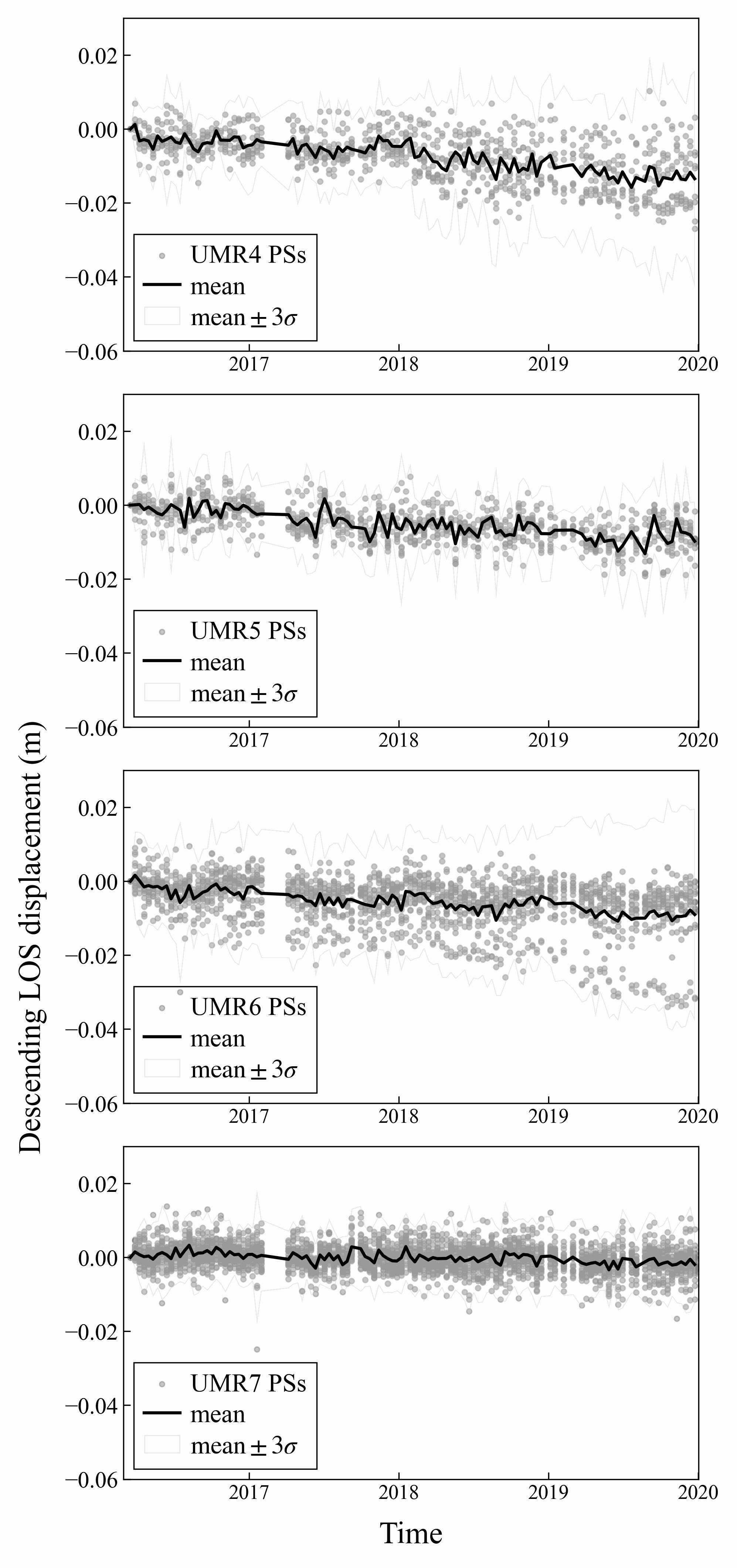}
        \caption{}
        \label{fig:los_hout_d4}
    \end{subfigure}
    \caption{\revise{LOS displacements of PSs in UMRs 4-7 for Houtmankade East, corresponding to the older section that was selected for replacement from: (a) ascending and (b) descending track.}}
    \label{fig:los_hout4}
\end{figure*}

\revise{The LOS displacement time series of individual PS observations, together with the mean displacement and the corresponding uncertainty range, are presented in Figures \ref{fig:los_hout3} and \ref{fig:los_hout4} for the UMRs in the safe part and the replaced part, respectively. Considerable variability was observed among the individual PS observations within each UMR. This large variability suggests that using only the mean LOS displacement may lead to misinterpretation of the actual condition of the structure. Therefore, the LOS displacements were modelled as a Gaussian distribution by incorporating all PSs within each UMR, with the mean and variance estimated using Equations \ref{eq:unbiased_mean} and \ref{eq:unbiased_var}.}

\begin{figure*}
    \centering
    \begin{subfigure}{0.4\linewidth}
        \centering
        \includegraphics[width=\linewidth]{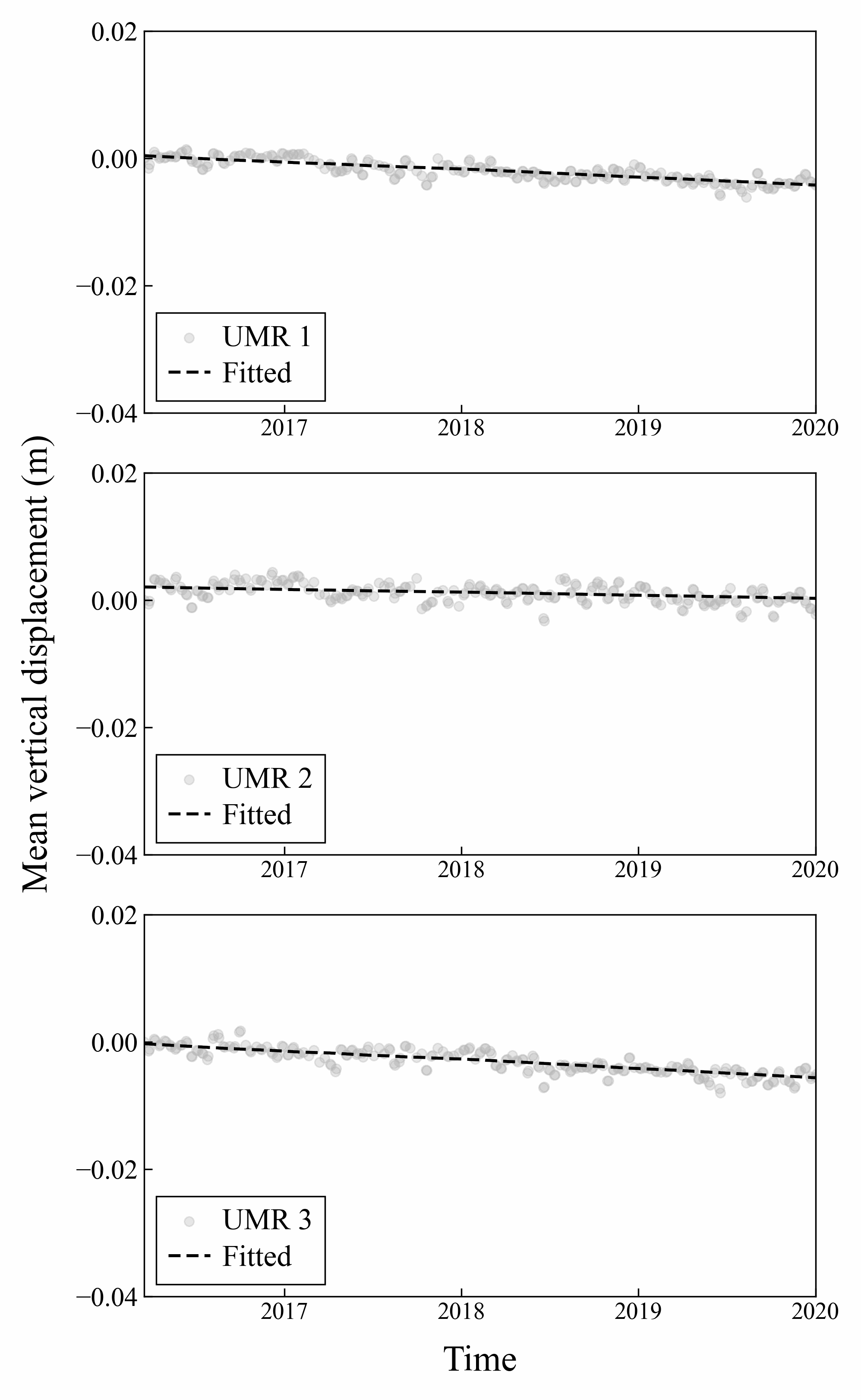}
        \caption{}
        \label{fig:hout_results_ts3}
    \end{subfigure}
    \hfill
    \begin{subfigure}{0.4\linewidth}
        \centering
        \includegraphics[width=\linewidth]{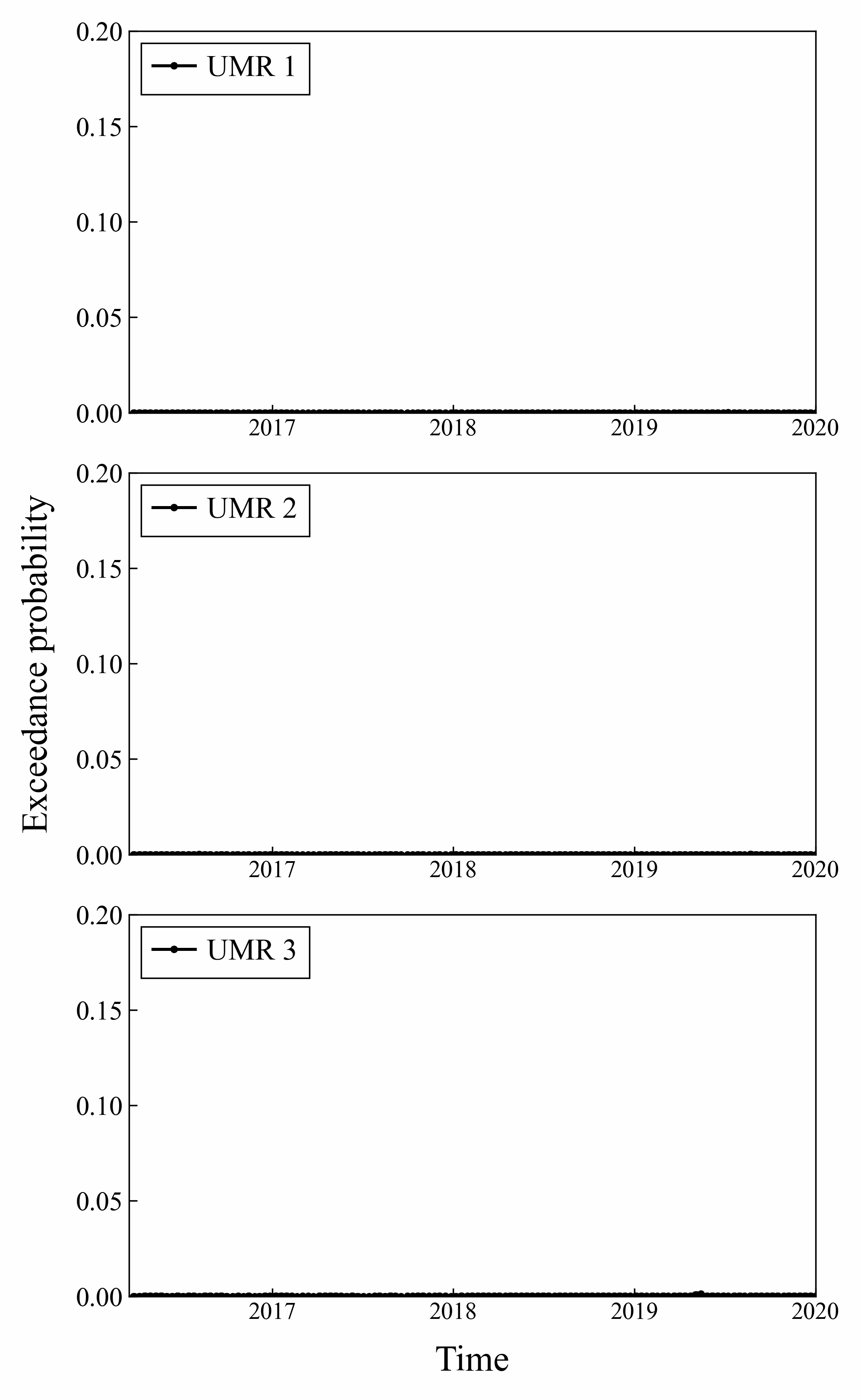}
        \caption{}
        \label{fig:hout_results_pf3}
    \end{subfigure}
    \caption{\revisetwo{Houtmankade East results for UMRs 1-3, corresponding to the section that was reported to be in a safe condition: (a) mean vertical displacement time series and (b) exceedance probability over time.}}
    \label{fig:hout_results3}
\end{figure*}

\begin{figure*}
    \centering
    \begin{subfigure}{0.4\linewidth}
        \centering
        \includegraphics[width=\linewidth]{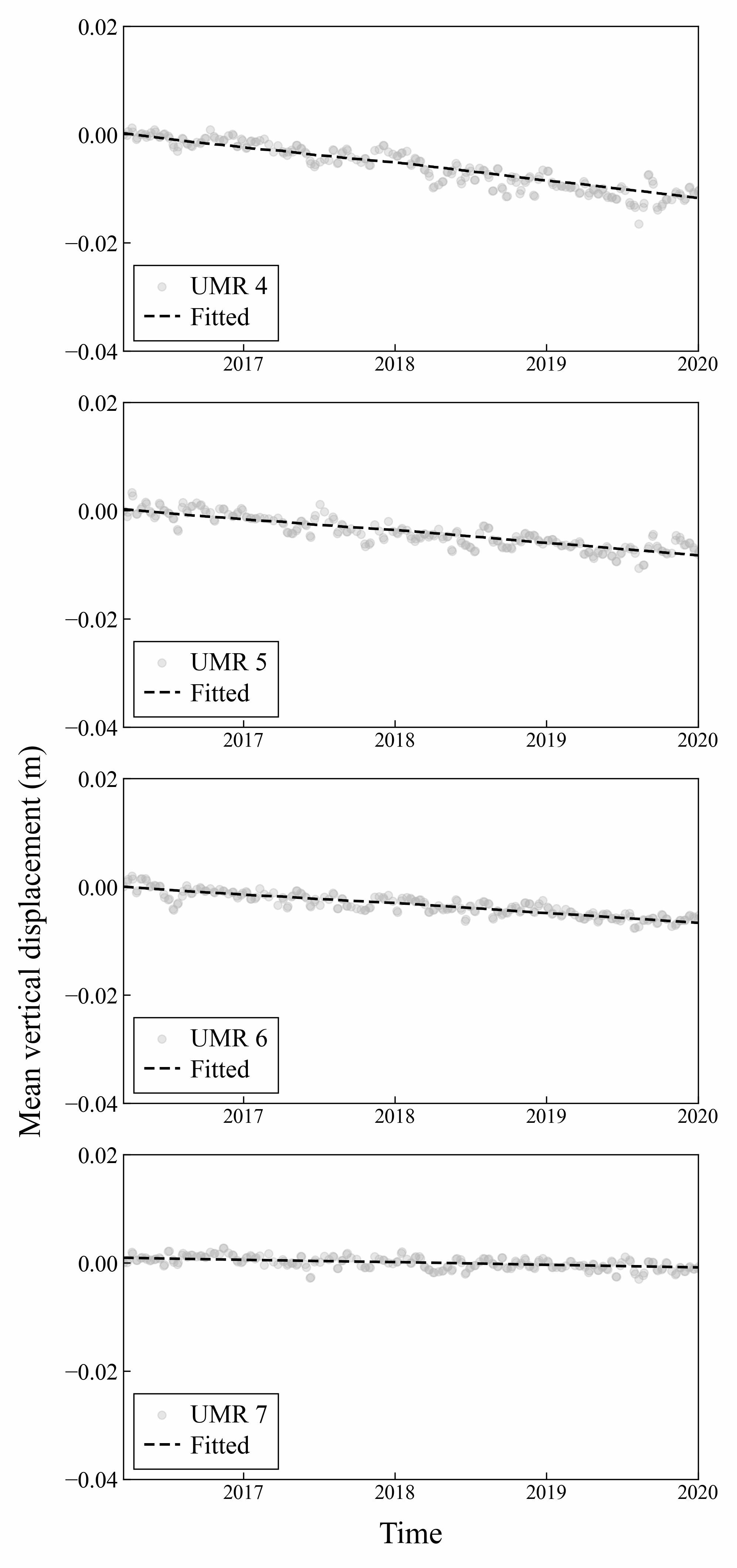}
        \caption{}
        \label{fig:hout_results_ts4}
    \end{subfigure}
    \hfill
    \begin{subfigure}{0.4\linewidth}
        \centering
        \includegraphics[width=\linewidth]{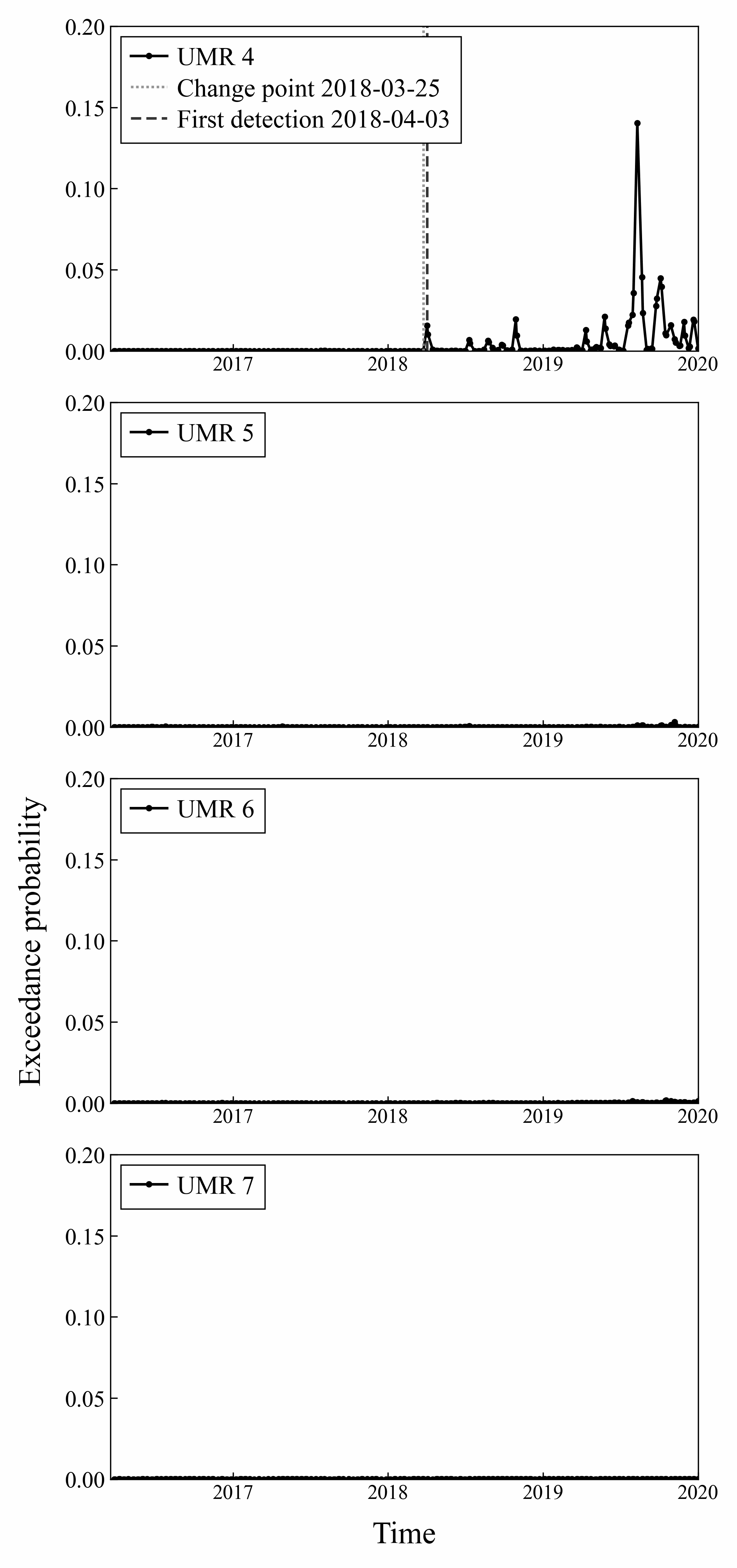}
        \caption{}
        \label{fig:hout_results_pf4}
    \end{subfigure}
    \caption{\revisethree{Houtmankade East results for UMRs 4-7, corresponding to the older section that was selected for replacement: (a) mean vertical displacement time series and (b) exceedance probability over time.}}
    \label{fig:hout_results4}
\end{figure*}

\revise{As shown in Figure \ref{fig:hout_UMRs_regions}, the longitudinal axis of Houtmankade East is approximately aligned in the north-south direction, which is close to the satellite flight direction. Consequently, the contribution of the longitudinal component to the LOS displacement was assumed to be negligible. The vertical and transverse displacement components were therefore derived using Equations \ref{eq:sl_dVdT}. Figures \ref{fig:hout_results_ts3} and \ref{fig:hout_results_ts4} show the vertical displacement time series, together with a linear regression fitted to the entire dataset, for the safe part and the replaced part, respectively.}

\revise{As in the previous cases, the maximum allowable vertical displacement was set to 1\% of the retaining height, corresponding to 25 mm in this case. It can be observed that the displacements for all seven UMRs remain well below this threshold and would therefore all be classified as safe under traditional threshold-based approaches.}

\revisetwo{To further evaluate the condition of this quay, the limit state function was applied, and Equation \ref{eq:pf} was used to calculate the probability of exceedance. Figure \ref{fig:hout_results_pf3} presents the temporal evolution of the probability of exceedance for UMRs 1–3, which corresponds to the safe part. The probability of exceedance remained consistently close to zero, and no abrupt change points were detected by the PELT algorithm.}

\revisetwo{Figure \ref{fig:hout_results_pf4} shows the probability of exceedance over time for UMR 4-7. No warning was issued for UMRs 5–7, which is considered reasonable. These sections were not initially identified as problematic and were only selected for replacement following subsequent underwater inspections. Notably, the initial assessment by Sweco revealed no abnormalities in these locations. 
As the proposed framework was designed to detect precursory behaviour associated with deformation anomaly, no warning was expected for UMRs that did not exhibit such behaviour.}
\revisethree{In contrast, UMR 4 exhibited a clear warning signal. PELT first detected a change on 2018-04-03, triggering a warning, while the corresponding change-point date was estimated as 2018-03-25. This warning preceded the alarm issued by Sweco by approximately one year. Remarkably, the spatial location of this warning coincides precisely with the area flagged by Sweco.}

\section{Discussion}

\revisetwo{The results demonstrate that the proposed probabilistic automated MT-InSAR warning framework can successfully provide timely warnings of deformation anomaly by addressing three limitations of existing methods. 
First, to address the limitation that LOS displacements are not along the principal structural axes, the automatic UMR definition allows ascending and descending PSs within the same UMR to be combined, thereby enabling LOS displacements to be projected into the structure's local coordinate system. Meanwhile, this automated process eliminates the need for prior manual structure information and improves the scalability of MT-InSAR-based SHM.
Second, to overcome the limitations of deterministic approaches that neglect uncertainties, the framework modelled LOS displacements as Gaussian distributions rather than single mean values for each UMR, thereby enabling characterisation of the uncertainty due to lack of knowledge associated with a fixed LOS displacement characterising a UMR.
Finally, to address the limitation caused by the relative temporal referencing of MT-InSAR measurements and the lack of pre-acquisition displacement history, the framework used abrupt increases in the probability of exceedance as the warning indicator rather than a fixed reliability threshold.}

These methodological improvements also resulted in a computationally efficient framework, making it suitable for large-scale monitoring of infrastructure networks. Nevertheless, several aspects of the framework require further discussion, particularly with respect to its practical implementation and application.

The buffer was defined as a spatial extension of the structure’s boundary polygon, serving as the operational zone for PS selection and mitigating positional uncertainties of PSs. In this study, the buffer width was determined by considering both the geocoding accuracy of the InSAR data and the specific structural context. For instance, in the Entrepotdok case, the buffer was determined by the road width because the quay wall and the adjacent road exhibited a similar displacement pattern. This strategy ensures that all PSs representing the same region are considered. However, in other cases, particularly for more complex structures or in densely built environments, this strategy may introduce unrelated PSs into the buffer, potentially compromising the performance of the warning.

Within each UMR, the LOS displacements obtained from individual PSs were used to determine the parameters of a Gaussian distribution representing the uncertainty on an unknown uniform displacement value characterising a single UMR, whose validity and limitations are examined in \ref{sec:app_b}. \revisethree{However, these contributions may still exhibit spatial variation in some cases, due to for example to localised deformation within a UMR. These effects may challenge the underlying UMR modelling assumptions, and affect the estimation of the distribution describing the LOS displacements within a UMR. In such cases, the UMR definition could be refined to separate regions exhibiting uniform displacements. However, further refinement may not be feasible for highly localised deformation mechanisms due to insufficient spatial coverage of PS observations. Moreover, if information about spatial correlations of PS displacement data (for example caused by atmospheric errors) were available, this could be used to improve the description of LOS displacements distribution.}

\revise{The contribution of one horizontal displacement to the LOS measurement was assumed to be negligible based on the geometric relationship between the structure and the satellite track. To assess the validity of this assumption, a sensitivity analysis was performed (see Appendix C), which shows that its influence on the warning results is limited for the analysed cases. However, this assumption may not hold for other structures or conditions, where a more rigorous treatment may be required, for example by explicitly incorporating the neglected component through probabilistic modelling.} \revisethree{Furthermore, although the warning performance of the proposed framework was evaluated against documented case histories, the accuracy of the projected vertical displacements could not be independently verified due to the lack of in-situ measurements. This limitation also highlights a key advantage of MT-InSAR: its ability to provide large-scale deformation observations of infrastructure without dedicated in-situ monitoring systems, which makes it particularly valuable for preliminary anomaly screening.}

\revisetwo{The framework is intended to analyse the vertical displacement derived from MT-InSAR and identify associated deformation anomalies. The limit state function was based on maximum allowable vertical displacement. Settlement is a common and often unavoidable process and is increasingly exacerbated by global land subsidence issues \citep{herrera-garciaMappingGlobalThreat2021, ohenhenLandSubsidenceRisk2025}. For the quay walls investigated in this study, abnormal vertical displacement may be associated with potentially critical structural failure due to the structural characteristics and local site conditions. In the Netherlands, land subsidence is a significant challenge, and various typical failure mechanisms of quay walls can induce vertical displacement \citep{ARKAmsterdamseRisicobeoordeling2024, sharmaAnalysisProceduresAccounting2024}. \revisethree{For other infrastructure types, where different displacement components may dominate, the limit state function can be adapted to incorporate the relevant displacement components.} However, the proposed framework is not intended to predict future structural states, remaining service life, or the absolute probability of collapse based solely on MT-InSAR observations. Instead, its primary objective is to provide a large-scale deformation-anomaly screening capability using MT-InSAR observations, enabling the rapid identification of structures exhibiting abnormal deformation patterns and prioritising them for further investigation. Once a potentially critical structure is identified, additional information, such as structural inspections, in-situ monitoring data, or other sensing techniques, should be integrated to achieve a more comprehensive assessment of structural condition and failure risk. \revisethree{All three case studies come from the same city and structure type, therefore further validation using a larger number of both failure and non-failure cases across different locations and infrastructure types is required to assess its operational reliability, including false alarm and detection performance.}}

The temporal resolution and acquisition intervals of MT-InSAR may influence the framework’s responsiveness. \revise{The projection of LOS displacements into vertical components requires the combination of ascending and descending PSs, which are acquired at different times.} This necessitates temporal interpolation, which may introduce errors, particularly during periods of rapid displacement. \revise{However, a sensitivity analysis based on FE-derived displacement demonstrated that temporal interpolation had a limited influence on the PELT-based warning results within the investigated sampling range. The details of this analysis are presented in \ref{sec:app_d}.} \revisethree{Nevertheless, the temporal interpolation may involve observations after the corresponding dates, while the MT-InSAR displacement time series were obtained using the full image stack. Consequently, although PELT was implemented sequentially, the overall analysis remained retrospective.}

\revisethree{In this study, the PELT penalty was set to $1 \times 10^{-5}$ and the maximum vertical allowable displacement was set to 1\% of the retaining height adapting from a horizontal displacement criterion, following \citep{bettsPortBotanyExpansion2011}. The selection of these two parameters may influence the warning dates. While large variations in these parameters may result in shifts in the warning dates, the sensitivity analyses based on the three case studies (\ref{app:appe} and \ref{app:appf}) showed that warnings were consistently identified across the investigated parameter ranges. }

\section{Conclusion}

\revisetwo{This \revisefour{paper} addressed key challenges in using InSAR data for automated anomaly warning of infrastructure deformation, where conventional deterministic methods often struggled with measurement uncertainties, geometric constraints and the relative nature of observations. To achieve this, we proposed a probabilistic MT-InSAR warning framework that automatically defines UMRs based on structural geometry and the spatial distribution of available PSs.} Within each UMR, ascending and descending PSs were \revise{grouped}. \revisethree{Their LOS displacements were assumed to consist of an actual deformation component that is identical for all PSs within the same UMR and residual error contributions, which introduce uncertainty in the estimation of this unknown fixed displacement.} Accordingly, they were modelled as Gaussian distributions. By exploiting the geometric relationship between LOS displacements and the displacements in the local coordinate system, \revise{the contribution of one horizontal displacement component in the local coordinate system to the LOS displacement was assumed to be negligible} and the probabilistic distribution of vertical displacements was subsequently derived. \revisetwo{We then applied a limit state function based on allowable vertical displacement to obtain an evolution of exceedance probability over time, where abrupt increases in this probability serve as indicators for anomaly warning. }

The effectiveness of the proposed framework was evaluated in \revise{three} case studies in Amsterdam. \revisetwo{The results demonstrate that the framework can \revisethree{characterise the deformation of each UMR} and provide timely warnings of deformation anomaly.} Compared with traditional MT-InSAR-based SHM methods, the proposed framework provides a more automated and scalable solution, while explicitly accounting for uncertainties, geometric constraints, and the relative nature of InSAR measurements. \revisetwo{As such, it enables probabilistic assessment across large infrastructure networks and supports the generation of actionable warnings, with good computational efficiency.} 
\revisefour{Future research could evaluate the performance of the framework across different infrastructure types and structural conditions. Its application to infrastructure networks could also be investigated to examine how the detected warnings can inform inspection and maintenance decisions.}

\section*{CRediT authorship contribution statement}
\revisefour{\textbf{Hao Kuai:} Writing - original draft, Conceptualization, Methodology, Formal analysis, Visualization, Software, Validation, Data curation. \textbf{Na Hao:} Writing - review \& editing, Conceptualization, Methodology, Software. \textbf{Pantelis Karamitopoulos:} Writing - review \& editing, Data curation. \textbf{Alice Cicirello:} Writing - review \& editing, Conceptualization, Supervision, Methodology, Validation. \textbf{Giorgia Giardina:} Writing - review \& editing, Validation, Supervision, Project administration, Funding acquisition, Conceptualization.} 

\section*{Funding}
This work is part of the Vidi project InStruct, project number 18912, financed by the Dutch Research Council (NWO).

\section*{Acknowledgments}
The authors would like to acknowledge SkyGeo Netherlands B.V. for providing the InSAR data and the Municipality of Amsterdam for supplying technical reports on Entrepotdok, Grimburgwal and Houtmankade East. 

\section*{Declaration of generative AI and AI-assisted technologies in the manuscript preparation process}
\revisefour{The authors used ChatGPT (OpenAI) to improve the clarity and readability of the text. All AI-assisted revisions were reviewed and edited by the authors, who take full responsibility for the content of the manuscript.}

\section*{Declaration of competing interest}
\revisefour{The authors declare no known competing financial interests or personal relationships that could have appeared to influence the work reported in this paper.}

\section*{Data availability}
\revisefour{Non commercial data can be made available on request.}

\appendix
\section{Expressions of projection coefficients}
\label{sec:app_a}
\revise{The detailed scalar formulations for the projection coefficients in Equation \ref{eq:func_los_scalar} are expressed as follows:}

\begin{equation}
    \begin{split}
    A_a &= \sin\theta_{asc} \sin\alpha_{d,asc} \cos\beta \cos\gamma_t - \sin\theta_{asc}\cos\alpha_{d,asc}\sin\beta\cos\gamma_t \\
    &+ \sin\theta_{asc}\sin\alpha_{d,asc}\sin\beta\sin\gamma_l\sin\gamma_t + \sin\theta_{asc}\cos\alpha_{d,asc}\cos\beta\sin\gamma_l\sin\gamma_t \\
    &- \cos\gamma_l\cos\theta_{asc}\sin\gamma_t \\
    B_a &= \sin\theta_{asc} \sin\alpha_{d,asc} \sin\beta \cos\gamma_l + \sin\theta_{asc} \cos\alpha_{d,asc}\cos\beta\cos\gamma_l + \sin\gamma_l\cos\theta_{asc} \\
    C_a &= \sin\theta_{asc} \sin\alpha_{d,asc} \cos\beta \sin\gamma_t - \sin\theta_{asc} \cos\alpha_{d,asc} \sin\beta \sin\gamma_t \\
    &- \sin\theta_{asc} \sin\alpha_{d,asc} \sin\beta \sin\gamma_l \cos\gamma_t - \sin\theta_{asc} \cos\alpha_{d,asc} \cos\beta \sin\gamma_l \cos\gamma_t \\
    &+ \cos\gamma_l \cos\theta_{asc} \cos\gamma_t \\
    A_d &= \sin\theta_{des} \sin\alpha_{d,des} \cos\beta \cos\gamma_t - \sin\theta_{des}\cos\alpha_{d,des}\sin\beta\cos\gamma_t \\
    &+ \sin\theta_{des}\sin\alpha_{d,des}\sin\beta\sin\gamma_l\sin\gamma_t + \sin\theta_{des}\cos\alpha_{d,des}\cos\beta\sin\gamma_l\sin\gamma_t \\
    &- \cos\gamma_l\cos\theta_{des}\sin\gamma_t \\
    B_d &= \sin\theta_{des} \sin\alpha_{d,des} \sin\beta \cos\gamma_l + \sin\theta_{des} \cos\alpha_{d,des}\cos\beta\cos\gamma_l + \sin\gamma_l\cos\theta_{des} \\
    C_d &= \sin\theta_{des} \sin\alpha_{d,des} \cos\beta \sin\gamma_t - \sin\theta_{des} \cos\alpha_{d,des} \sin\beta \sin\gamma_t \\
    &- \sin\theta_{des} \sin\alpha_{d,des} \sin\beta \sin\gamma_l \cos\gamma_t - \sin\theta_{des} \cos\alpha_{d,des} \cos\beta \sin\gamma_l \cos\gamma_t \\
    &+ \cos\gamma_l \cos\theta_{des} \cos\gamma_t \\
    \end{split}
    \label{eq:coeff_projection}
\end{equation}

\section{Empirical evaluation of the Gaussian assumption for LOS displacements}
\label{sec:app_b}

\revise{In Section \ref{sec:uq}, at each acquisition time $t_i$, the LOS displacement $d_{LOS,t_i}^j$ from the $j$-th PS (from either ascending or descending track) was assumed to represent independent and identically distributed (i.i.d.) samples drawn from a Gaussian distribution $\mathcal{N}(\mu_{t_i}, \sigma_{t_i}^2)$. To empirically evaluate this assumption, the Grimburgwal case study was analysed using PSs from both ascending and descending tracks.}

\revise{After assigning PSs to their corresponding UMRs (Figure \ref{fig:grim_UMRs_regions}), the mean and standard deviation of the LOS displacements within each UMR at acquisition time $t_i$ were computed as:}

\begin{equation}
     {\mu}_{t_i} = \frac{1}{n} \sum_{j=1}^n d_{LOS, t_i}^j
\end{equation}

\begin{equation}
    {\sigma}_{t_i} = \sqrt{\frac{1}{n-1} \sum_{j=1}^n \left(d_{LOS, t_i}^j - {\mu}_{t_i}\right)^2}
\end{equation}
\revise{where $d_{LOS,t_i}^{j}$ denotes the LOS displacement of the $j$-th PS within a UMR at acquisition time $t_i$, and $n$ is the number of PSs in that UMR.}

\revise{In the Grimburgwal case, a total of 67 PSs from the ascending track and 43 PSs from the descending track were available across all five UMRs. However, the number of PSs within individual UMRs was limited (in UMR 3 as low as three), which prevents a reliable estimation of the underlying distribution using histograms at the UMR level.}
\revise{Furthermore, since each UMR was characterised by its own mean and variance, directly aggregating the raw LOS displacements across UMRs would result in a mixture of distributions. To overcome this limitation, the LOS displacements within each UMR were standardised using their corresponding mean and standard deviation:}

\begin{equation}
    z_{j,t_i} = \frac{d_{LOS, t_i}^j - {\mu}_{t_i}}{{\sigma}_{t_i}}
\end{equation}

\revise{This standardisation removes location and scale differences among UMRs. Under the Gaussian assumption, the standardised variables $z_{j,t_i}$ should follow a common standard normal distribution $\mathcal{N}(0,1)$.}

\revise{For each acquisition time, the standardised samples from all UMRs were combined to construct histograms separately for ascending and descending tracks. The number of bins $k$ was determined using Sturges' rule \citep{sturgesChoiceClassInterval1926}:}

\begin{equation}
    k = 1 + \log_2(m)
\end{equation}
\revise{where $m$ is the number of samples, equal to 67 for the ascending track and 43 for the descending track. This yields $k \approx 7.07$ and $k \approx 6.43$ for ascending and descending track, respectively. For consistency and comparability, $k=7$ was adopted for both datasets.}

\begin{figure*}
    \centering
    \begin{subfigure}{0.42\textwidth}
        \includegraphics[width=\linewidth]{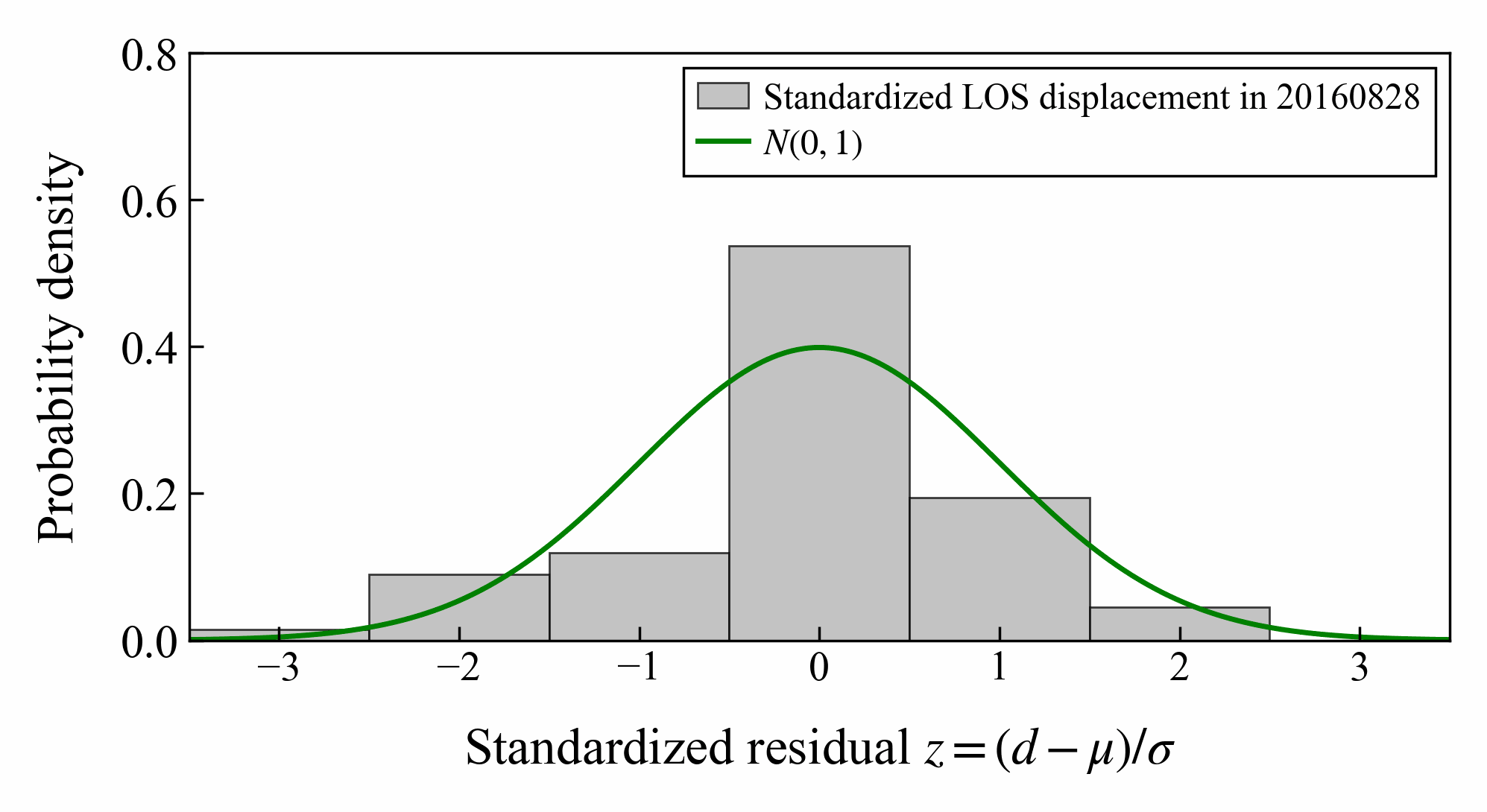}
        \caption{}
    \end{subfigure}
    \hfill
    \begin{subfigure}{0.42\textwidth}
        \includegraphics[width=\linewidth]{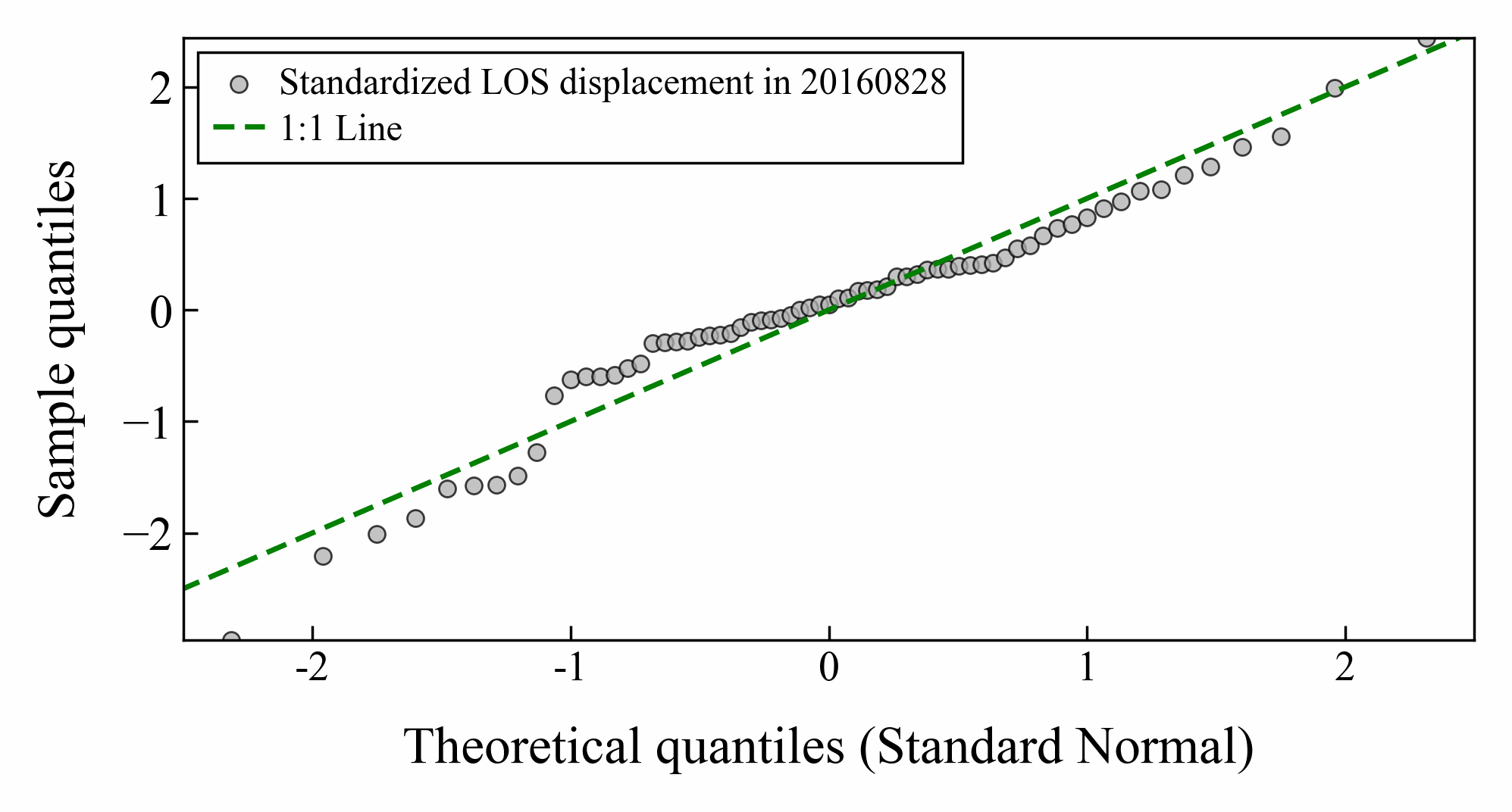}
        \caption{}
    \end{subfigure}

    \begin{subfigure}{0.42\textwidth}
        \includegraphics[width=\linewidth]{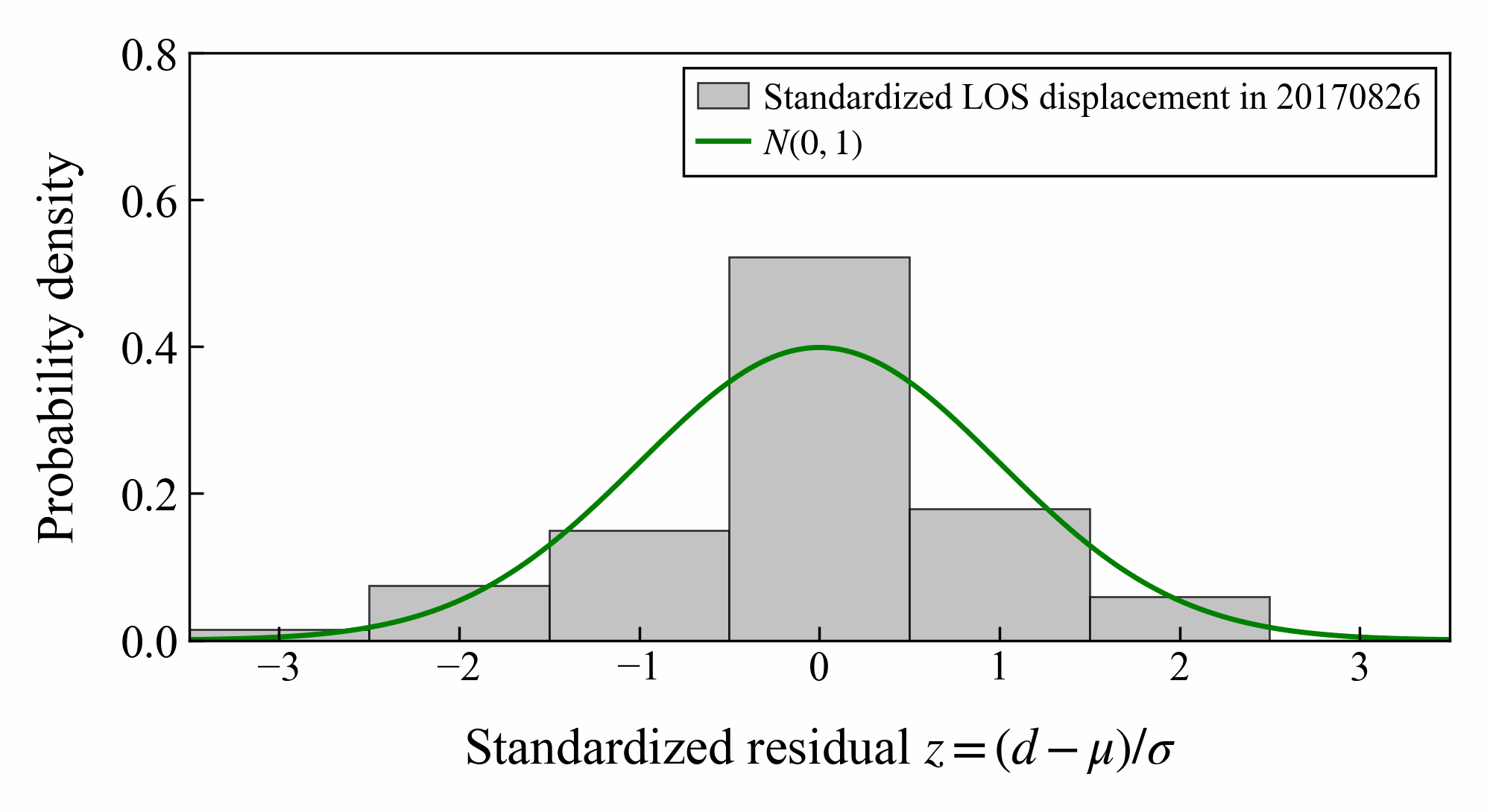}
        \caption{}
    \end{subfigure}
    \hfill
    \begin{subfigure}{0.42\textwidth}
        \includegraphics[width=\linewidth]{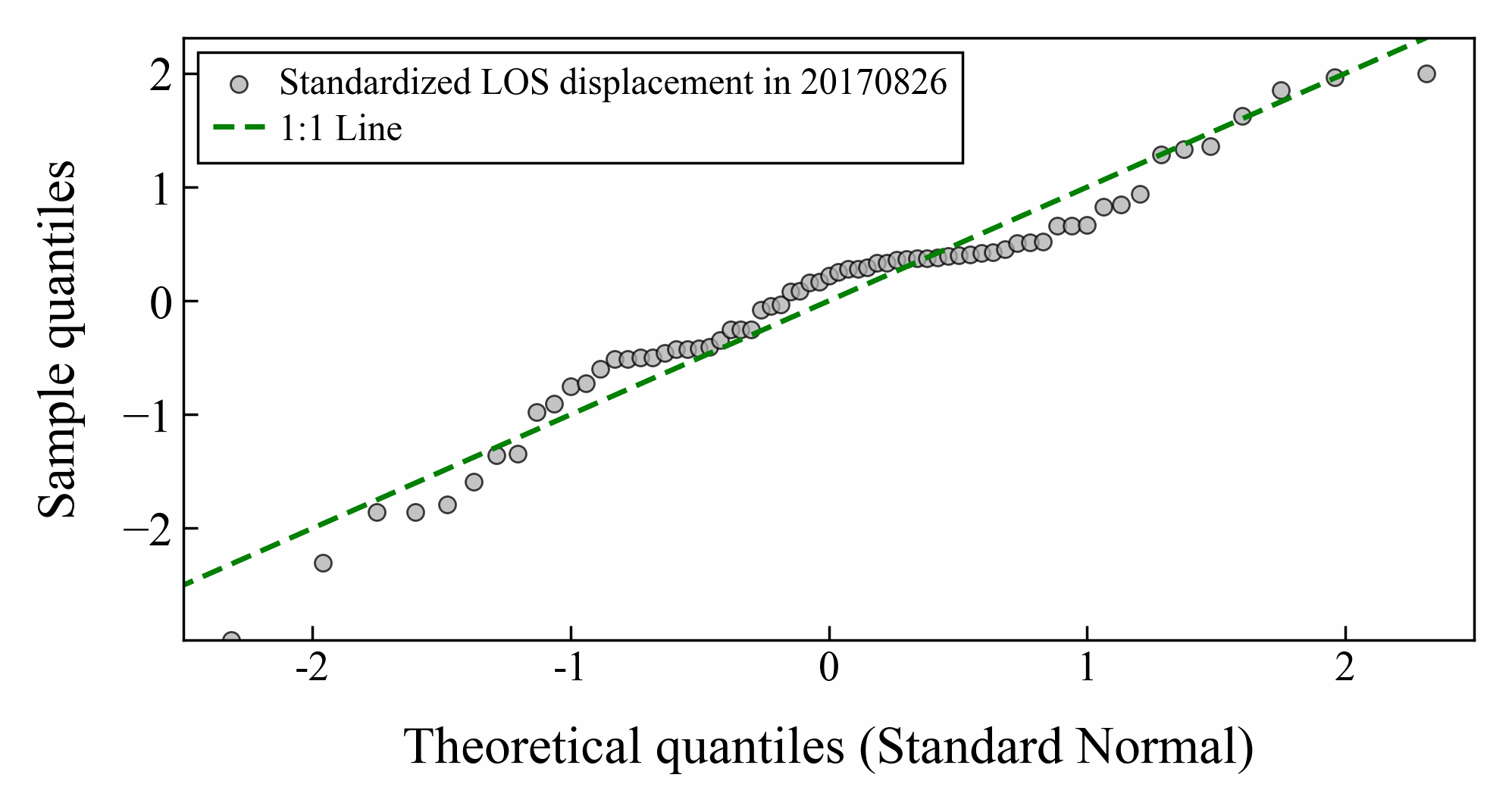}
        \caption{}
    \end{subfigure}

    \begin{subfigure}{0.42\textwidth}
        \includegraphics[width=\linewidth]{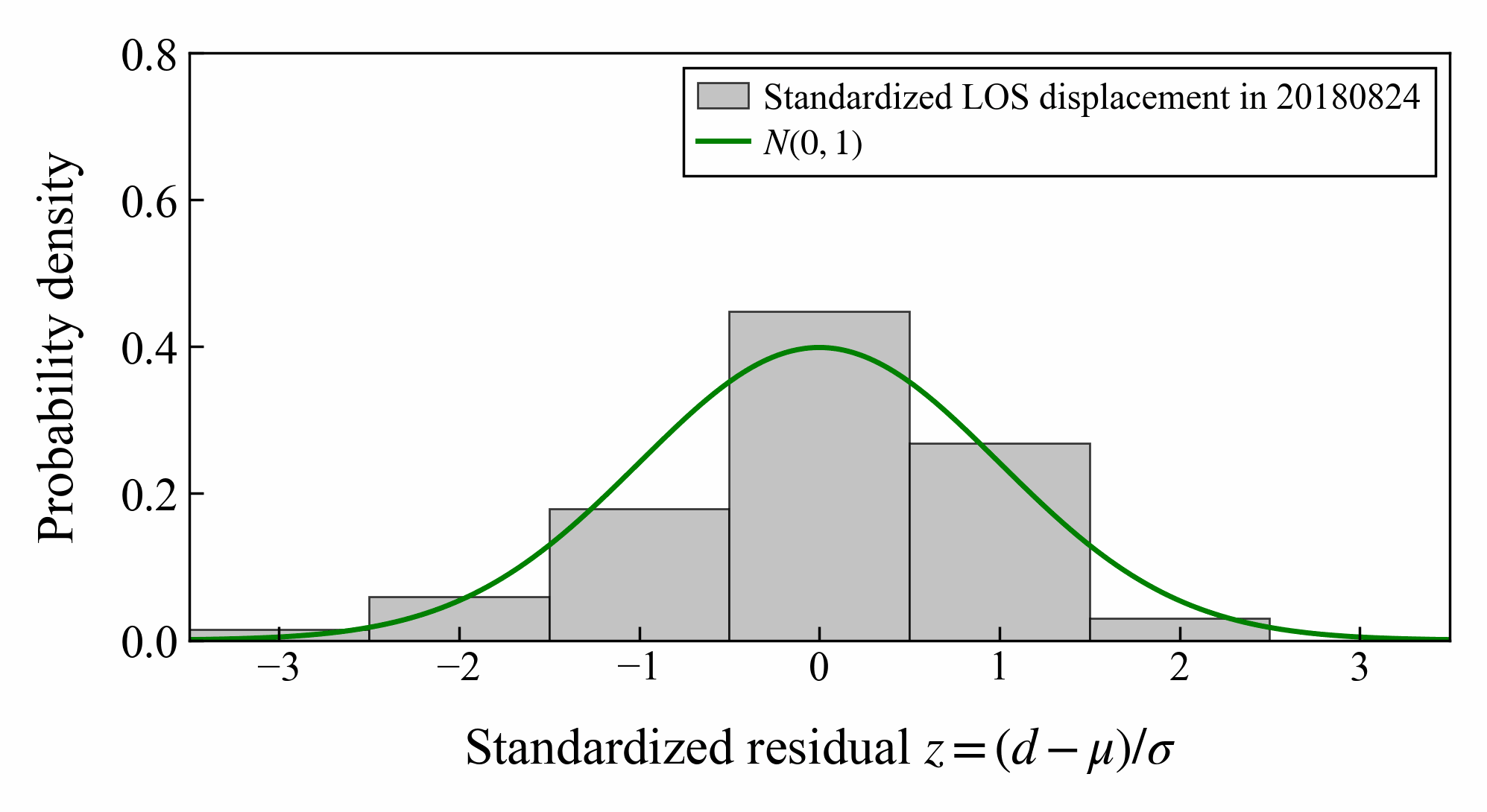}
        \caption{}
    \end{subfigure}
    \hfill
    \begin{subfigure}{0.42\textwidth}
        \includegraphics[width=\linewidth]{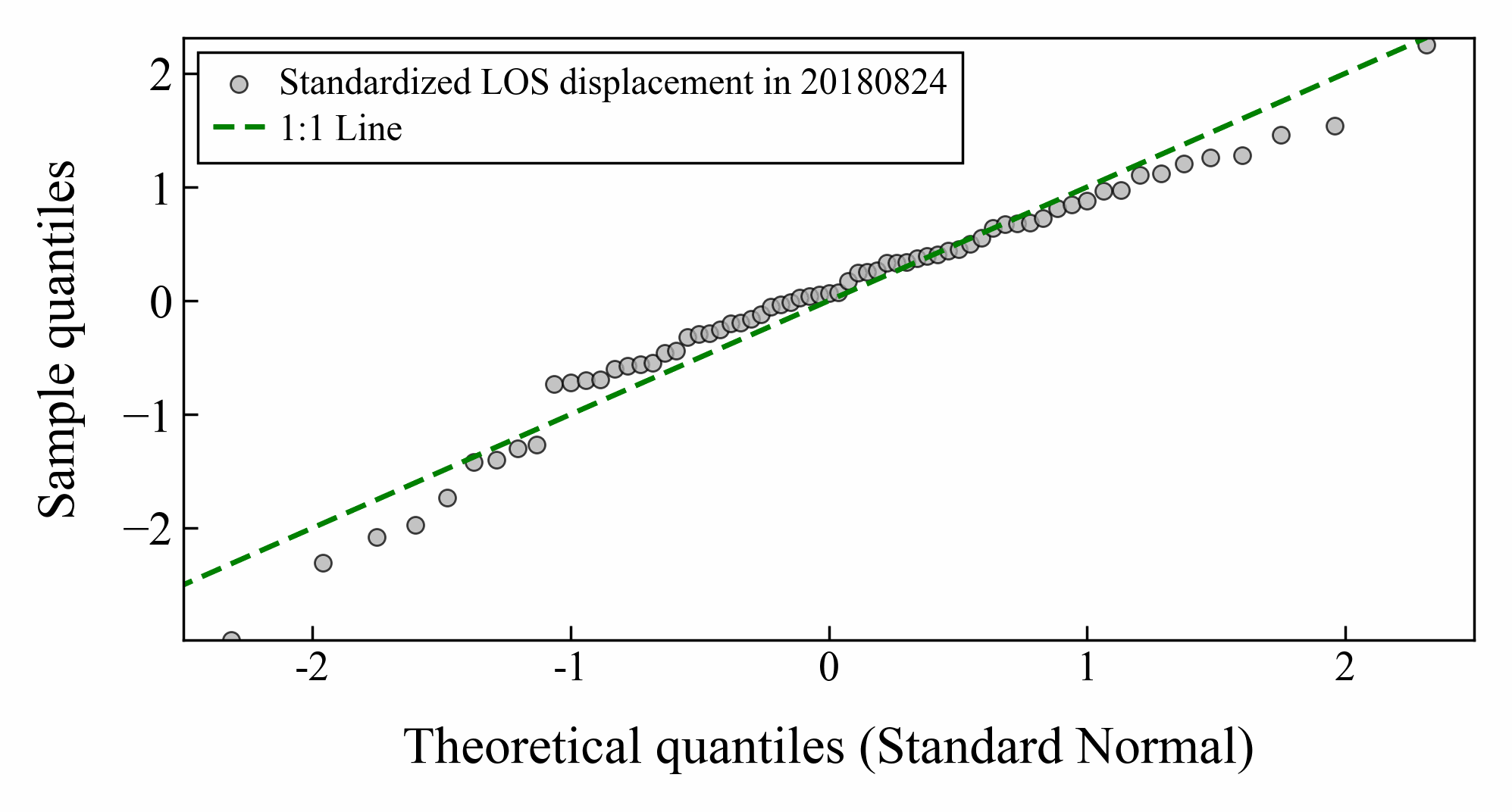}
        \caption{}
    \end{subfigure}

    \begin{subfigure}{0.42\textwidth}
        \includegraphics[width=\linewidth]{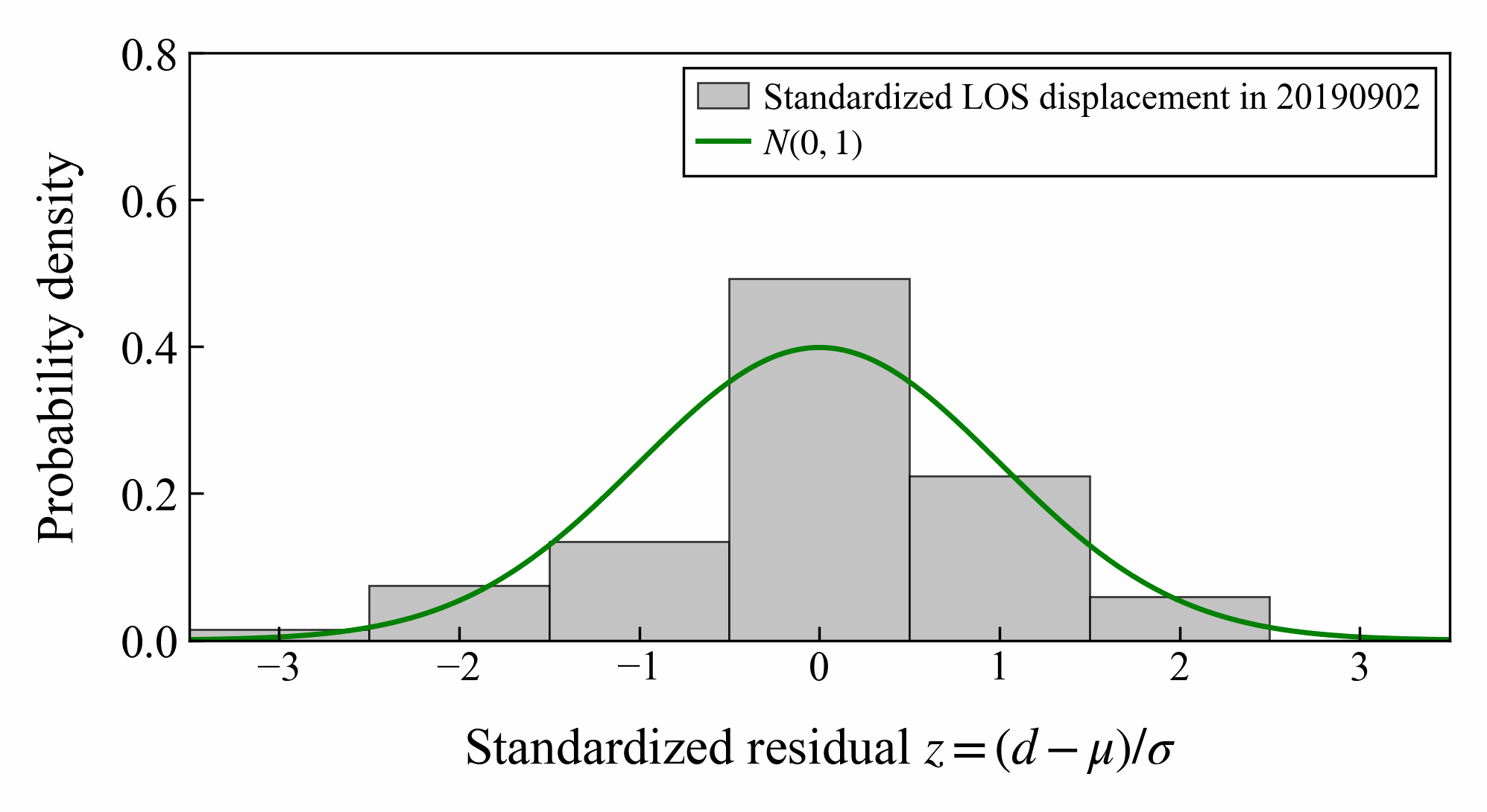}
        \caption{}
    \end{subfigure}
    \hfill
    \begin{subfigure}{0.42\textwidth}
        \includegraphics[width=\linewidth]{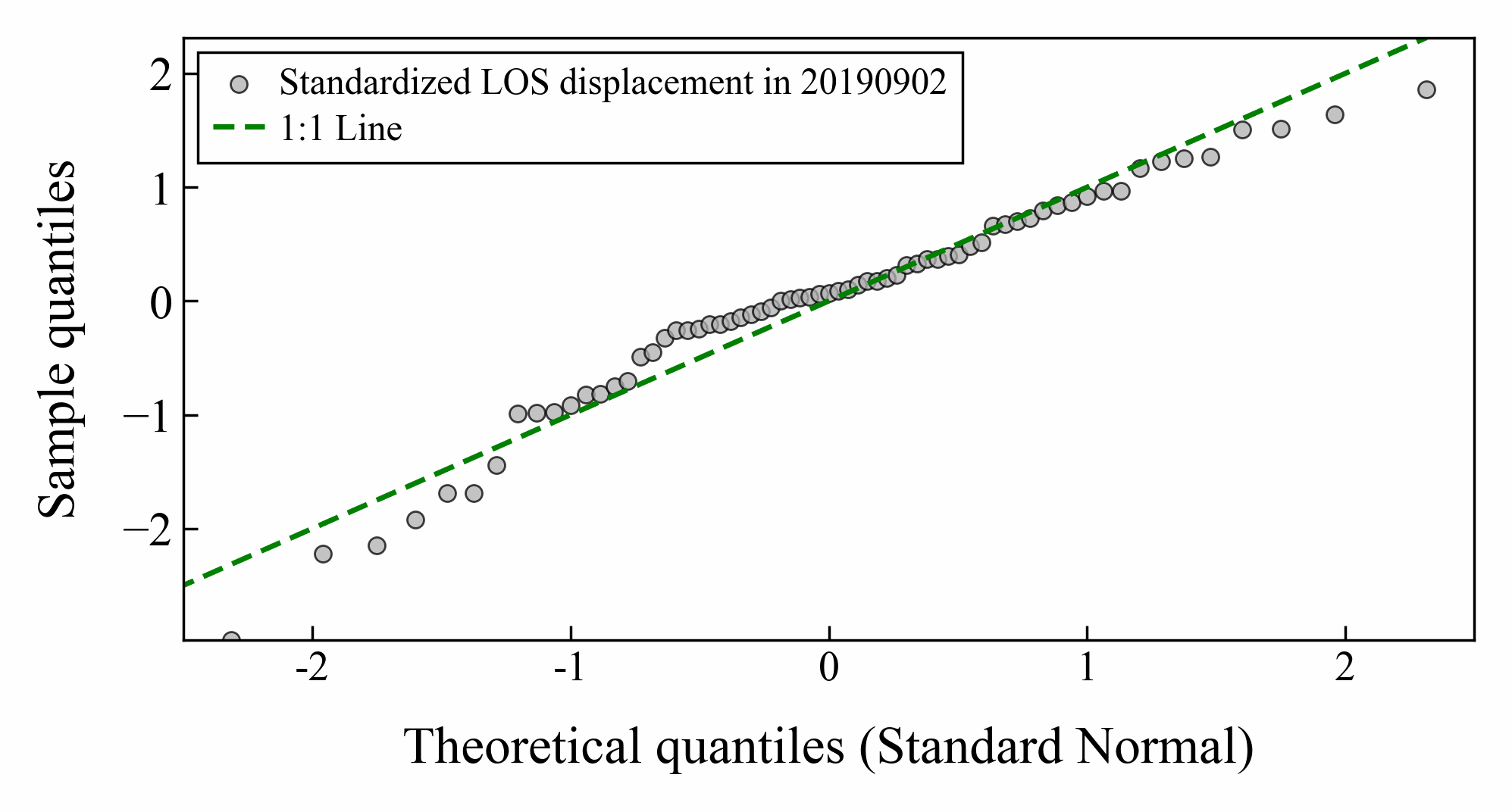}
        \caption{}
    \end{subfigure}

    \begin{subfigure}{0.42\textwidth}
        \includegraphics[width=\linewidth]{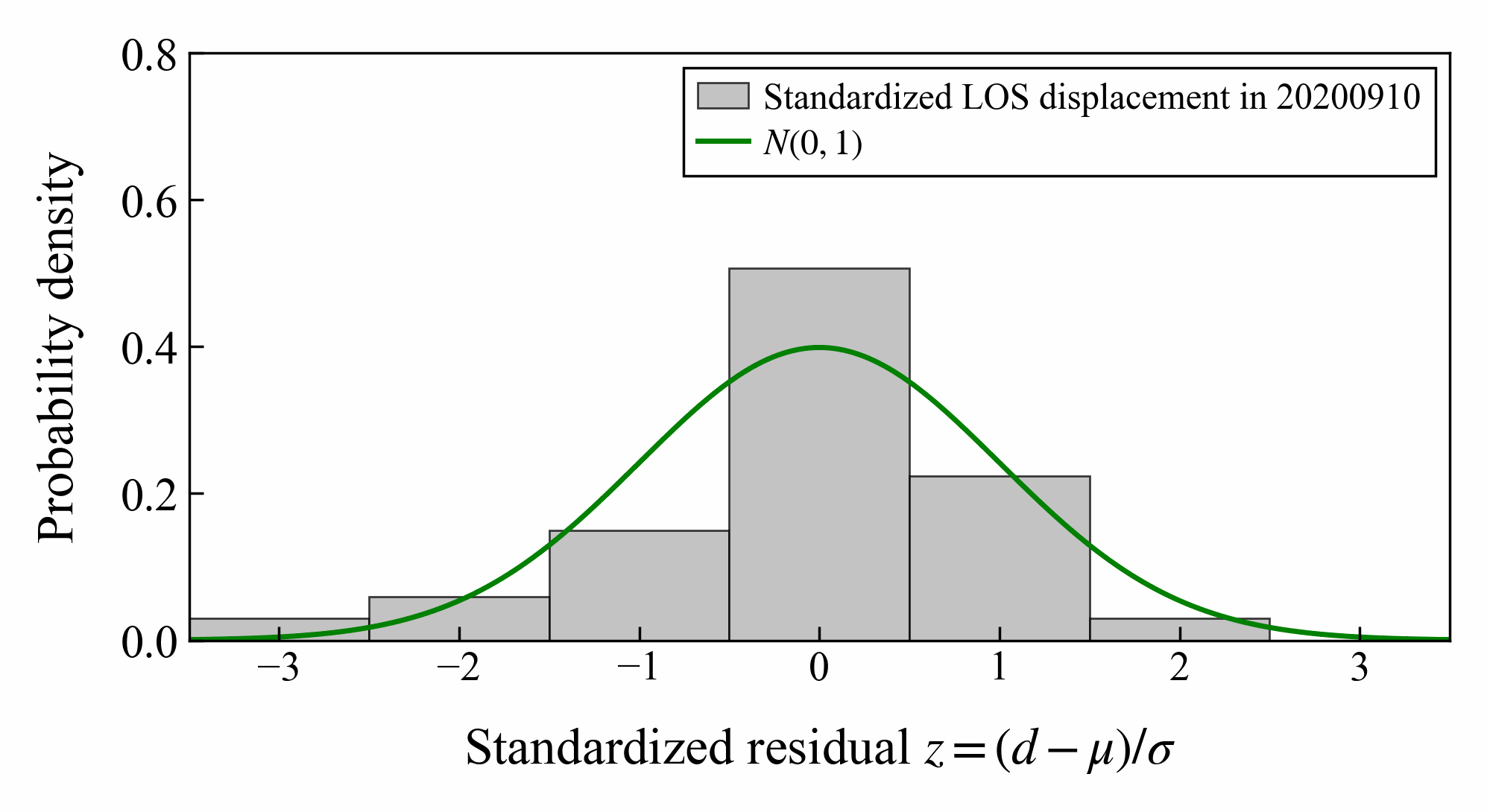}
        \caption{}
    \end{subfigure}
    \hfill
    \begin{subfigure}{0.42\textwidth}
        \includegraphics[width=\linewidth]{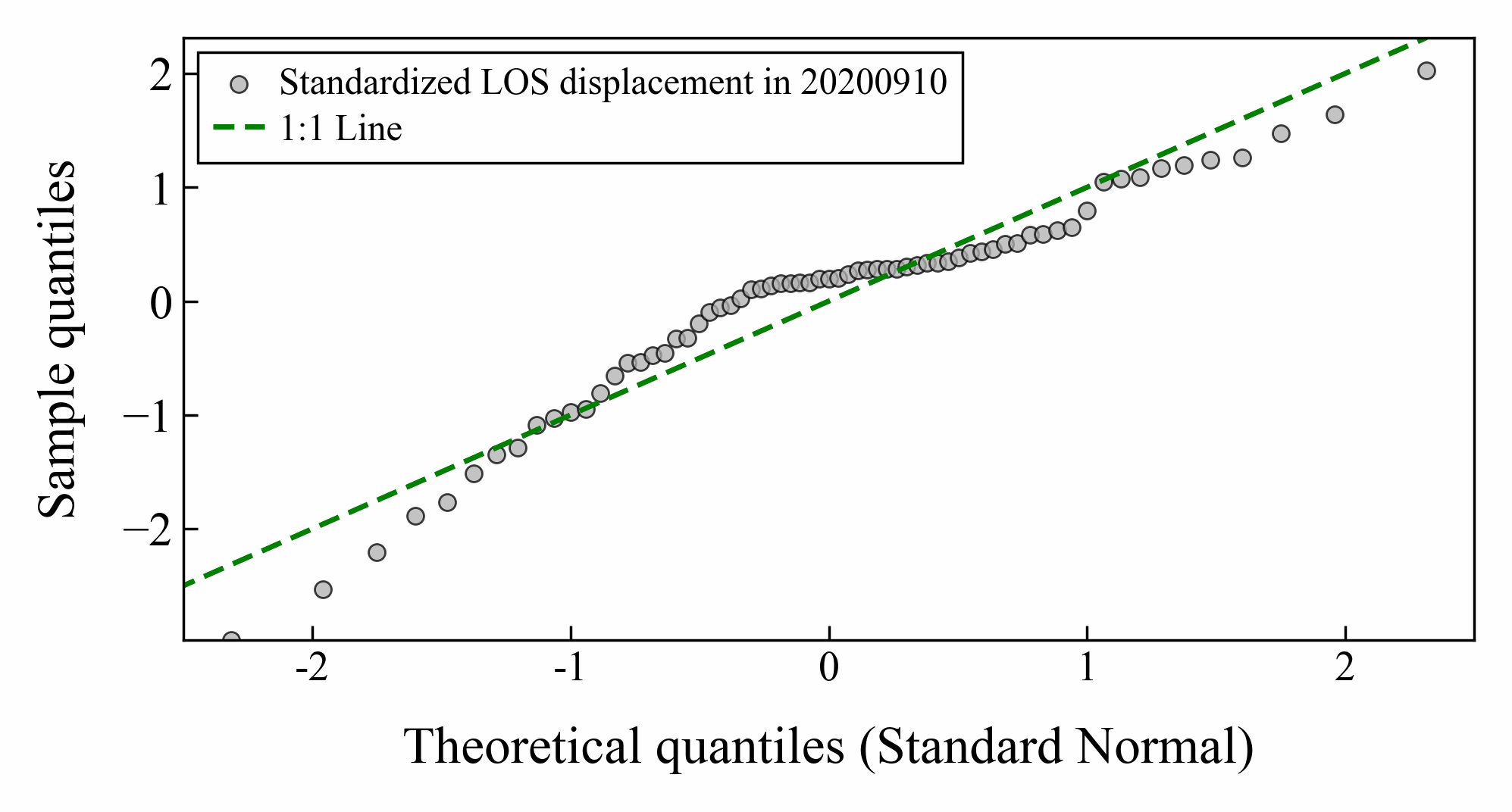}
        \caption{}
    \end{subfigure}
    \caption{Validation of the Gaussian assumption for the standardised LOS displacement of the ascending track. (a, c, e, g, i) probability density histograms of the empirical standardised LOS displacements with the theoretical standard normal curve $N(0,1)$ (solid green); (b, d, f, h, j) quantile-quantile (Q-Q) plots against a standard normal distribution, with the dashed green line representing the 1:1 reference. The rows correspond to the acquisition dates 2016-08-28 (a, b), 2017-08-26 (c, d), 2018-08-24 (e, f), 2019-09-02 (g, h), and 2020-09-10 (i, j).}
    \label{fig:gaussian_val_asc}
\end{figure*}

\begin{figure*}
    \centering
    \begin{subfigure}{0.42\textwidth}
        \includegraphics[width=\linewidth]{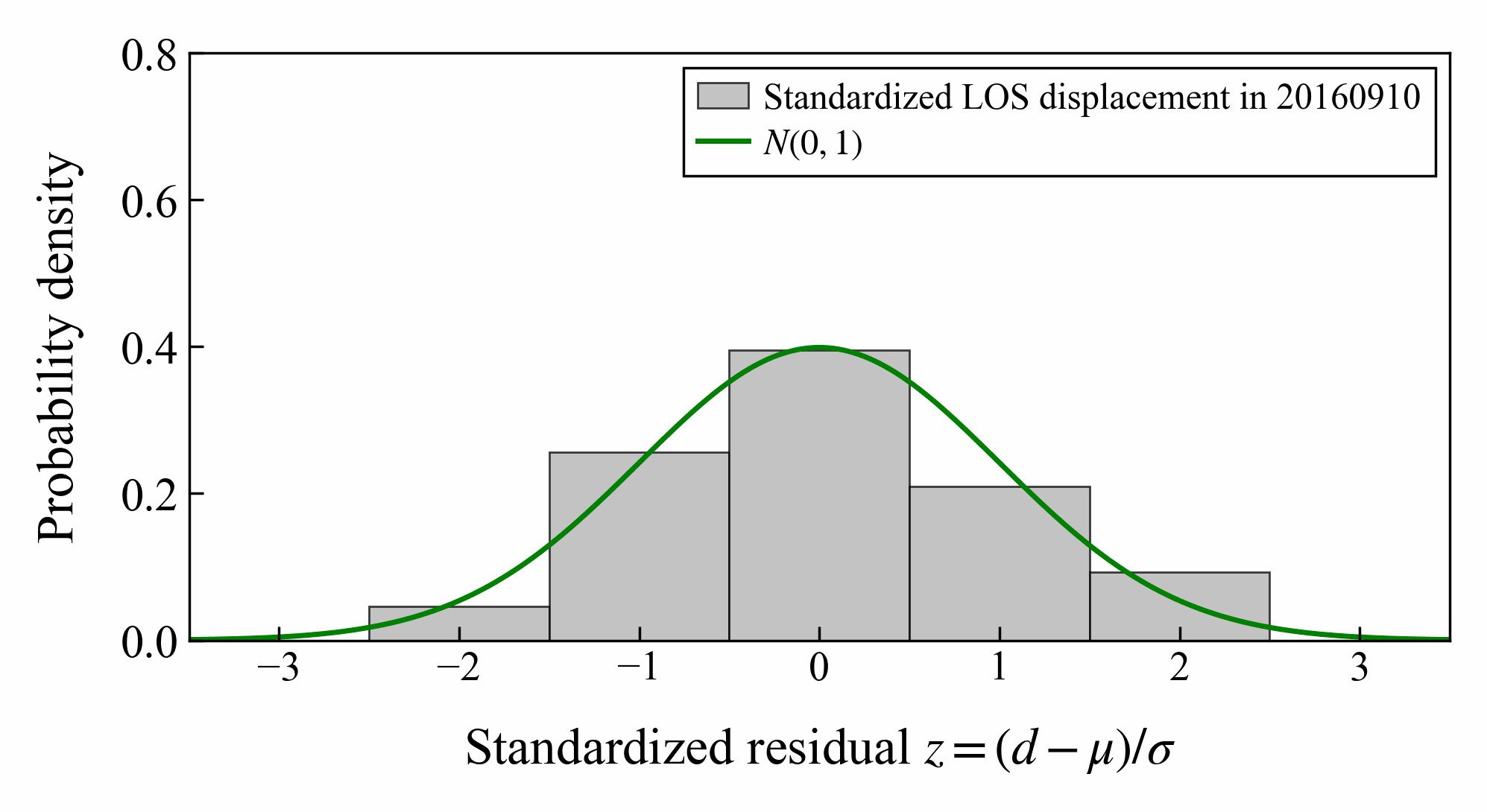}
        \caption{}
    \end{subfigure}
    \hfill
    \begin{subfigure}{0.42\textwidth}
        \includegraphics[width=\linewidth]{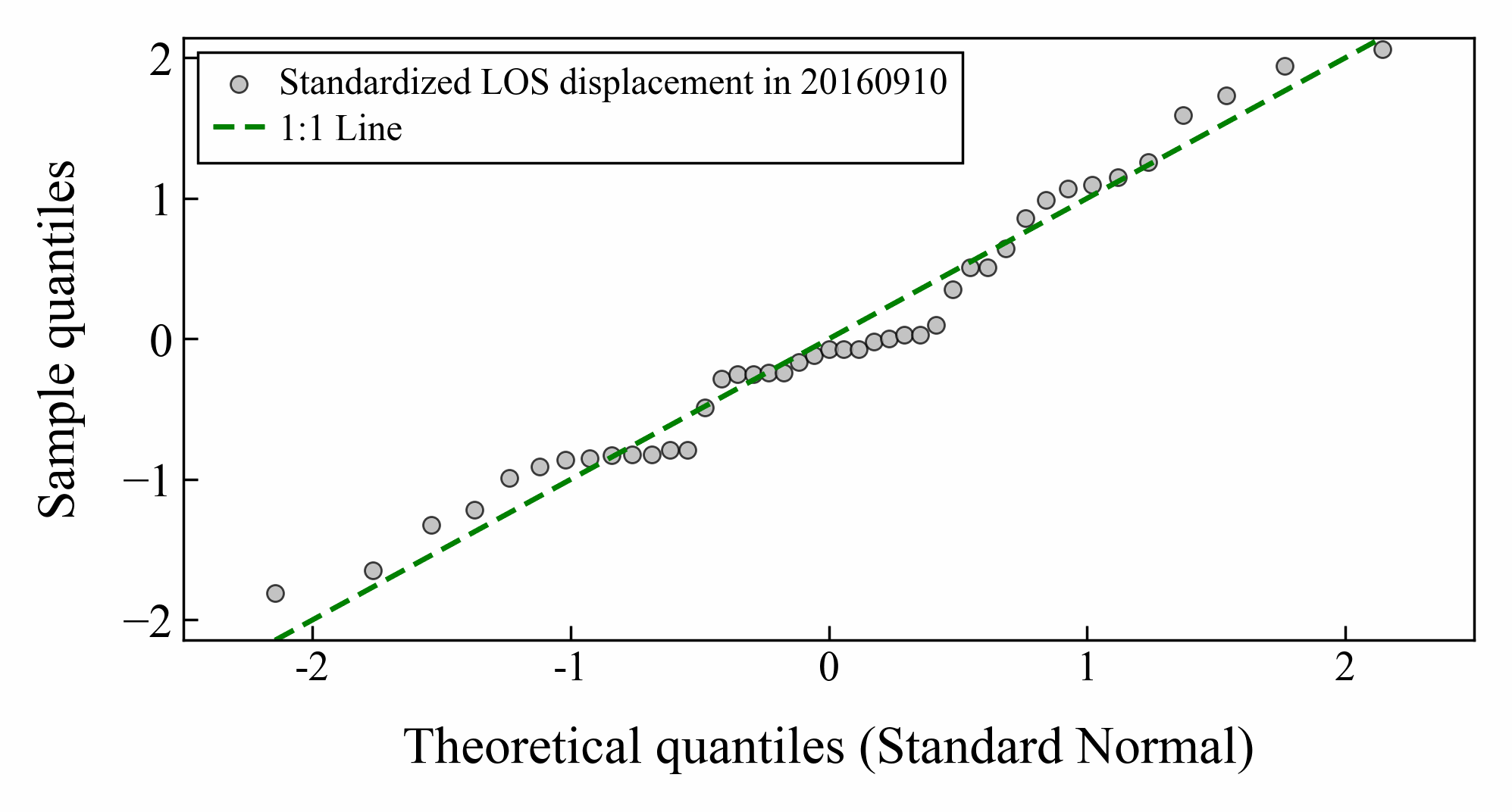}
        \caption{}
    \end{subfigure}

    \begin{subfigure}{0.42\textwidth}
        \includegraphics[width=\linewidth]{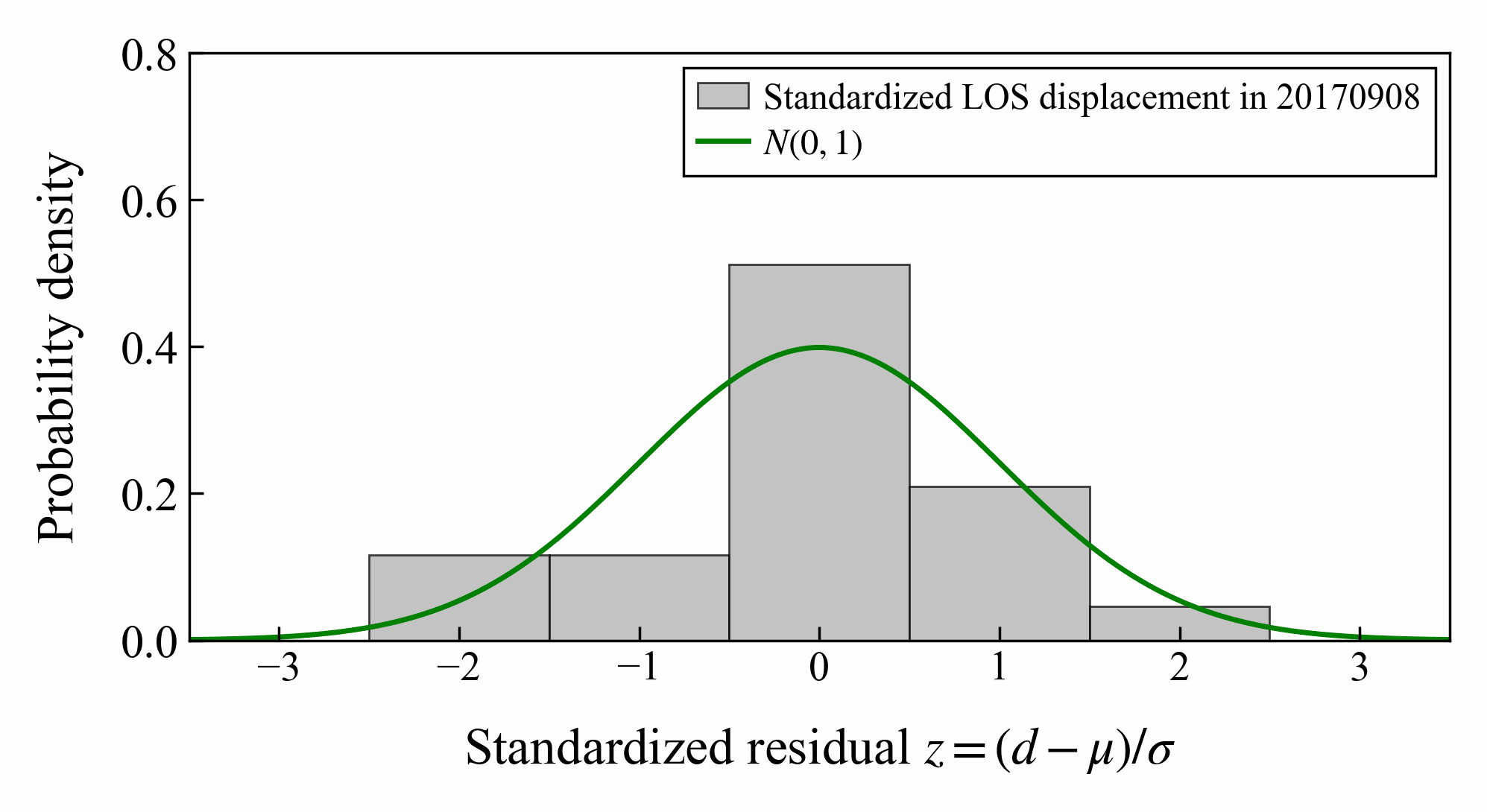}
        \caption{}
    \end{subfigure}
    \hfill
    \begin{subfigure}{0.42\textwidth}
        \includegraphics[width=\linewidth]{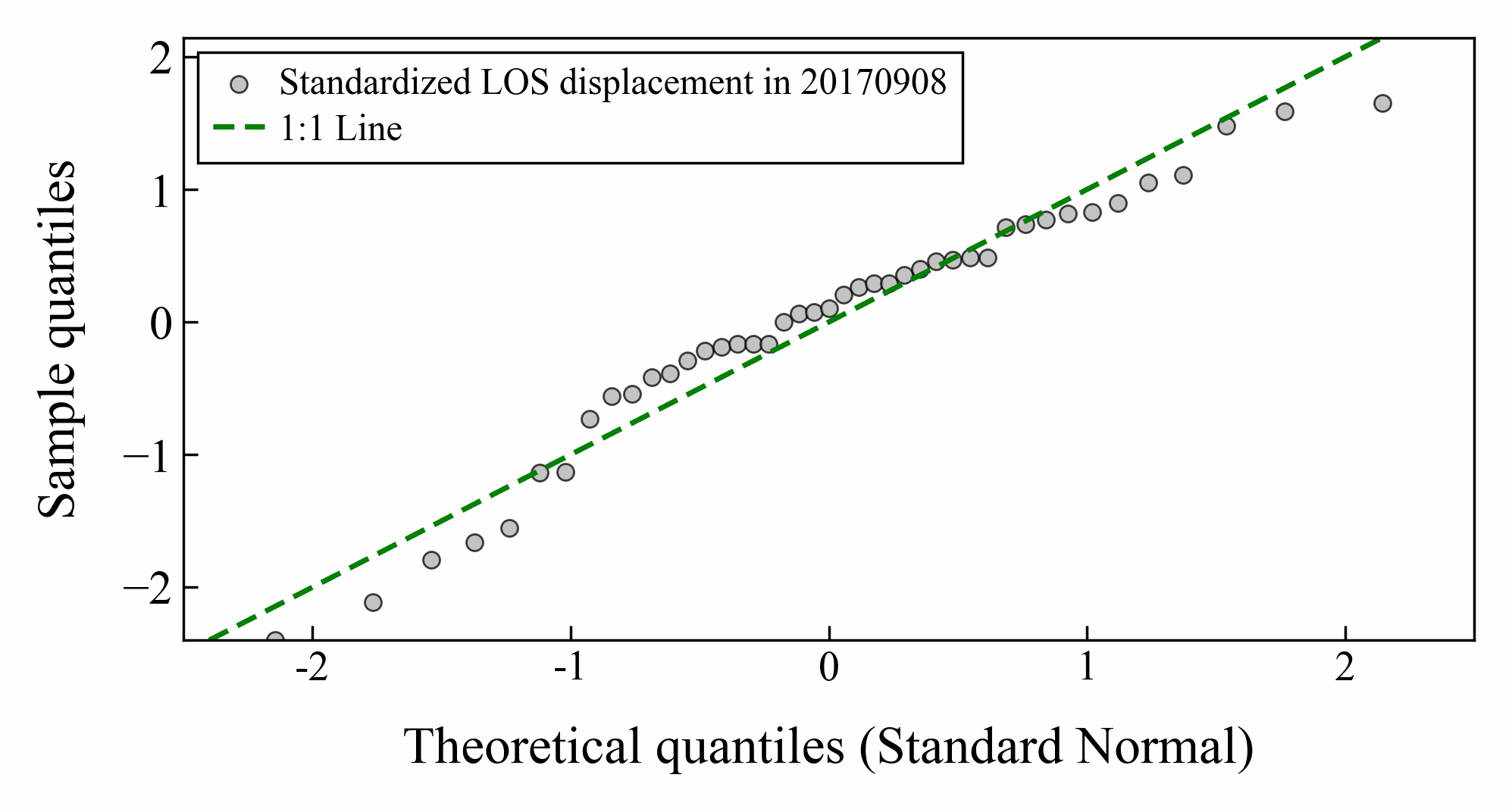}
        \caption{}
    \end{subfigure}

    \begin{subfigure}{0.42\textwidth}
        \includegraphics[width=\linewidth]{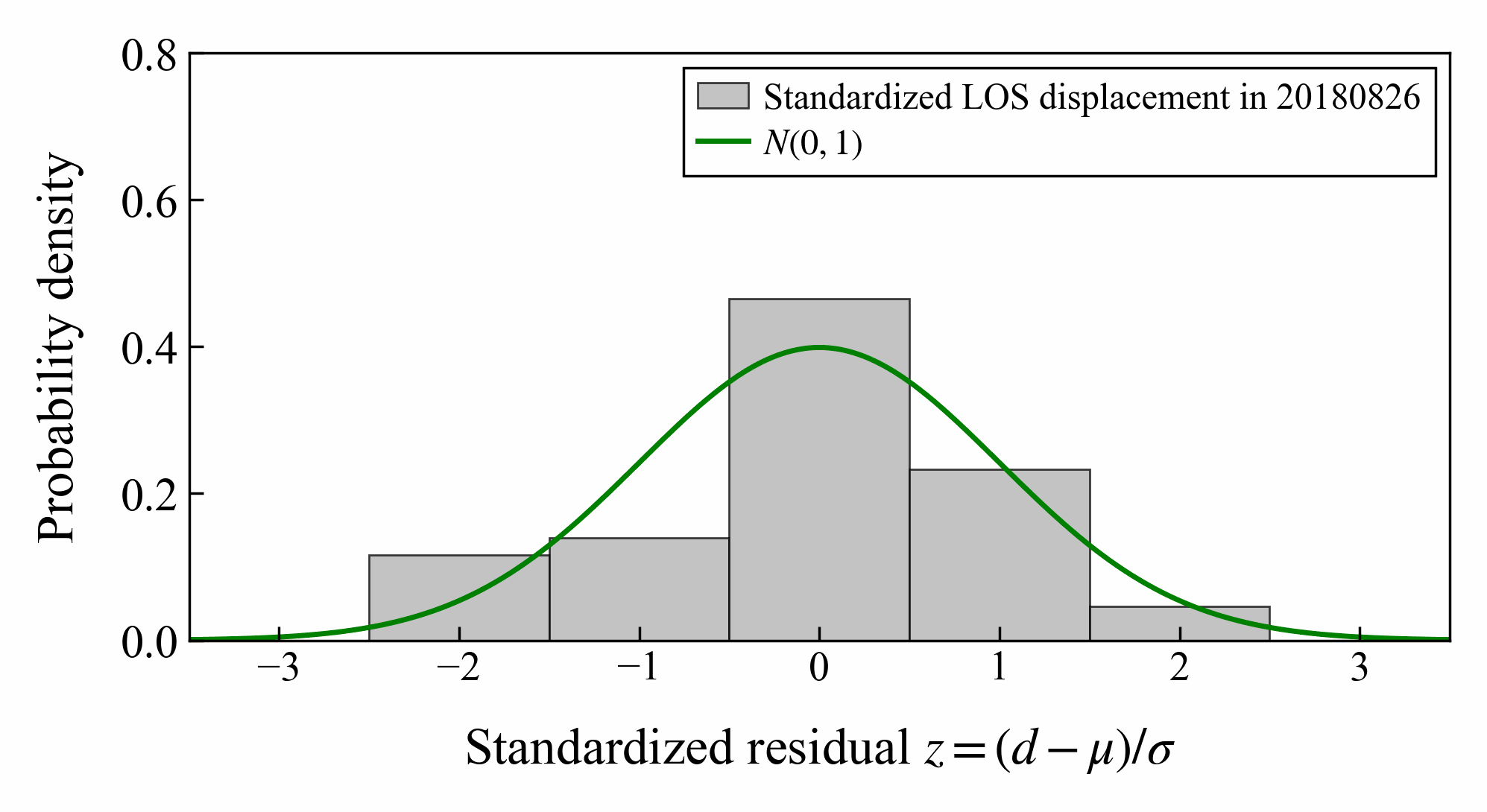}
        \caption{}
    \end{subfigure}
    \hfill
    \begin{subfigure}{0.42\textwidth}
        \includegraphics[width=\linewidth]{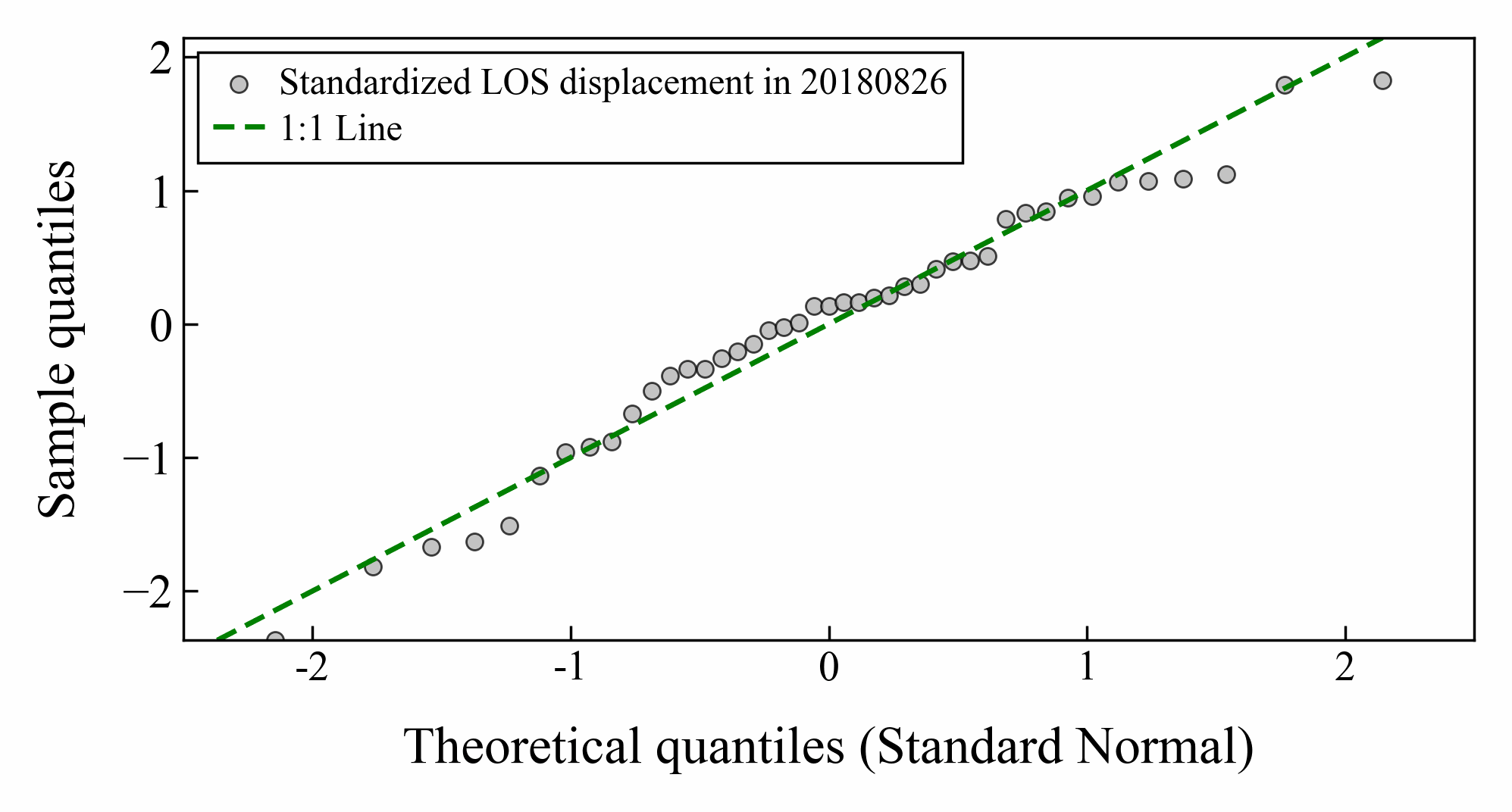}
        \caption{}
    \end{subfigure}

    \begin{subfigure}{0.42\textwidth}
        \includegraphics[width=\linewidth]{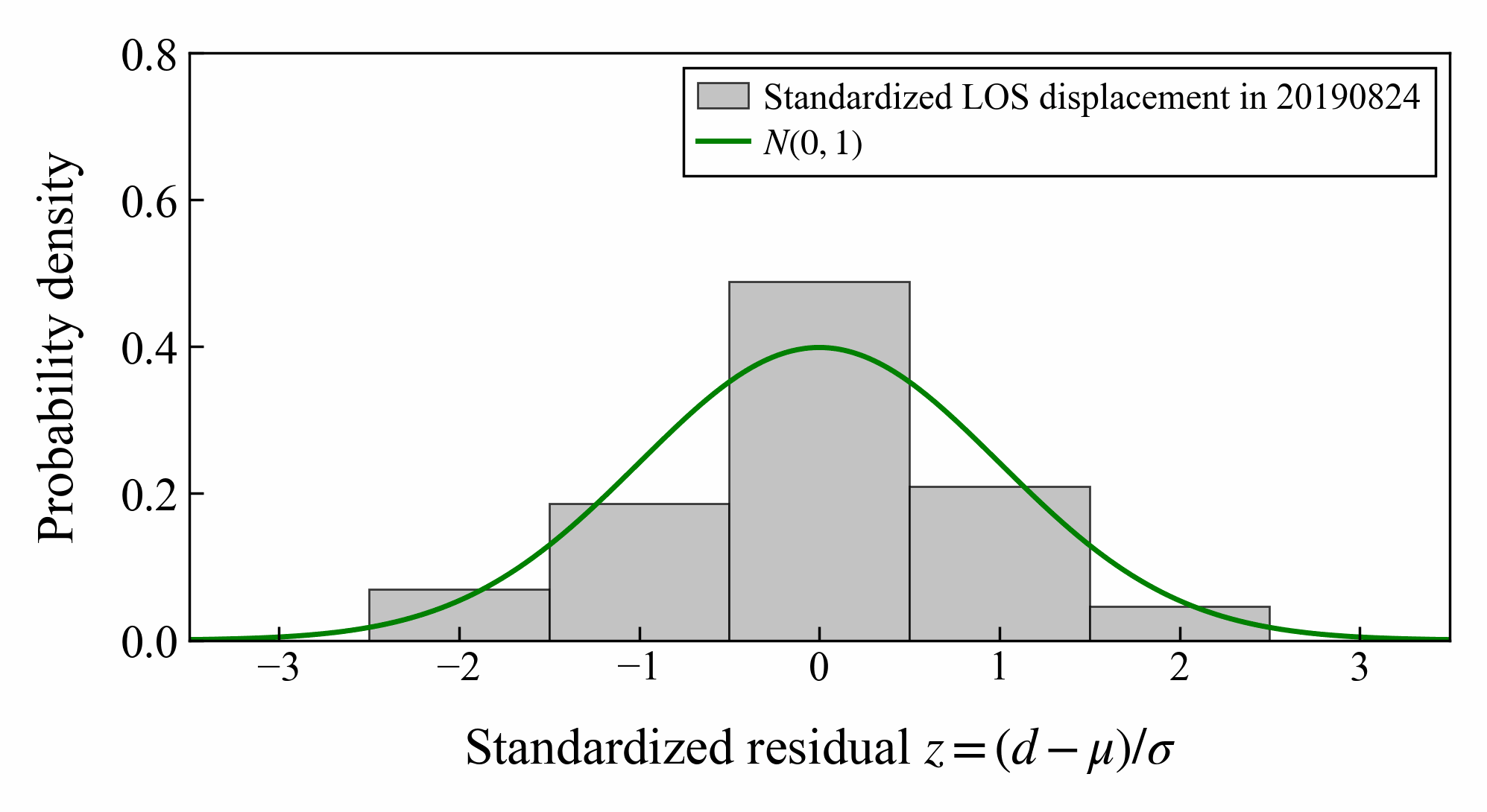}
        \caption{}
    \end{subfigure}
    \hfill
    \begin{subfigure}{0.42\textwidth}
        \includegraphics[width=\linewidth]{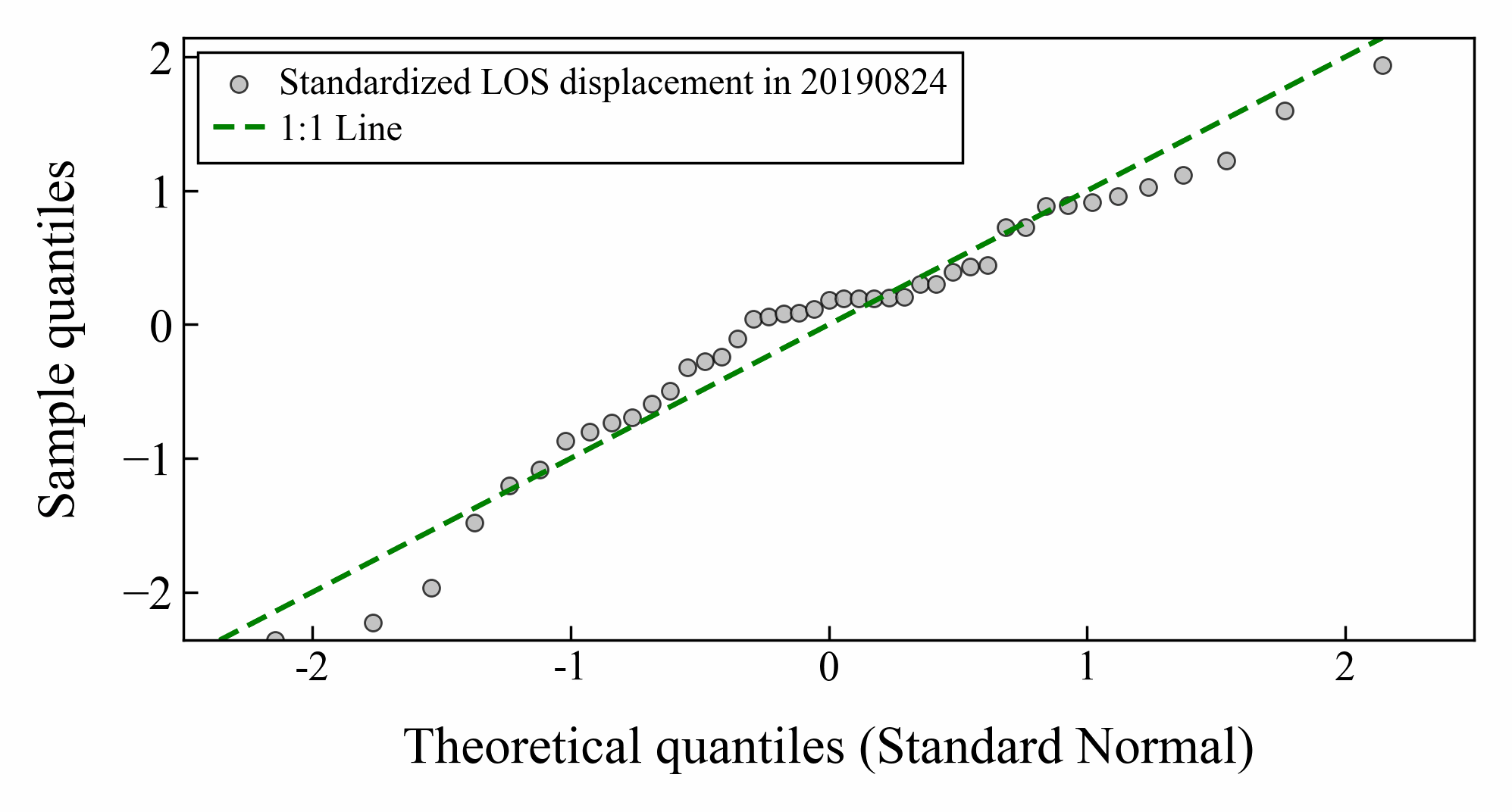}
        \caption{}
    \end{subfigure}

    \begin{subfigure}{0.42\textwidth}
        \includegraphics[width=\linewidth]{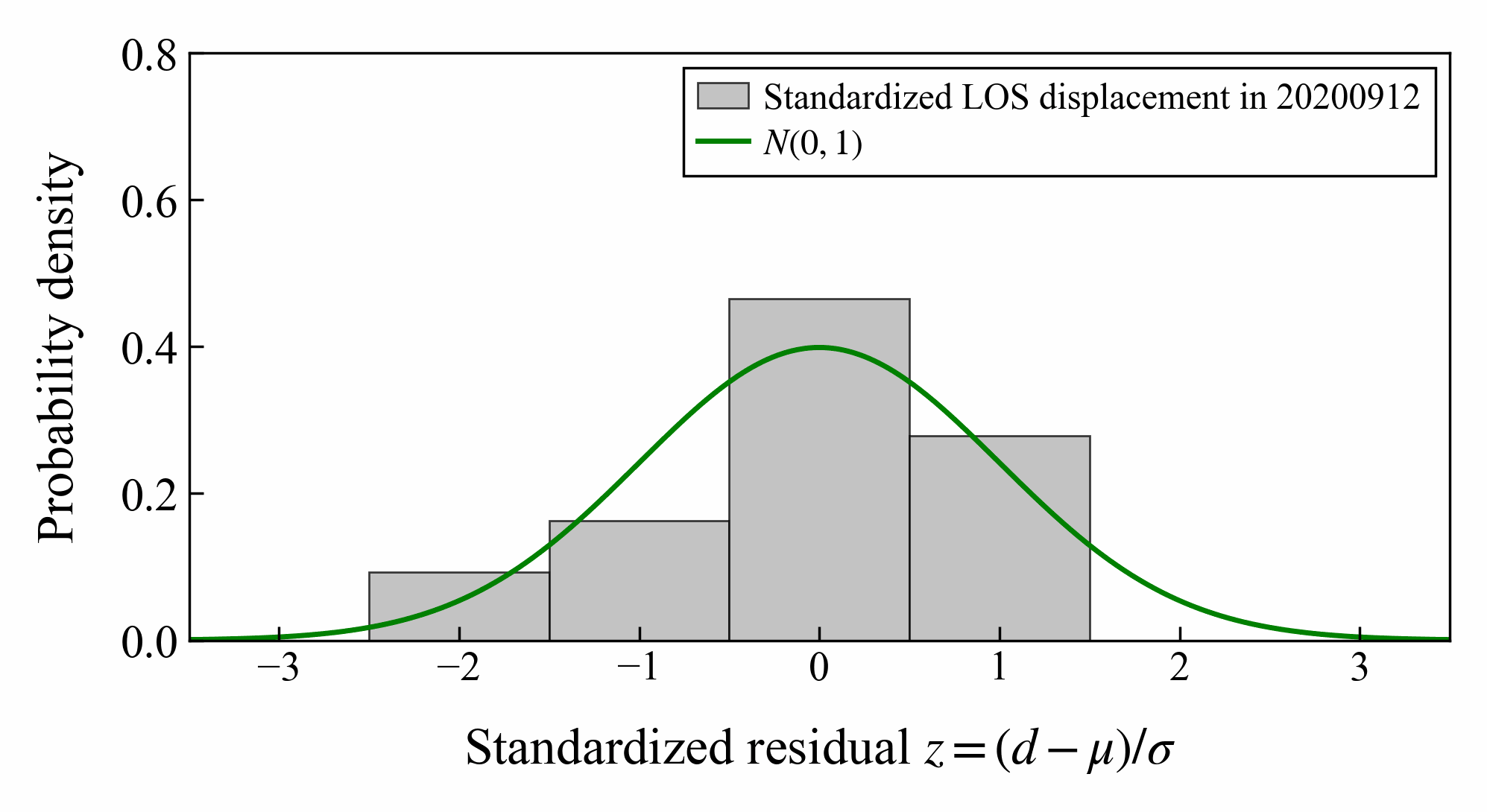}
        \caption{}
    \end{subfigure}
    \hfill
    \begin{subfigure}{0.42\textwidth}
        \includegraphics[width=\linewidth]{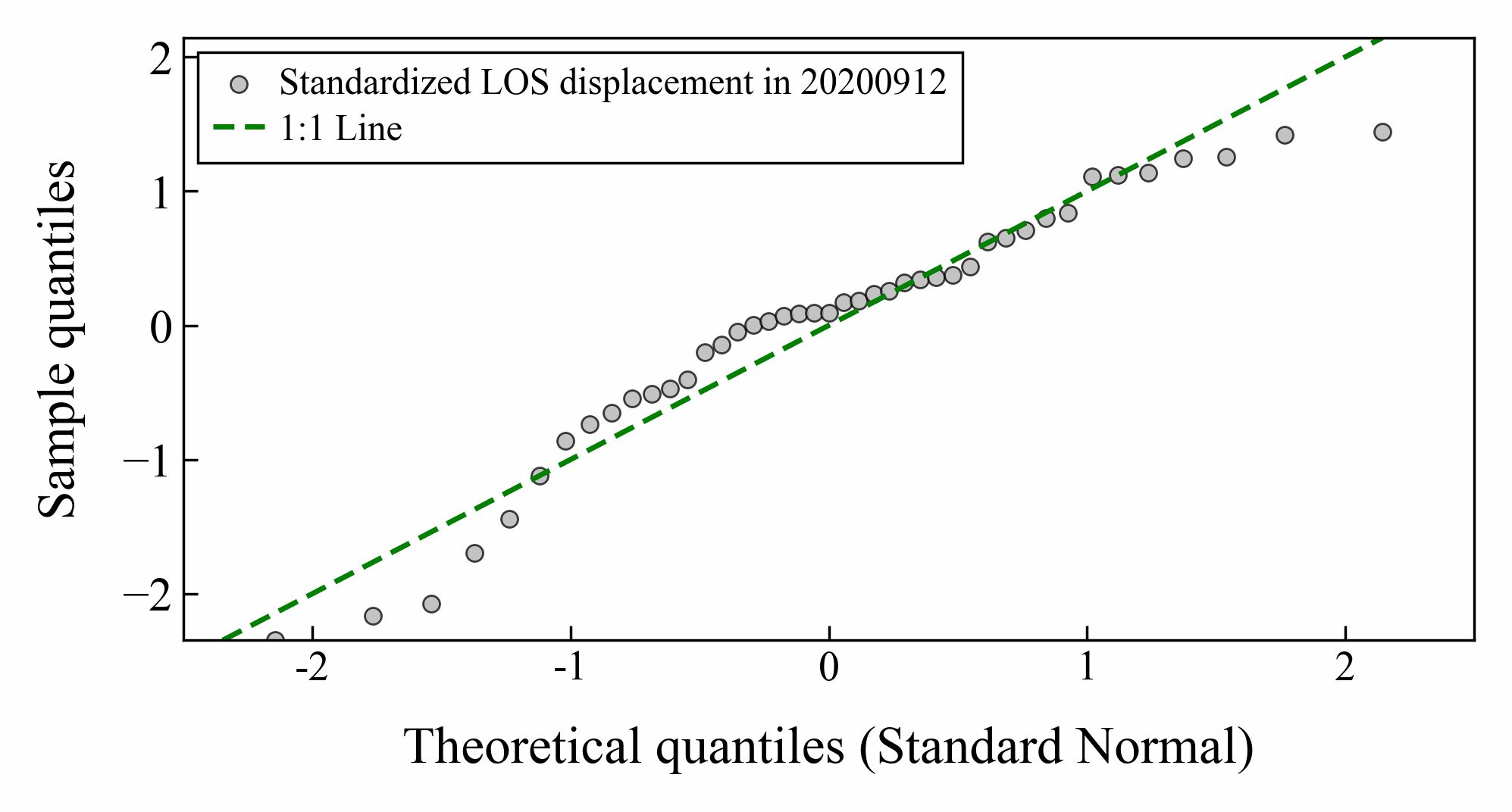}
        \caption{}
    \end{subfigure}
    \caption{Validation of the Gaussian assumption for the standardised LOS displacement of the descending track: (a, c, e, g, I) probability density histograms of the empirical standardised LOS displacements with the theoretical standard normal curve $N(0,1)$ (solid green); (b, d, f, h, j) quantile-quantile (Q-Q) plots against a standard normal distribution, with the dashed green line representing the 1:1 reference. The rows correspond to the acquisition dates 2016-09-10 (a, b), 2017-09-08 (c, d), 2018-08-26 (e, f), 2019-08-24 (g, h), and 2020-09-12 (i, j).}
    \label{fig:gaussian_val_des}
\end{figure*}

\revise{Histograms and corresponding Quantile-Quantile (Q-Q) plots were generated for each acquisition date. Figures \ref{fig:gaussian_val_asc} and \ref{fig:gaussian_val_des} present representative results for selected dates (chosen as the closest dates to September 1st of each year).}
\revise{The histograms (left column) illustrate the empirical distribution of the standardised LOS displacements. Across the selected dates, the distributions exhibit a symmetric bell-shaped pattern centred around zero, consistent with the theoretical standard normal distribution $\mathcal{N}(0,1)$.}
\revise{To further assess normality, Q-Q plots are presented (right column). A Q-Q plot compares the empirical quantiles of the sample with the theoretical quantiles of a standard normal distribution. If the data follow a Gaussian distribution, the points are expected to align along the 1:1 reference line. As observed, the sample quantiles closely follow the 1:1 line over the central region, with only minor deviations at the extreme tails. These deviations are limited and do not indicate significant non-Gaussian behaviour.}
\revise{Overall, the consistent agreement observed in both the histograms and Q-Q plots supports the assumption that the LOS displacement can be reasonably approximated as Gaussian at each acquisition time.}

\addtolength{\footskip}{1cm}

\addtolength{\footskip}{1cm}



\revise{Potential spatially correlated errors, such as residual atmospheric phase in InSAR measurements and failure mechanisms involving local structural deformation gradients, may introduce dependencies among neighbouring PS points. These effects are not explicitly considered in the current independent Gaussian assumption and may result in an underestimation of the uncertainty. Further investigation is required to quantify and incorporate such dependencies.}

\section{Sensitivity analysis of the projection assumption in local coordinates}

\revise{For the projection from LOS displacement to the local coordinate system, it was assumed that the contribution of the displacement component along the structural axis (longitudinal or transversal), which was more closely aligned with the satellite track, to the LOS displacement was negligible, following \citep{macchiaruloStructuralEvaluationUrban2024}.}
\revise{To evaluate the validity of this assumption, a sensitivity analysis was conducted for all three case studies. The neglected displacement component (i.e., transversal displacement for Grimburgwal, and longitudinal displacement for Entrepotdok and Houtmankade East) was assigned non-zero values of 1 mm, 2 mm, 5 mm, and 10 mm.}

\begin{figure*}
    \centering
    \begin{subfigure}{0.4\linewidth}
        \centering
        \includegraphics[width=\linewidth]{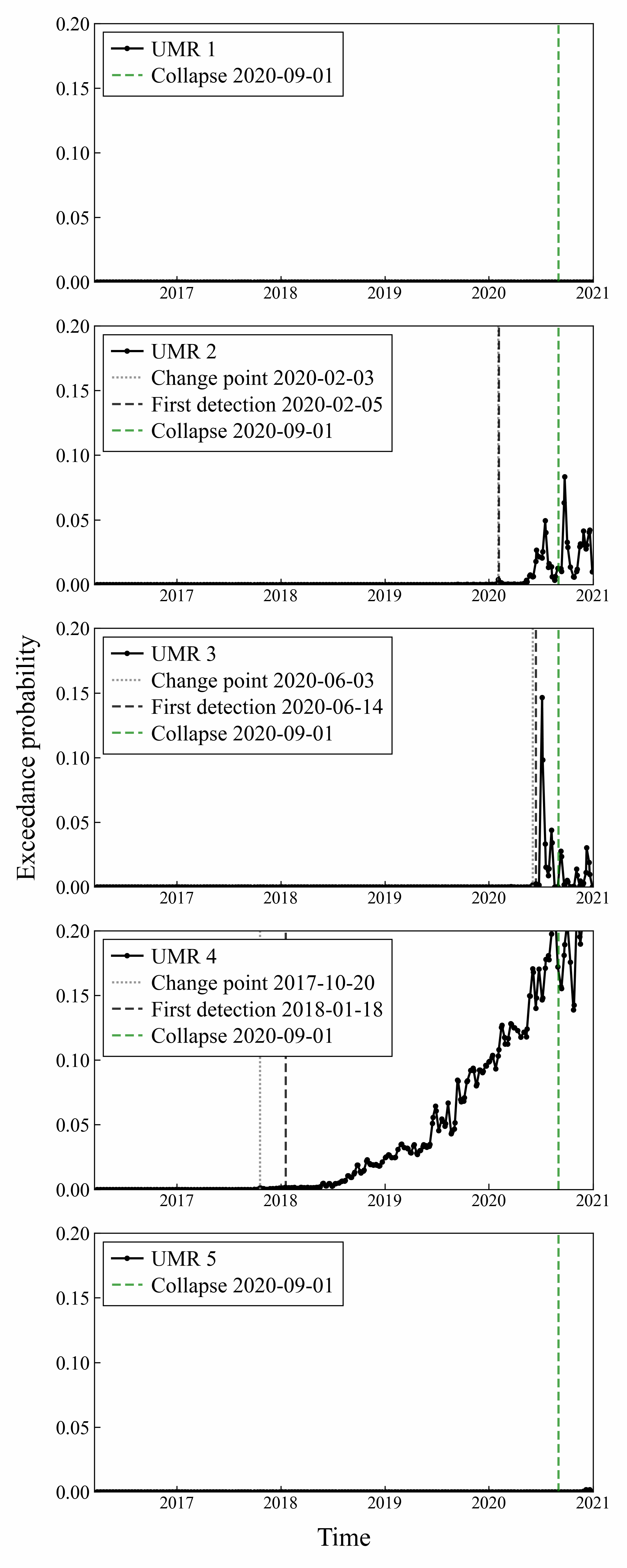}
        \caption{}
        \label{fig:sensi_10_grim}
    \end{subfigure}
    \hfill
    \begin{subfigure}{0.4\linewidth}
        \centering
        \includegraphics[width=\linewidth]{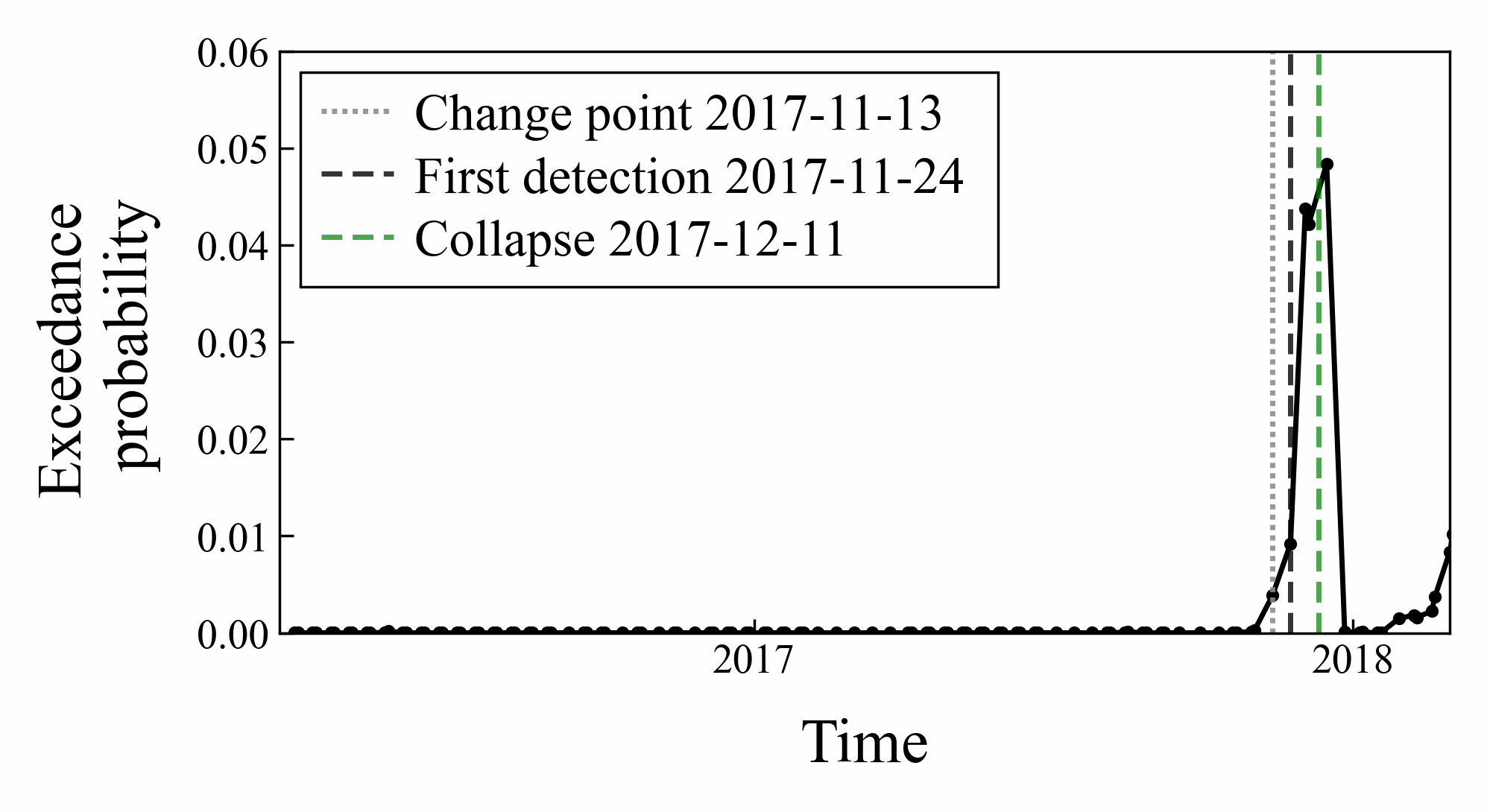}
        \caption{}
        \label{fig:sensi_10_entre}
    \end{subfigure}
    \caption{\revisethree{The probability of exceedance and the issued date of warning after adding 10 mm neglected displacement component: (a) Grimburgwal and (b) Entrepotdok.}}
    \label{fig:sensi_10_ge}
\end{figure*}

\begin{figure*}
    \centering
    \begin{subfigure}{0.4\linewidth}
        \centering
        \includegraphics[width=\linewidth]{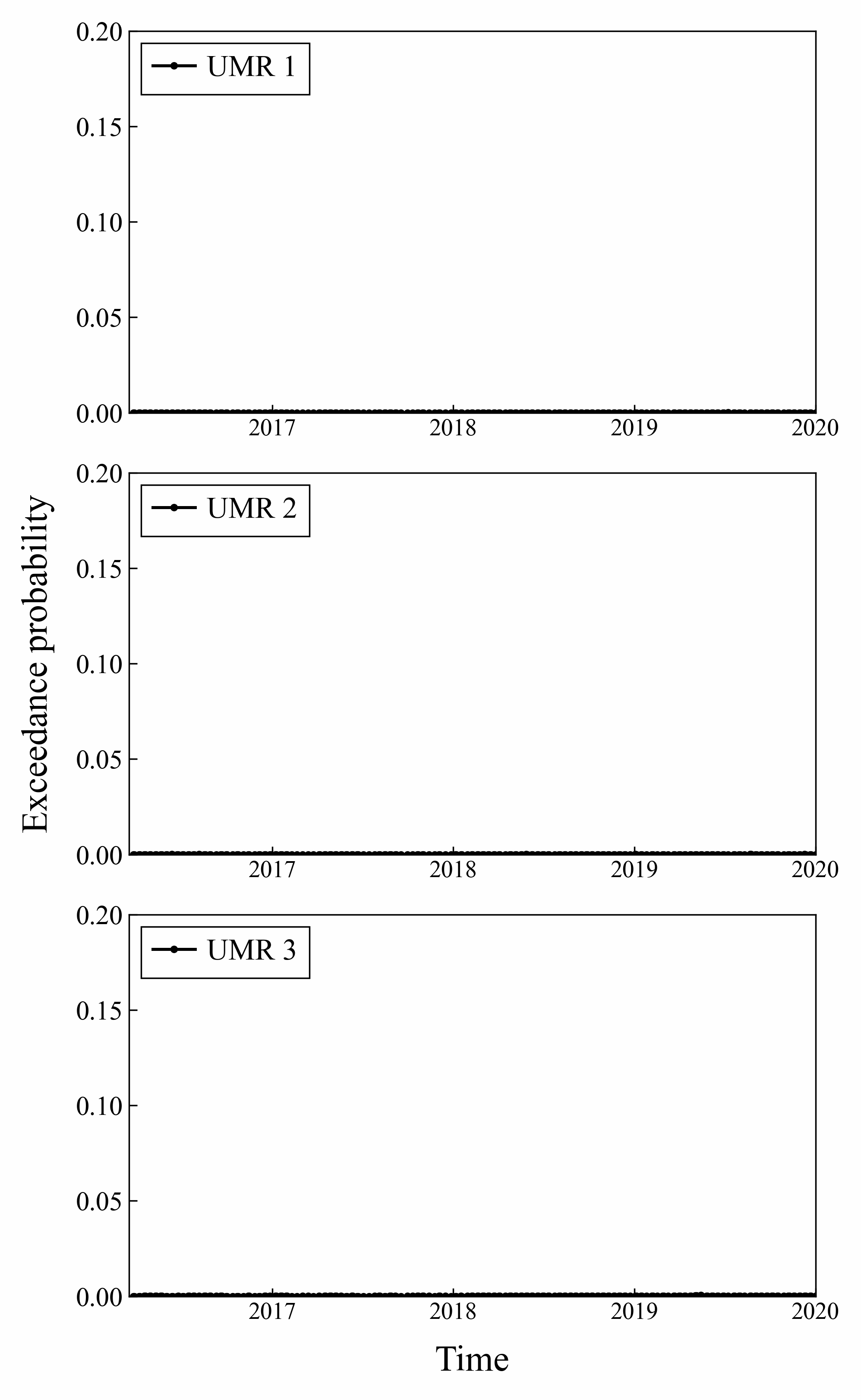}
        \caption{}
        \label{fig:sensi_10_hout_u}
    \end{subfigure}
    \hfill
    \begin{subfigure}{0.4\linewidth}
        \centering
        \includegraphics[width=\linewidth]{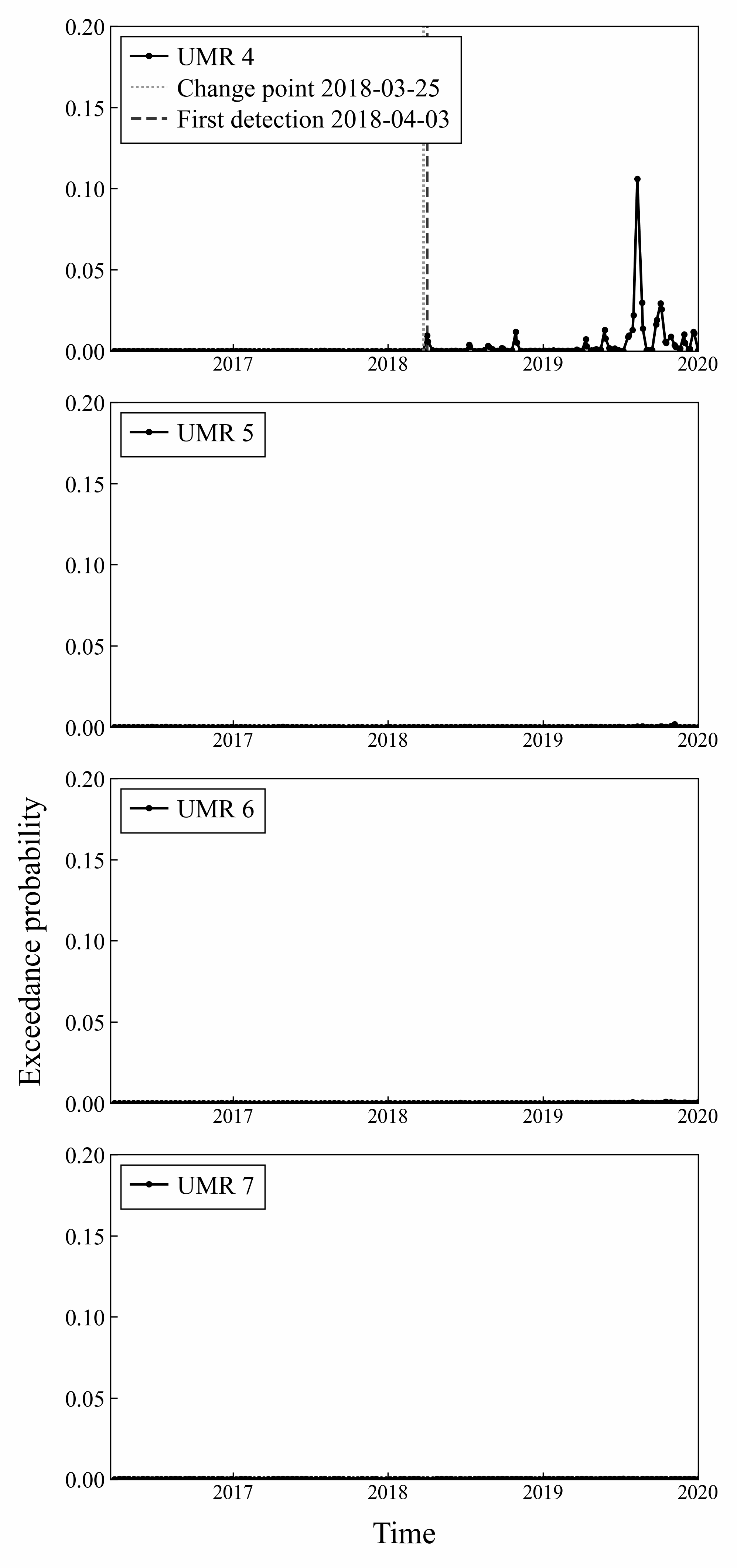}
        \caption{}
        \label{fig:sensi_10_hout_d}
    \end{subfigure}
    \caption{\revisethree{The probability of exceedance and the issued date of warning after adding 10 mm neglected displacement component for Houtmankade East: (a) safe section and (b) replaced section.}}
    \label{fig:sensi_10_ho}
\end{figure*}

\revisetwo{For each scenario, the LOS contribution of the assumed neglected component was first calculated and removed from the original LOS displacement. The corrected LOS displacement was then used to recompute the vertical displacement, exceedance probability, and warning results. The results show that the neglected displacement component had minimal influence. Even in the extreme case of 10 mm (Figures \ref{fig:sensi_10_ge} and \ref{fig:sensi_10_ho}), the issued warning dates remained mostly consistent with the original results. A minor shift was observed for the Grimburgwal case. However, this change did not alter the overall warning performance or lead to any false alarms. Therefore, the results suggest that the assumed neglected displacement component does not have a substantial impact on the warning assessment.}

\section{Sensitivity analysis of temporal interpolation}
\label{sec:app_d}

\revise{To evaluate the influence of temporal interpolation on the results, a sensitivity analysis was conducted using the physically based displacement evolution obtained from the FE model presented in Figure \ref{fig:marnix_a}.}
\revise{The original FE simulation provided a continuous displacement evolution consisting of 99 calculation steps. To simulate the effect of different temporal sampling intervals, 19 scenarios were created by retaining only every $n$-th displacement step (where $n$ ranges from 2 to 20). For each scenario, the missing displacement values were then reconstructed using linear interpolation, representing the temporal resampling procedure applied when combining ascending and descending MT-InSAR observations.}
\revisetwo{Subsequently, these interpolated displacement time series were used to calculate the corresponding exceedance probability and detect the change point using the PELT algorithm. The detected warning dates were then compared with those obtained from the original continuous displacement evolution.}

\begin{figure}
    \centering
    \includegraphics[width=0.5\linewidth]{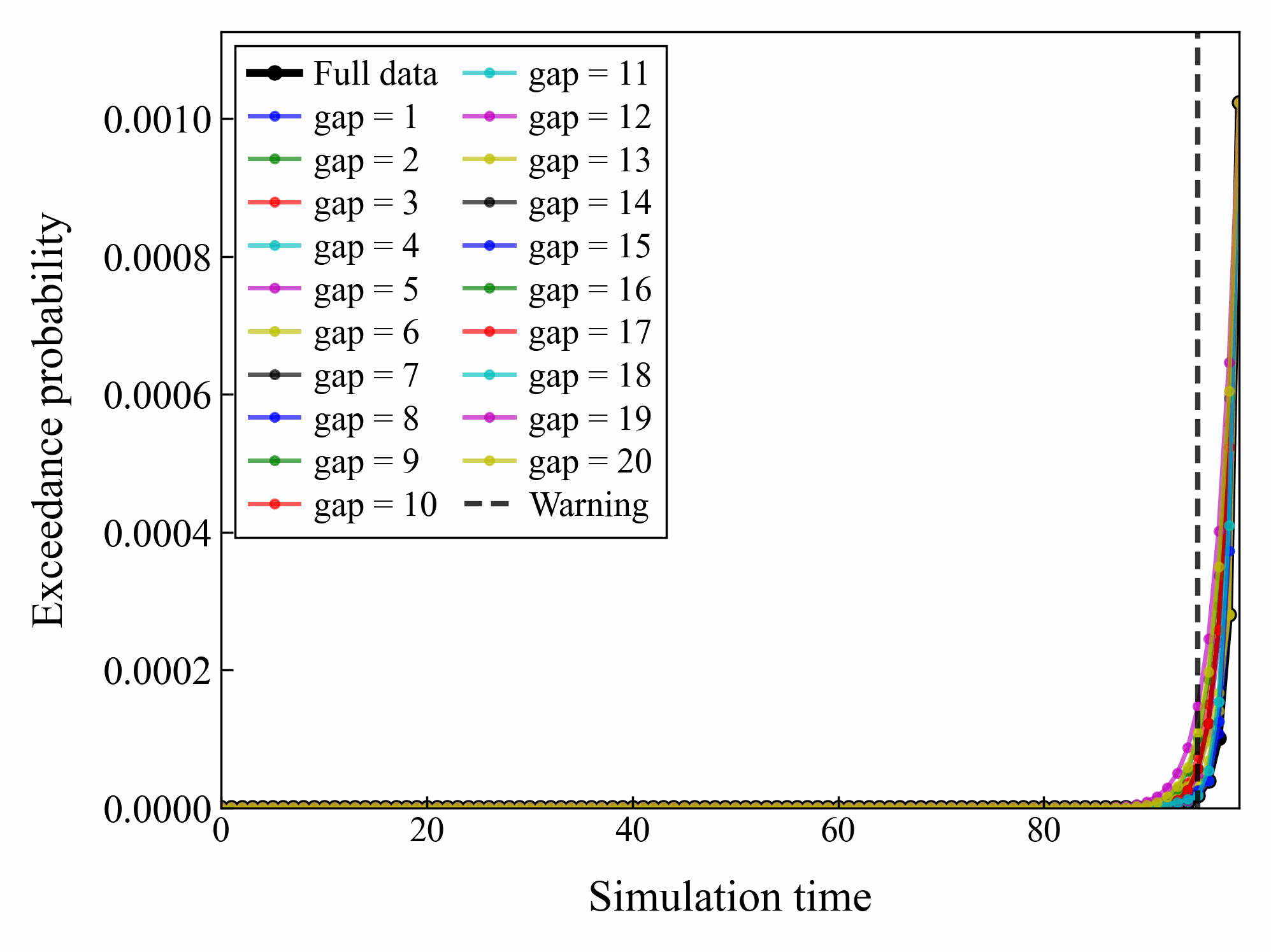}
    \caption{\revisetwo{Influence of temporal sampling intervals on exceedance probability estimation and warning issued using the PELT algorithm}}
    \label{fig:pf_sensi_inter}
\end{figure}

\revise{The results (Figure \ref{fig:pf_sensi_inter}) show that the temporal interpolation has a limited influence on the warning results. Even under the most severe sampling scenario, where only every twentieth displacement step was retained, the issued warning date remained identical to those obtained from the original displacement evolution.}
\revise{It should be noted that retaining only every twentieth simulation step represents an extremely conservative temporal sampling scenario. Under this extreme case, the displacement difference between two consecutive retained observations reaches approximately 15.7 mm. TerraSAR-X operates in the X-band with a wavelength $\lambda$ of approximately 31 mm \citep{werninghausTerraSARXMissionSystem2010}; therefore, this displacement difference far exceeds the theoretical maximum detectable limit of $\lambda/4$ (7.75 mm) \citep{macchiaruloMultiTemporalInSARTransport2023}. Consequently, this scenario represents a more severe condition than normally encountered in reliable MT-InSAR measurements. The results indicate that, within realistic acquisition intervals, the temporal interpolation required to combine ascending and descending observations does not significantly affect the anomaly warning capability of the proposed framework.}

\section{Sensitivity analysis of the PELT penalty}
\label{app:appe}
\revisethree{The PELT penalty $\zeta$ controls the trade-off between detecting changes in the exceedance probability time series and avoiding excessive segmentation \citep{killickOptimalDetectionChangepoints2012}. The penalty was empirically selected as $1\times10^{-5}$ considering the magnitude and variability of the exceedance probability time series. To test the influence of this choice, a sensitivity analysis was performed using the three case studies. The penalty was varied according to:}

\revisethree{
\begin{equation}
    \zeta' = f_\zeta ~ \zeta
\end{equation}
where $f_\zeta$ is the factor, ranging from $0.1$ to $10$. All other PELT parameters were kept constant. For each penalty, the first detection date was determined using the same procedure. Table \ref{tab:peltpenalty} presents the results for six selected penalty factors.}

\begin{table*}
    \centering
    \begin{tabular}{c|c|c|c|c|c|c}
        \hline
        Factor $f_\zeta$ & 0.10 & 0.50 & 1.00 & 1.50 & 2.00 & 10.00\\
        \hline
        Grimburgwal & 2017-11-02 & 2018-02-07 & 2018-04-03 & 2018-05-19 & 2018-05-30 & 2018-08-13\\
        \hline
        Entrepotdok & 2017-11-13 & 2017-11-24 & 2017-11-24 & 2017-11-24 & 2017-12-03 & 2017-12-03\\
        \hline
        Houtmankade East & 2018-04-03 & 2018-04-03 & 2018-04-03 & 2018-04-03 & 2018-04-03 & 2018-04-03\\
        \hline
    \end{tabular}
    \caption{The warning date of sensitivity analysis of the PELT penalty}
    \label{tab:peltpenalty}
\end{table*}

\revisethree{The results show that the anomalies were consistently identified across the investigated penalty range, although the first detection dates varied, particularly for Grimburgwal. In general, a larger penalty resulted in a later first detection date. Grimburgwal exhibited the greatest sensitivity to the penalty; nevertheless, even when the penalty was increased tenfold, the anomaly was still detected more than two years before the collapse.}

\section{Sensitivity analysis of the maximum vertical allowable displacement}
\label{app:appf}
\revisethree{In this study, the maximum vertical allowable displacement was set to 1\% of the retaining height following \cite{bettsPortBotanyExpansion2011}. To investigate the influence of this choice, a sensitivity analysis was performed using the three case studies. The maximum vertical allowable displacement was varied as follows:}

\revisethree{
\begin{equation}
    d'_{V,max} = f_V ~ d_{V,max}
\end{equation}
where $f_V$ is the factor, ranging from $0.75$ to $1.25$. For each allowable vertical displacement, the first detection date was determined using the same procedure. Table \ref{tab:allow} presents the results.}

\begin{table*}
    \centering
    \begin{tabular}{c|c|c|c|c|c}
        \hline
        Factor $f_V$ & 0.75 & 0.90 & 1.00 & 1.10 & 1.25\\
        \hline
        Grimburgwal & 2017-09-30 & 2017-10-22 & 2018-04-03 & 2018-08-13 & 2018-10-29\\
        \hline
        Entrepotdok & 2017-08-17 & 2017-11-13 & 2017-11-24 & 2017-12-03 & 2017-12-05\\
        \hline
        Houtmankade East & 2016-07-26 & 2018-04-03 & 2018-04-03 & 2018-04-03 & 2019-08-11\\
        \hline
    \end{tabular}
    \caption{The warning date of sensitivity analysis of the maximum vertical allowable displacement}
    \label{tab:allow}
\end{table*}

\revisethree{The results show that increasing the maximum allowable vertical displacement leads to later warning dates, while decreasing it results in earlier warnings. The sensitivity varies among the three cases, with shifts in warning dates observed for some threshold choices. Nevertheless, warnings were identified before collapses. Although lower maximum vertical displacements provide earlier warnings, they may also lead to unnecessarily early interventions.}

\end{document}